\documentclass[reprint,showpacs,superscriptaddress,english,twocolumn,notitlepage,longbibliography,nofootinbib]{revtex4-1}
\usepackage[utf8]{inputenc}
\usepackage{color}
\usepackage[dvipsnames]{xcolor}
\usepackage{array}
\usepackage{verbatim}
\usepackage{multirow}
\usepackage{amsmath}
\usepackage{amssymb}
\usepackage{dsfont}
\usepackage{graphicx}
\usepackage{esint}
\usepackage[unicode=true,
 bookmarks=true,bookmarksnumbered=true,bookmarksopen=false,
 breaklinks=true,pdfborder={0 0 0},backref=false,colorlinks=true]
 {hyperref}
\usepackage{enumitem}
\usepackage{titlesec}
\usepackage{ntheorem}
\usepackage{cleveref}
\usepackage{longtable}
\usepackage{bbm}
\usepackage{pifont}
\newcommand{\xmark}{\ding{55}}%
\usepackage[normalem]{ulem}

\hypersetup{
citecolor={blue},
urlcolor={magenta}
}

\newcommand{\PreserveBackslash}[1]{\let\temp=\\#1\let\\=\temp}
\newcolumntype{C}[1]{>{\PreserveBackslash\centering}p{#1}}
\newcolumntype{R}[1]{>{\PreserveBackslash\raggedleft}p{#1}}
\newcolumntype{L}[1]{>{\PreserveBackslash\raggedright}p{#1}}

\usepackage{cleveref}
\crefname{appsec}{Appendix}{Appendices}
\crefname{equation}{Eq.}{Eqs.}
\crefname{figure}{Fig.}{Figs.}
\crefname{table}{Table}{Tables}
\crefname{section}{Section}{Sections}
\crefname{mythe}{Theorem}{Theorems}
\crefname{mydef}{Definition}{Definitions}
\crefname{appendix}{Appendix}{Appendices}

\renewcommand{\paragraph}[1]{\vspace{0.2cm}{\bf \textit{#1}}}
\def\ie{{\it i.e.},\ }

\def\etc{{\it etc.}\ }
\newcommand{\uemph}[1]{\textit{\uline{#1}}}

\def\BAB#1{}
\def\LE#1{}
\def\SZD#1{}

\newcommand{\mbf}{\mathbf}
\newcommand{\mbb}{\mathbb}
\newcommand{\mcl}{\mathcal}

\newcommand{\mrm}{\mathrm}

\newcommand{\td}{\widetilde}
\newcommand{\ovl}{\overline}

\def\pare#1{\left( #1 \right)}
\def\brak#1{\left[#1\right]}
\def\brace#1{\left\{#1\right\}}
\def\bra#1{\langle #1 |}
\def\ket#1{| #1 \rangle}

\def\nono{\nonumber}

\def\up{\uparrow}
\def\down{\downarrow}
\def\pr{\prime}
\def\prpr{{\prime\prime}}
\def\Tr{\mrm{Tr}}

\def\Re{\mrm{Re}}

\def\mmod{\;\mrm{mod}\;}

\def\RR{\mbf{R}}

\def\rr{\mbf{r}}
\def\tt{\mbf{t}}
\def\aa{\mbf{a}}
\def\bb{\mbf{b}}
\def\kk{\mbf{k}}
\def\TR{\mcl{T}} 

\def\TRS{\mcl{T}}
\def\EBR{{EBR}}
\def\dt{\delta}
\def\hH{\hat{H}}
\def\hV{\hat{V}}

\begin{document}
\title{Real Space Invariants: Twisted Bulk-Boundary Correspondence of Fragile Topology}

\author{Zhi-Da Song}
\affiliation{Department of Physics, Princeton University, Princeton, New Jersey 08544, USA}

\author{Luis Elcoro}
\affiliation{Department of Condensed Matter Physics, University of the Basque Country UPV/EHU, Apartado 644, 48080 Bilbao, Spain}

\author{B. Andrei Bernevig }
\email{bernevig@princeton.edu}
\affiliation{Department of Physics, Princeton University, Princeton, New Jersey 08544, USA}
\affiliation{Physics Department, Freie Universitat Berlin, Arnimallee 14, 14195 Berlin, Germany}
\affiliation{Max Planck Institute of Microstructure Physics, 06120 Halle, Germany}

\date{\today}

\begin{abstract}
In this paper, we propose a new type of bulk-boundary correspondence as a generic approach to theoretically and experimentally detect fragile topological states. When the fragile phase can be written as a difference of a trivial atomic insulator and the so-called obstructed atomic insulator, the gap between the fragile phase and other bands must close under a specific novel twist of the boundary condition of the system.
We explicitly work out \uemph{all} the twisted boundary conditions (TBC) that can detect all the 2D fragile phases implied by symmetry eigenvalues in \uemph{all} wallpaper groups.
We develop the concept of real space invariants - local good quantum numbers in real space - which fully characterize the eigenvalue fragile phases.
We show that the number of unavoidable level crossings under the twisted boundary condition is completely determined by the real space invariants.
Possible realizations of the TBC of the fragile band in metamaterial systems are  discussed.
\end{abstract}

\maketitle

\section{Introduction}
The theories of topological quantum chemistry \cite{Bradlyn2017,Elcoro2017,Vergniory2017,Jennifer2018}, symmetry-based indicator \cite{Po2017,watanabe_indicator_2018} and band combinatorics \cite{kruthoff_topological_2017} have developed systematic methods for searching new topological materials \cite{Vergniory_nature_2019,zhang_catalogue_2019,Tang_nature_2019, Song_NC_2018,Song_PRX_2018} and have discovered new types of topological states\cite{Benalcazar2017,Benalcazar2017b,Schindler2018,Langbehn2017,Song2017,Fang2017,Ezawa2018,khalaf_SC_2018,Geier2018,Khalaf2018,Song_NC_2018,Song_PRX_2018,po_fragile_2018,cano_topology_2018,Barry2019EBR,manes_fragile_2019,de_paz_engineering_2019,song_fragile_2019,hwang_fragile_2019}. 
Among these breakthroughs, higher-order topological states (HOTI) \cite{Benalcazar2017,Benalcazar2017b,Schindler2018,Langbehn2017,Song2017,Fang2017,Ezawa2018,khalaf_SC_2018,Geier2018} and fragile topological states \cite{po_fragile_2018,cano_topology_2018,Barry2019EBR,manes_fragile_2019,de_paz_engineering_2019,song_fragile_2019,hwang_fragile_2019} have drawn tremendous attention.
Fragile bands have topological obstructions for Wannier functions (WFs): one cannot construct symmetric and localized WFs for them, but this obstruction can be removed by coupling the fragile band to certain \uemph{trivial} bands.
Remarkably, the low energy electron states in twisted bilayer graphene  \cite{Bistritzer2011,Kim2017-TBG,cao_TBG1,cao_TBG2,Huang2018-TBG,Fu2018Graphene,kang2018,po_origin_2018,yuan2018,Yankowitz2019-TBG,tarnopolsky2018,Leon2018TBG,liu_pseudo_2019} are predicted to exhibit a fragile topology \cite{po_faithful_2018,Ahn2018b,Song2018}, which was suggested to be responsible for the relatively high $T_c$ of superconductivity \cite{xie_topology_2019}.
Different from the stable topological states,\cite{Benalcazar2017,Benalcazar2017b,Schindler2018,Langbehn2017,Song2017,Fang2017,Ezawa2018,Khalaf2018,Song_NC_2018,khalaf_SC_2018,Geier2018}, the fragile phases generally do not have symmetry protected gapless surface or hinge states, and are hence experimentally elusive. Here, we show that the fragile phases have a new type-of bulk- boundary correspondence and have protected spectral flow under twisted boundary conditions (TBC).
We focus on the so-called eigenvalue fragile phase (EFP) \cite{Barry2019EBR,Song2018,hwang_fragile_2019}, which have recently been fully classified \cite{song_fragile_2019}. 

TBCs were introduced in Ref. \cite{niu_quantized_1985} to prove the quantization of Hall conductance. On a torus, a particle under U(1) TBC gains a phase $e^{i\theta_{x,y}}$ whenever it undergoes through a period in the $x/y$ direction.
This phase factor was then generalized to a complex number $\lambda = r e^{i\theta}$ ($0\le r \le 1$)  \cite{qi_general_2006} for a gapped state with two pairs of helical edge states, with unclear results. 
We generalize this idea to the case of fragile phases by considering a slow deformation of the boundary condition controlled by a single parameter $\lambda$. If the EFP can be written as a difference of a trivial atomic insulator and an obstructed atomic insulator (with WFs away from atoms) the energy gap between the fragile bands and other bands must close as we tune $\lambda$ in a particular path.

To quantitatively describe the EFP and its gap closing under TBCs, we introduce the concept of real space invariant (RSI): local good quantum numbers protected by PG (PG) symmetries. 
With translation symmetry, the RSIs can be calculated through symmetry eigenvalues of the band structure.
After a properly designed evolution of the boundary condition, where the symmetry of some Wyckoff position is preserved, the system goes through a gauge transformation, which does not commute with the symmetry operators. The symmetry eigenvalues and the RSIs at this Wyckoff position also go through a transformation: If the RSIs change, a gap closing happens during the process.
We find that EFPs always have nonzero RSIs: therefore, TBCs generally detect fragile topology.

We derive all the RSIs of all the 2D PGs with and without spin-orbit coupling(SOC) and/or time-reversal symmetry (TRS) (\cref{sec:RSI}).
For each 2D PG, we introduce a set of TBCs to detect the RSIs (\cref{sec:TBC}) and we derive their momentum space formulae in all wallpaper groups (\cref{sec:RSI-WG,tab:WGRSI}).
Symmetry eigenvalue criteria for \uemph{both} stable and fragile phase can be written as equations/ inequalities of  RSIs (\cref{sec:FC-RSI}). We use a spinless model as an example.

\section{Fragile phase in Topological Quantum Chemistry}
The symmetry (not Berry's phase) property of bands is fully described by its decompositions into irreducible representations (irreps) of little groups at momenta in the first Brillouin Zone (BZ).
For a gapped bands, the multiplicities of the irreps at different $\kk$ are not independent: they satisfy so-called compatibility relations \cite{Bradley2010,Bradlyn2017,Vergniory2017,Elcoro2017}, and only  the irreps  at maximal momenta are needed \cite{Bradlyn2017,Po2017,Elcoro2017,Vergniory2017,cano_topology_2018} 
to define the \emph{symmetry data vector} $B$
\begin{equation}
B = (m(\rho_{G_{K_1}}^1), m(\rho_{G_{K_1}}^2),\cdots,m(\rho_{G_{K_2}}^1),m(\rho_{G_{K_2}}^2),\cdots)^T, \label{eq:B-main}
\end{equation}
where $m(\rho_{G_{K_i}}^j)$ is the multiplicity of the $j$'th irrep of the little group $G_{K_i}$ at the maximal momentum $K_i$.

Band representations (BRs) are space group representations formed by decoupled symmetric atomic orbitals, representing atomic insulators. 
Generators of BRs are called elementary BRs (EBRs) \cite{Bradlyn2017,Elcoro2017,Vergniory2017,Jennifer2018}.
The EBRs in all space groups are available at the on the \href{http://www.cryst.ehu.es/cgi-bin/cryst/programs/bandrep.pl}{Bilbao Crystallographic Server (BCS)} \cite{Bradlyn2017,Elcoro2017,Vergniory2017,Jennifer2018,BCS1,*BCS2,*BCS3}.  We denote the symmetry data vector of the $i$th EBR as $\EBR_i$.
A symmetry data vector that satisfies the compatibility relations is a linear combination of $\EBR_i$ \cite{Luis_induction,Po2017}, \ie
\begin{equation}
B = \sum_i \EBR_i\ p_i, \label{eq:B-EBR-main}
\end{equation}
where $p_i$ are positive or negative rational or integer numbers.
In most space groups the EBRs are \uemph{not} linearly independent and the choice of $p$-vector for a fixed $B$-vector is not unique. 
If a choice where all the components of $p$ are nonnegative and integer exists, \ie $B$ is a multiple of EBRs, $B$ is consistent with an atomic insulator.
Otherwise, the corresponding band structure is topological.
There are two distinct categories of topological $B$-vectors (\cref{sec:TQC}).
{(I)} If some $p_i$ is fractional, the band structure has a stable topology \cite{Po2017,Bradlyn2017}. Examples include the inversion symmetric topological insulator \cite{Fu2007,Alexandradinata2014}, HOTIs \cite{Khalaf2018,Song_NC_2018}, and topological semimetals (without SOC) \cite{Fang2012,Song_PRX_2018}, \etc
{(II)} If $p$ can be chosen as an integer vector but some $p_i$ are \uemph{necessarily} negative, then the band structure has fragile topology. 
Examples include twisted bilayer graphene \cite{po_faithful_2018,Ahn2018b,Song2018} and other materials \cite{song_fragile_2019}. We refer to such phase as eigenvalue fragile phase (EFP). 
EFP may also have a stable topology un-diagnosed from symmetry eigenvalues \cite{Barry2019EBR,song_fragile_2019}. 

There are two distinct categories of non-topological $B$-vectors.
{(III)} If $p$ can be chosen as nonnegative integer vectors but some $p_i$ corresponding to an EBR at an empty Wyckoff position, where no atom exists, is \uemph{necessarily} nonzero, then $B$ is consistent with an obstructed atomic insulator \cite{Bradlyn2017,Vergniory2017,Elcoro2017,cano_topology_2018}. 
We refer to such phase as eigenvalue obstructed atomic phase (EOAP). 
EOAP may also have stable or fragile topology un-diagnosed from symmetry eigenvalues.
{(IV)} If $p$ can be chosen as nonnegative integer vectors where all nonzero $p_i$ correspond to atomically occupied Wyckoff positions, then $B$ is consistent with a trivial atomic insulator.

\section{A tight-binding model for EFP}\label{sec:model}
We build a spinless tight-binding model whose bands split into an EFP branch and an EOAP branch. Consider a $C_4$ symmetric square lattice (wallpaper group $p4$). Its BZ has three maximal momenta $\Gamma\ (0,0)$, $M\ (\pi,\pi)$ and $X\ (\pi,0)$.
The little group of $\Gamma$ and $M$ is PG $4$, and the little group of $X$ is PG $2$, with irreps  tabulated in \cref{tab:char24}. The irreps form co-irreps when we impose TRS.
Following  Ref. \cite{song_fragile_2019}, we are able to find all the so-called EFP roots -- minimal EFP generators that cannot be written as sums of other EFPs and EBRs.
As shown in \cref{tab:roots}, $p4$ has 12 EFP roots.
Here we consider the root $2\Gamma_1 + 2M_2 + 2X_1$, a state of two bands where each band forms the irreps $\Gamma_1$, $M_2$,  $X_1$ at  $\Gamma, M, X$.
These bands are topological: they cannot decompose into a sum of EBRs. We tabulate all the EBRs of $p4$ with TRS in \cref{tab:EBR-p4}.  We write the EFP root as (necessarily) a difference of EBRs as $ 2 (A)_b \up G  \oplus ({^1E^2E})_b \up G \ominus ({^1E^2E})_a \up G$.
(See \cref{sec:p4-model} for more details.)

\begin{table}[t]
\centering
\begin{tabular}{c|c|c|cc}
\multicolumn{3}{c|}{PG 2} & 1 & $2$ \\
\hline
$A$ & $X_1$ & $s$ & 1 & 1 \\
$B$ & $X_2$ & $p_x$ & 1 & -1 \\
\end{tabular}
~~~~~
\begin{tabular}{c|c|c|cccc}
\multicolumn{3}{c|}{PG $4$} & 1 & $4^+$ & $2$ & $4^-$\\
\hline
$A$ & $\Gamma_1$, $M_1$ & $s$ & 1 & 1 & 1 & 1\\
$B$ & $\Gamma_2$, $M_2$ & $d_{xy}$ & 1 & -1 & 1 & -1\\
$^1E$ & $\Gamma_3$, $M_3$ & $p_x+ip_y$ & 1 & $-i$ & -1 & $i$ \\
$^2E$ & $\Gamma_4$, $M_4$ & $p_x-ip_y$ & 1 & $i$ & -1 & $-i$ \\
\end{tabular}
\caption{Character tables of irreps of PGs $2$ and $4$. 
The first column: \href{http://www.cryst.ehu.es/cgi-bin/cryst/programs/representations_point.pl?tipogrupo=dbg}{BCS notations} \cite{Elcoro2017} of the PG irreps. 
The second column: notations of momentum space irreps at $X$, $\Gamma$, and $M$ for wallpaper group $p4$.
The third column are the atomic orbitals forming the corresponding irreps.
In presence of time-reversal symmetry, the two irreps $^1E$ ($\Gamma_3$, $M_3$) and $^2E$ ($\Gamma_4$, $M_4$) of PG $4$ form the co-irrep $^1E^2E$ ($\Gamma_3\Gamma_4$, $M_3M_4$).
 \label{tab:char24}}
\end{table}

\begin{table}[t]
\centering
\begin{tabular}{c|c||c|c|c}
WP & irrep & $\Gamma$ & $M$ & $X$ \\
\hline
\multirow{3}{*}{$a$ ($4$)} & $A$ & $\Gamma_1$ & $M_1$ & $X_1$ \\
\cline{2-5}
& $B$ & $\Gamma_2$ & $M_2$ & $X_1$ \\
\cline{2-5}
& $^1E^2E$ & $\Gamma_3\Gamma_4$ & $M_3M_4$ & $2X_2$ \\ 
\hline
\multirow{3}{*}{$b$ ($4$)} & $A$ & $\Gamma_1$ & $M_2$ & $X_2$ \\
\cline{2-5}
& $B$ & $\Gamma_2$ & $M_1$ & $X_2$ \\
\cline{2-5}
& $^1E^2E$ & $\Gamma_3\Gamma_4$ & $M_3M_4$ & $2X_1$ \\ 
\hline
\multirow{3}{*}{$c$ ($2$)} & $A$ & $\Gamma_1 \oplus \Gamma_2$ & $M_3M_4$ & $X_1\oplus X_2$ \\
\cline{2-5}
& $B$ & $\Gamma_3\Gamma_4$ & $M_1\oplus M_2$ & $X_1\oplus X_2$ \\
\end{tabular}
\caption{EBRs of wallpaper group $p4$ without SOC with TRS. \href{http://www.cryst.ehu.es/cgi-bin/cryst/programs/bandrep.pl}{BCS} \cite{Elcoro2017}.
$a\ (0,0)$, $b\ (\frac12,\frac12)$, $c\ (0,\frac12),\ (\frac12,0)$ are maximal Wyckoff positions. 4, 4, 2 in the following parethenesses are the corresponding site-symmetry groups. The second column: the irreps at the Wyckoff positions. The fourth to sixth columns:  momentum space irreps of the corresponding EBRs. \label{tab:EBR-p4} }
\end{table}


\begin{figure*}[t]
\centering
\includegraphics[width=0.8\linewidth]{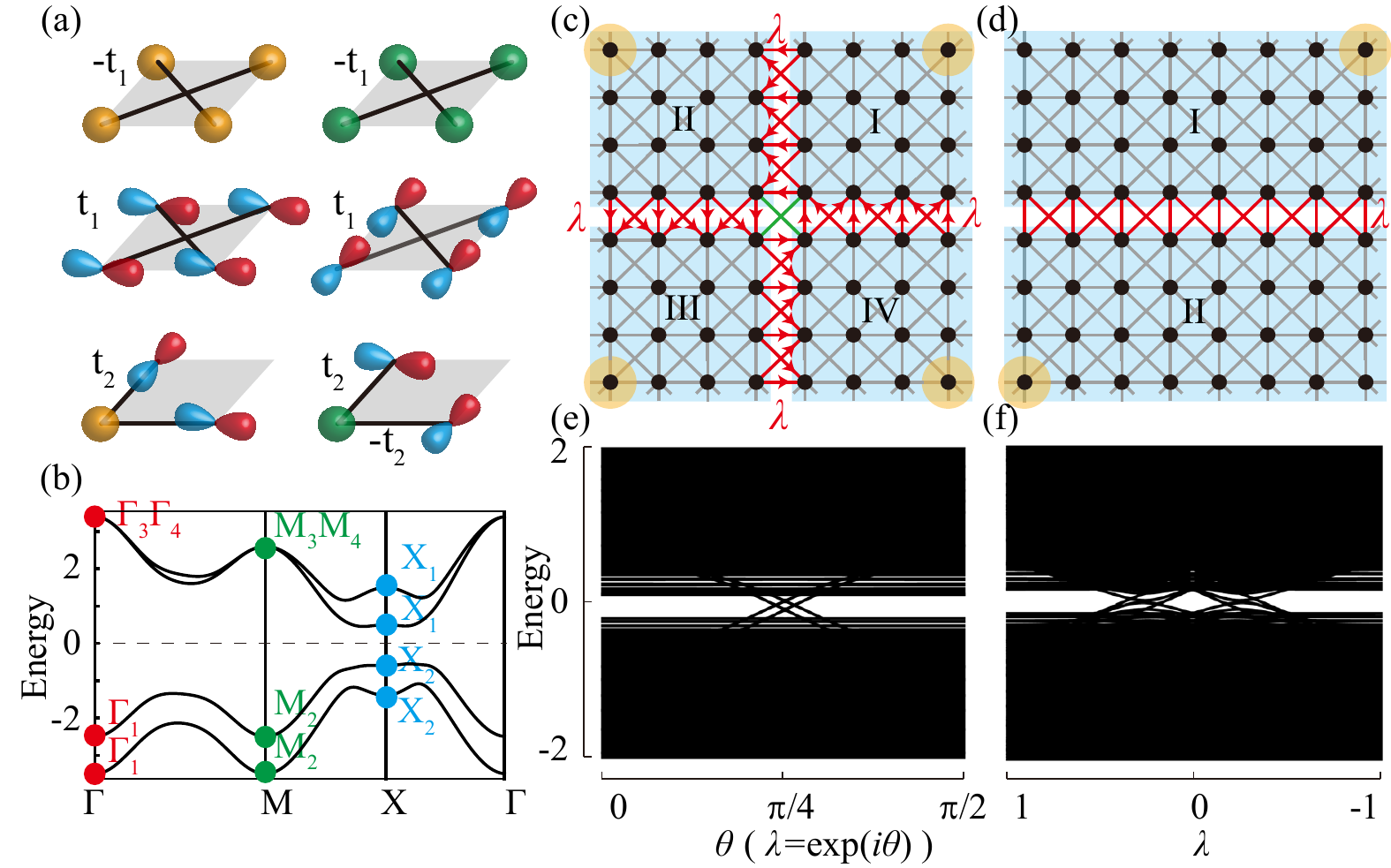}    
\caption{
(a) Fragile phase model (wallpaper group $p4$ with TRS).
The yellow, green, red and blue orbitals are the two $s$ and the $p_{x,y}$ orbitals.
The grey parallelogram is the unit cell, black lines are the hoppings.
(b) Band structure of the fragile phase.
(c) The $C_4$-symmetric TBC of a finite size system.
Black dots are the atoms; bonds are hoppings; four yellow circles are corner states of the fragile state.
Four shaded regions ($\mu=$I, II, III, IV) transform to each other under $C_4$ action.
Hoppings from the $\mu$th part to the $(\mu+1)$th part (red bonds)/from $\mu$th part to the $(\mu+2)$th part (green bonds)/from the $\mu$th part to the $(\mu-1)$th part (red bonds) are multiplied by a complex $\lambda$/real $\Re(\lambda^2)$/complex $\lambda^*$.
(d) The $C_2$ and TRS symmetric TBC.
The two shaded regions ($\mu=$I, II) transform to each other under $C_2$ rotation.
The hoppings between I, II (red bonds) are multiplied by a real number $\lambda$.
(e) Spectral flow under $C_4$-symmetric TBC.
(f) Spectral flow under $C_2$ and TRS symmetric TBC.
\label{fig:TB-TBC}}
\BAB{Space. Why?}
\SZD{It's latex's fault.}
\end{figure*}

Consider a four-band model of two $s$ ($s_1$ and $s_2$), one $p_x$, and one $p_y$ orbitals at the $b$ position (\cref{fig:TB-TBC}(a)).
According to \cref{tab:char24}, $s_{1,2}$ orbitals (irrep $A$) induce the BR $2 (A)_b \up G = 2\Gamma_1+ 2M_2+ 2X_2$; $p_{x,y}$ orbitals (irrep $^1E^2E$)  induce the EBR $({}^1E{}^2E)_b \up G = \Gamma_3\Gamma_4+ M_3M_4+ 2X_1$.
Let the $p_{x,y}$ orbitals have a higher energy than the $s_{1,2}$. We create a band inversion at the $X$ point such that the upper two bands become $\Gamma_3\Gamma_4+ M_3M_4+ 2X_2$, which is the EBR $({}^1E{}^2E)_a\up G$, and the lower two bands become the EFP root $2\Gamma_1 + 2M_2 + 2X_1$.
The model thus obtained  (\cref{sec:p4-model}) in the basis ($p_x, p_y, s_1,s_2$) is
\begin{align}
H(\kk) =& \tau_z\sigma_0 (E + 2t_1\cos(k_x+k_y) + 2t_1\cos(k_x-k_y)) \nono\\
        &+\tau_y \sigma_z t_2 \sin(k_x) + \tau_y \sigma_x t_2 \sin(k_y). \label{eq:Hk-main}
\end{align}
$E$ ($-E$) is the onsite energy for the $p_{x,y}$ ($s_{1,2}$) orbitals, $t_1$ introduces band inversion at $X$, $t_2$ guarantees a full gap between the upper and lower two bands. We introduce a term $\Delta H(\kk)$ (\cref{eq:DHk}) to break two accidental symmetries - $M_z$ ($z\to -z$) $=\tau_z\sigma_y$,  chiral  $\tau_x \sigma_0$.  The band structure of $H(\kk) + \Delta H(\kk)$ is plotted in \cref{fig:TB-TBC}(b).


We construct a finite-size ($30\times 30$) TRS  Hamiltonian with $C_4$ rotation symmetry preserved at the coordinate origin on the $a$-site (\cref{fig:TB-TBC}(c)). The spectrum consists of 1798 occupied states, 4 degenerate partially occupied levels at the Fermi level, and 1798 empty levels; they form the representations $450A \oplus 450 B \oplus 449 ({}^1E{}^2E)$, $A\oplus B\oplus {}^1E{}^2E$, and $449A \oplus 449 B \oplus 450 ({}^1E{}^2E)$, respectively.
The presence of partially occupied states are due to the corner states, or, the ``filling anomaly'' \cite{wieder2018axion} of fragile topology \cref{fig:TB-TBC}(c).
The gap between  the four partially occupied levels and the the occupied/empty levels is about $0.3/0.01$(\cref{sec:p4-model}), as $\Delta H(\kk)$ (\cref{eq:DHk}) breaks the accidental chiral symmetry.
Every four states forming the irreps $A \oplus B \oplus {}^1E{}^2E$ can be recombined as 
$\ket{1}=(\ket{A}+\ket{B}+\ket{^1E}+\ket{^2E})/2$, 
$\ket{2}=(\ket{A}-\ket{B}-i\ket{^1E}+i\ket{^2E})/2$, 
$\ket{3}=(\ket{A}+\ket{B}-\ket{^1E}-\ket{^2E})/2$, 
$\ket{4}=(\ket{A}-\ket{B}+i\ket{^1E}-i\ket{^2E})/2$, which transform to each other under the $C_4$ rotation and with Wannier centers away from the $C_4$-center: we hence move the occupied and empty states, both of which form the representation $449 A \oplus 449 B \oplus 449 ({}^1E{}^2E)$, away from the $C_4$-center. We are left with two occupied states, $A \oplus B$, and two empty states, $({}^1E{}^2E)$, at the $C_4$-center. These  four states form a level crossing under TBC evolution.

\section{Twisted boundary condition}\label{sec:TBC-main}
 We divide (\cref{fig:TB-TBC}(c)) the system into four parts ($\mu=$I, II, III, IV) which transform into each other under $C_4$. The boundaries between the four parts do not cut the corner states (\cref{fig:TB-TBC}(c)). We introduce the TBC by multiplying the hoppings between different parts by specific factors such that the twisted and original Hamiltonians are equivalent up to a \uemph{gauge} transformation. Specifically, the multiplication factors on hoppings from $\mu$th part to $(\mu+1)$th part/from $\mu$th part to $(\mu+2)$th part/from $\mu$th part to $(\mu-1)$th part are $i,-1,-i$. The twisted/untwisted Hamiltonians $\hH(i), \hH(1)$ satisfy:
\begin{equation}
\bra{\mu,\alpha} \hat{H}(i) \ket{\nu,\beta} \equiv (i)^{\nu-\mu} \bra{\mu,\alpha} \hat{H}(1) \ket{\nu,\beta}. \label{eq:TBC-main1}
\end{equation}
 $\ket{\mu,\alpha}$ is the $\alpha$th orbital in the $\mu$th part, and $H(\lambda)$ is the Hamiltonian with multiplier $\lambda$. We introduce the twisted basis $\hV \ket{\mu,\alpha}=(-i)^{\mu-1} \ket{\mu,\alpha}$. The elements of $\hH(i)$ on twisted basis  equal those of $\hH(1)$ on untwisted basis:
$\bra{\mu,\alpha} \hV^\dagger \hat{H}(i) \hV \ket{\nu,\beta} = \bra{\mu,\alpha} \hat{H}(1) \ket{\nu,\beta}$.
$C_4$ transforms the $\mu$th part into the $(\mu+1)$th part: the twisting phases of $\ket{\mu,\alpha}$ and $\hat{C}_4\ket{\mu,\alpha}$ under $\hV$ are $(-i)^{(\mu-1)}$ and $(-i)^\mu$, implying $\hV \hat{C}_4  = -i \hat{C}_4 \hV$.
If $\ket{\psi}$ is an eigenstate of $\hH(1)$ with $C_4$ eigenvalue $\xi$, then, $\hV\ket{\psi}$ will be an eigenstate of $\hH(i) = \hV \hH(1) \hV^\dagger$ of equal energy but different $C_4$ eigenvalue $i\xi$. The irreps $A$, $B$, $^1E$, $^2E$ will become $^2E$, $^1E$, $A$, $B$ under the gauge transformation (\cref{tab:char24}) .
Therefore two of the irreps $A\oplus B$ in the occupied states interchange with two of the irreps $^1E \oplus {}^2E$ in the empty states after the gauge transformation; all other irreps, ($449 A\oplus 449 B \oplus 449(^1E) \oplus 449(^2E)$) remain unchanged.  We generalize the $C_4$ symmetric TBC:

\begin{align}
& \bra{\mu,\alpha} \hat{H}(\lambda) \ket{\nu,\beta} = \nono \\
&\begin{cases}
\bra{\mu,\alpha} \hat{H}(1) \ket{\nu,\beta},\qquad & \nu = \mu\\
\lambda \bra{\mu,\alpha} \hat{H}(1) \ket{\nu,\beta},\qquad & \nu =\mu+1 \mod 4\\
\lambda^* \bra{\mu,\alpha} \hat{H}(1) \ket{\nu,\beta},\qquad & \nu = \mu-1 \mod 4\\
\Re(\lambda^2) \bra{\mu,\alpha} \hat{H}(1) \ket{\nu,\beta},\qquad & \nu = \mu+2 \mod 4\\
\end{cases}, \label{eq:TBC-main2}
\end{align}
$\Re(\lambda^2)$ is the real part of the complex $\lambda^2$. The factor between the $\mu$th and $(\mu+2)$th part is real due to $C_2$ (\cref{sec:TBC-2D,app:p4-model-TBC}).
\cref{eq:TBC-main1} is the $\lambda=i$  case of \cref{eq:TBC-main2}. Under continuous tuning of $\lambda$ from $1$ to $i$,  two occupied irreps $A\oplus B$ interchange with two empty irreps $^1E \oplus {}^2E$. Their level crossings are protected by $C_4$ symmetry (\cref{fig:TB-TBC}(e)).  Two other $C_4$ symmetric evolution paths are discussed in \cref{app:p4-model-TBC}.

\begin{figure*}[t]
\centering
\includegraphics[width=1\linewidth]{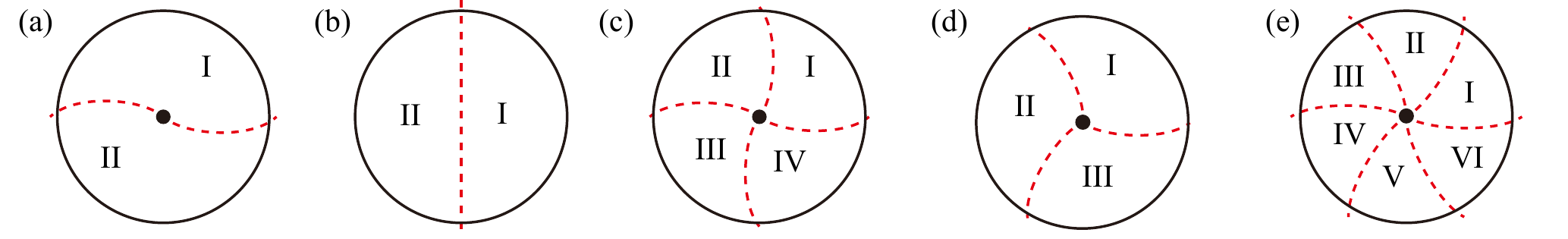}    
\caption{Illustration of TBCs for 2D PGs $2$, $m$, $4$, $3$, $6$. The systems are divided into 2, 2, 4, 3, 6 parts, respectively. Hoppings between different parts are multiplied by factors determined by a single parameter $\lambda$. 
These boundary conditions are used to detect spectral flow protected by $2mm$, $4mm$, $3m$, $6mm$. 
See \cref{sec:TBC-2D} for complete discussion.
\label{fig:TBC-2D-open}}
\end{figure*}

Now we consider $C_4$-breaking but $C_2$ and TRS preserving TBC. Divide the system into two parts (I,II), transforming  into each other under $C_2$ (\cref{fig:TB-TBC}(d)), and multiply all hoppings between orbitals in part I and  II by  a real $\lambda$.
 The gauge transformation relating the twisted and untwisted Hamiltonians anti-commutes with $C_2$: $\{\hV,\hat{C}_2\}=0$. $\hV$ transforms between eigenstates of $\hH(1), \hH(-1)$ with equal energy but opposite $C_2$ eigenvalue.  Under a continuous tuning of $\lambda$ from $1$ to $-1$, the two final occupied (empty) states must have the $C_2$ eigenvalue $-1$ ($1$). This inconsistency implies $C_2$-protected gap closing, as shown in \cref{fig:TB-TBC}(f). The unitary transformation relating $H(-1)$ to $H(1)$ also maps $H(-\lambda)$ to $H(\lambda)$ and the gap must close as $\lambda$ changes from 1 to 0. In \cref{sec:TBC-2D} we generalize the TBC to all the 2D PGs, illustrated in \cref{fig:TBC-2D-open}. The gapless states under TBC are the experimental consequences of the fragile states.

\begin{table} \scriptsize
\centering
\begin{tabular}{|c|c|c|c|c|c|c|c|c|c|c|}
\hline
SOC & TRS & $2$ & $m$ & $2mm$ & $4$ & $4mm$ & $3$ & $3m$ & $6$ & $6mm$ \\
\hline
\xmark & \xmark  & $\mbb{Z}$ & $\mbb{Z}$ & $\mbb{Z}$ & $\mbb{Z}^3$ & $\mbb{Z}^2$ & $\mbb{Z}^2$ & $\mbb{Z}$ & $\mbb{Z}^5$ & $\mbb{Z}^3$ \\
\hline
\xmark & \checkmark  & $\mbb{Z}$ & $\mbb{Z}$ & $\mbb{Z}$ & $\mbb{Z}^2$ & $\mbb{Z}$ & $\mbb{Z}$ & $\mbb{Z}$ & $\mbb{Z}^3$ & $\mbb{Z}^3$ \\
\hline
\checkmark & \xmark  & $\mbb{Z}$ & $\mbb{Z}$ & $\mbb{Z}_1$ & $\mbb{Z}^3$ & $\mbb{Z}$ & $\mbb{Z}^2$ & $\mbb{Z}$ & $\mbb{Z}^5$ & $\mbb{Z}^2$ \\
\hline
\checkmark & \checkmark  & $\mbb{Z}_2$ & $\mbb{Z}_2$ & $\mbb{Z}_2$ & $\mbb{Z}_2\times\mbb{Z}$ & $\mbb{Z}_2\times\mbb{Z}$ & $\mbb{Z}$ & $\mbb{Z}$ & $\mbb{Z}_2\times\mbb{Z}^2$ & $\mbb{Z}_2\times\mbb{Z}^2$ \\
\hline 
\end{tabular}
\protect\caption{\label{tab:RSI-group} The RSI groups of 2D PGs.}
\end{table}

\section{Real space invariant}\label{sec:RSI-main}

We introduce the RSI as an exhaustive description of the local states, pinned at the $C_4$-center, that  undergo gap closing under TBC. The Wannier centers of occupied states of a Hamiltonian can adiabatically move if their displacements preserve symmetry. Orbitals away from a symmetry center can move on it and form an \uemph{induced representation} of the site-symmetry group at the center. Conversely, orbitals at a symmetry center can move away from it symmetrically \uemph{iff}  they form a representation induced from the site-symmetry groups away from the center. The RSIs are defined (and proved \cref{sec:RSI}) as linear invariant - upon  such induction processes - functions of irrep multiplicities. For the PG $4$ with TRS, we assume a linear form RSI of the occupied levels $\delta = c_1 m(A) + c_2 m(B) + c_3 m({}^1E{}^2E)$. The induced representation at the $C_4$ center from four states at $C_4$-related positions is $A\oplus B\oplus {}^1E{}^2E$. (See the end of \cref{sec:TRS-irrep}.)
After the induction process, the irrep multiplicities at the $C_4$-center change as $m(A)\to m(A)\pm1$, $m(B)\to m(B)\pm1$, $m({}^1E{}^2E)\to m({}^1E{}^2E)\pm1$. 
The two linear combinations of irrep multiplicities that remain invariant are
\begin{equation}
\delta_1 = m({}^1E{}^2E) - m(A),
\quad
\delta_2 = m(B) - m(A). \label{eq:d1d2-main}
\end{equation}
In our model, the occupied states that can be moved away from the $C_4$-center form the representation $449A \oplus 449 B \oplus 449 ({}^1E{}^2E)$ and have vanishing RSIs. The states pinned at the $C_4$ center form $A\oplus B$ with RSIs $\delta_1=-1$, $\delta=0$.
If an RSI is nonzero, spectral flow exist upon a particular TBC (\cref{sec:TBC}). 
We calculate all the RSIs in all 2D PGs with and without SOC/TRS  (\cref{tab:RSI-formula} in \cref{sec:RSI}). The groups formed by RSIs are shown in \cref{tab:RSI-group}. PG $4$ with TRS has two integer-valued RSIs: the RSI group is $\mbb{Z}^2$. Most RSIs are $\mbb{Z}$-type; some groups with SOC and TRS have $\mbb{Z}_2$-type RSI, the parities of the number of occupied Kramer pairs (\cref{tab:RSI-formula}).

For the $C_4$-symmetric TBC (\ref{eq:TBC-main2}),  the occupied irrep multiplicities $m^\prime$ at $\lambda = i$ are determined by the multiplicities $m$ at $\lambda=1$ as $m^\prime(A)=m({}^1E)$, $m^\prime(B)=m({}^2E)$, $m^\prime({}^1E)=m(B)$, $m^\prime({}^2E)=m(A)$.
The changes of irreps in the evolution $\lambda=1\to i$ are $\Delta m(A) = m^\pr(A)-m(A)= m({}^1E) - m(A)=\delta_1$, $\Delta m(B) = \delta_1-\delta_2$, $\Delta m({}^1E) = \delta_2-\delta_1$, $\Delta m({}^2E) = -\delta_1$.
Therefore, there will be $|\delta_1|$ crossings formed by $A$ and $^2E$ and $|\delta_2-\delta_1|$ crossings formed by $B$ and $^1E$. 
This and the similar analysis for $C_2$ and TRS-symmetric TBC are given in \cref{eq:TBC-lambda-mainTOTAL} and expanded in \cref{app:p4-model-TBC}.
Our model ($\delta_1=-1$ and $\delta_2=0$), has 2 level crossings protected by $C_2$ in the process $\lambda=1\to -1$. 
\begin{equation} 
\begin{tabular}{c|c}
$C_4$: $\lambda$ & $1\to i$ \\\hline\hline
$\Delta m(A)$ & $\dt_1$ \\
$\Delta m({}^1E)$ & $ \dt_2- \dt_1$ \\
$\Delta m(B)$ & $\dt_1-\dt_2$ \\
$\Delta m({}^2E)$ & $-\dt_1$ \\
\end{tabular},\;\;\;\;\;\;\;  \begin{tabular}{c|c}
$C_2$: $\lambda \in \mbb{R}$ & $1\to -1$ \\ \hline\hline
$\Delta m(A)$ & $2\delta_1-\dt_2$ \\
$\Delta m(B)$ & $-2\delta_1+\dt_2$
\end{tabular} \label{eq:TBC-lambda-mainTOTAL}
\end{equation}

\section{RSI and band topology}\label{sec:RSI-top}
The advantage of the RSI is that it can be calculated from both the momentum space irreps of the band structure or from symmetry-center PG-respecting disordered configurations. We use wallpaper group $p4$ as an example  (for all wallpaper groups see \cref{tab:WGRSI}). We find that the RSI fully describe eigenvalue band topology: (eigenvalue) fragile topology is diagnosed by inequalities or mod-equations of RSIs, and stable topology is implied by fractional RSIs. 
We prove this in all the wallpaper groups by exhaustive calculations (\cref{sec:RSI-WG}).

We denote the site-symmetry group at a Wyckoff position $w$ as $G_w$. To obtain the RSIs at $w$ of a given band structure, we recombine the occupied Bloch states to form irreps of $G_w$ and then substitute these irreps for the RSIs. This is the definition of RSI in band structures under periodic boundary condition irrespective of the bands's Wannierizability. If we cut the periodic boundary to obtain a finite-size system with the PG symmetry $G_w$, the RSIs at the symmetry center ($w$) of the system remain unchanged: they are determined by \uemph{local} states around the symmetry center, and cannot be affected by the boundary condition. We sketch how to obtain the RSIs in terms of momentum space irreps.

We first \uemph{assume} (justified in \cref{sec:RSI-WG}) that RSIs can be calculated as linear combinations of multiplicities of momentum space irreps, \ie $\delta_{wi} = \sum_{K j} c_{Kj}^{wi} m(\rho_{G_K}^j)$, where $\delta_{wi}$ stands for the $i$th RSI at $w$, $K$ sums over all maximal momenta, $j$ sums over all irreps at $K$, and $m(\rho_{G_K}^j)$ is the multiplicity of the $j$th irrep at $K$. Since each EBR is induced by an irrep of the site-symmetry group of some Wyckoff position, we can directly calculate the RSIs from this real space irrep, which we then match with the linear expression of RSIs in terms of momentum space irreps. This match leads to a set of linear constraints of $c_{Kj}^{wi}$, which determine them. In \cref{sec:RSI-WG} we prove that, for $\mbb{Z}$-type RSIs, the linear function assumption is always true.
In the following we would only focus on $\mbb{Z}$-type RSIs.

For our example, SG $p4$ has two inequivalent $C_4$Wyckoff positions, $a$ and $b$, and one $C_2$  Wyckoff position, $c$ (\cref{tab:EBR-p4}). PG $4$ has two RSIs, $\delta_1$ and $\delta_2$ (\cref{eq:d1d2-main}), and PG $2$ has a single RSI (\cref{tab:RSI-formula})
\begin{equation}
\delta_{1}=m(B)-m(A), \label{eq:d1-main}
\end{equation} 
being $A$ and $B$ the $C_2$-even and $C_2$-odd irreps, respectively.
Thus we have two RSIs at $a$, two RSIs at $b$, and one RSI at $c$.
We denote the RSI vector as $\delta=(\delta_{a1},\delta_{a2},\delta_{b1},\delta_{b2},\delta_{c1})^T$ and the linear mapping matrix from irrep multiplicities of the $B$ symmetry data vector (\ref{eq:B-main}) to RSIs as $F$, \ie $\delta = F\cdot B$. The order of irreps in $B$ is $\Gamma_1$, $\Gamma_2$, $\Gamma_{3}\Gamma_4$, $M_1$, $M_2$, $M_3M_4$, $X_1$, $X_2$. Consider the band structure of the EBR $(A)_a\up G$, with the irreps $\Gamma_1 + M_1 + X_1$ and hence  $B=(10010010)^T$.
Since it is induced from a single $A$ irrep at the $a$ position, the EBRs RSIs at $a$ are $\delta_{a1}=-1$, $\delta_{a2}=-1$ (\cref{eq:d1d2-main}); the RSIs at other positions are zero.
Thus the $F$ matrix satisfies
\begin{equation}
(-1,-1,0,0,0)^T = F\cdot (1,0,0,1,0,0,1,0)^T.
\end{equation}
Solving the linear equations of $F$ obtained from the other EBRs we find
\begin{align}
\delta_{a1} =& - m(\mathrm{\Gamma_{1}}) -\frac{m(\mathrm{\Gamma_{2}})}{2}  - m(\mathrm{\Gamma_{3}\Gamma_{4}})\nono\\
&  + \frac{m(\mathrm{M_{2}})}{2} + m(\mathrm{M_{3}M_{4}}) + \frac{m(\mathrm{X_{2}})}{2},\label{eq:da1-main}
\end{align}
\begin{equation}
\delta_{a2} = - m(\mathrm{\Gamma_{1}}) - m(\mathrm{\Gamma_{3}\Gamma_{4}}) + m(\mathrm{M_{2}}) + m(\mathrm{M_{3}M_{4}}),\label{eq:da2-main}
\end{equation}
\begin{equation}
\delta_{b1} = \frac{1}{2} m(\mathrm{\Gamma_{2}}) + m(\mathrm{\Gamma_{3}\Gamma_{4}}) -\frac{1}{2} m(\mathrm{M_{2}}) -\frac{1}{2} m(\mathrm{X_{2}}),\label{eq:db1-main}
\end{equation}
\begin{equation}
\delta_{b2} = m(\mathrm{\Gamma_{2}}) + m(\mathrm{\Gamma_{3}\Gamma_{4}}) - m(\mathrm{M_{2}}) - m(\mathrm{M_{3}M_{4}}),\label{eq:db2-main}
\end{equation}
\begin{equation}
\delta_{c1} = m(\mathrm{\Gamma_{3}\Gamma_{4}}) - m(\mathrm{M_{3}M_{4}}). \label{eq:dc1-main}
\end{equation}
The solution of $F$ might be \uemph{not} unique, but all solutions give the same RSI. 
(See \cref{sec:RSI-WG}). The above formulas give the correct RSIs for our EBRs in \cref{tab:EBR-p4} (\cref{app:p4-model-RSI}).

Due to \cref{eq:da1-main,eq:db1-main}, $\dt_{a1}$ and $\dt_{b1}$ can take fractional values. 
For example, the band $\Gamma_1 + M_2 + X_1$ has $\dt_{a1}=-\frac12$, $\dt_{a2}=0$, $\dt_{b1}=-\frac12$, $\dt_{b2}=-0$, $\dt_{c1}=0$.
According to Eq. (38) in Ref. \cite{Song_PRX_2018}, the $\beta_2$ invariant of this state is $\beta_2 = m(\Gamma_2) + m(M_2) + m(X_2) \mod2 = 1$, implying that there is 1 (mod 2) Dirac point in each quadrant of the BZ. This state is a topological semimetal. If the RSI calculated from momentum space are fractional \emph{iff}  the corresponding band structure must have symmetry eigenvalue indicated \uemph{stable topology} (proof in \cref{sec:FC-RSI}).

We now relate the RSI  to fragile topology using an example.
First consider topologically trivial (Wannierizable) phases in wallpaper group $p4$ with TRS,
with $N_a=m( (A)_a )+m( (B)_a )+2m(({}^1E{}^2E)_a)$, $N_b=m( (A)_b )+m( (B)_b )+2m(({}^1E{}^2E)_b)$, $N_c=m( (A)_c )+m( (B)_c )$ states at $a,b,c$.
The number of bands is $N_{\rm occ}= N_a + N_b + N_c$. $N_a$, $N_b$, $N_c$ are \uemph{not} gauge invariant quantities: four states at $a$ forming the induced representation $(A)_a \oplus (B)_a \oplus ({}^1E{}^2E)_a$, can move to general Wyckoff positions, and then recombine at $b$ position, giving many different choices for Wannierization. In every choice, however, the multiplicity of each irrep at each position must be $\ge 0$, which constrains $N_w$. Specifically, assume $0\le m((A)_a)\le \min(m((B)_a),m(({}^1E{}^2E)_a))$.  We recombine $4m((A)_a)$ states forming the induced representation $m((A)_a) [(A)_a  \oplus (B)_a \oplus ({}^1E{}^2E)_a]$ as four states at the general Wyckoff position, and leaving only (a minimum) $m( (B)_a )+2m(({}^1E{}^2E)_a)-3 m((A)_a)=2\delta_{a1} + \delta_{a2}$ number of WFs at $a$.
Similarly, when $(B)_a$ and/or $({}^1E{}^2E)_a$ has the least multiplicity at $a$, the minimum number of Wannier functions is $m((A)_a) +2m(({}^1E{}^2E)_a) -3m( (B)_a )=2\delta_{a1} - 3\delta_{a2}$ and $m((A)_a) +m( (B)_a )-2m(({}^1E{}^2E)_a) =-2\delta_{a1} + \delta_{a2}$.
Thus, $N_a\ge \max(2|\delta_{a1}| + \delta_{a2},\ 2\delta_{a1} -3 \delta_{a2})$.
Similarly, for the $b$ and $c$ positions, we obtain $N_b\ge \max(2|\delta_{b1}| + \delta_{b2},\ 2\delta_{b1} -3 \delta_{b2})$ and $N_c \ge |\dt_{c1}|$. For a Wannierizable state, the number of bands satisfies
\begin{align}
N_{\rm occ} \ge & \max(2|\delta_{a1}| + \delta_{a2},\ 2\delta_{a1} -3 \delta_{a2}) \nono \\
 + & \max(2|\delta_{b1}| + \delta_{b2},\ 2\delta_{b1} -3 \delta_{b2}) + |\dt_{c1}|. \label{eq:non-FC-main}
\end{align} If \cref{eq:non-FC-main} is violated, the band structure must be topological.
If all RSIs are integers (band structure does not have stable topology), the violation implies EFP.
The EFP $2\Gamma_1 + 2M_2 + 2X_1$ considered in \cref{sec:model}, has the band number $N_{S}=2$ and the RSIs $\dt_{a1}=-1$, $\dt_{a2}=0$, $\dt_{b1}=-1$, $\dt_{b2}=-2$, $\dt_{c1}=0$,  and violates \cref{eq:non-FC-main}. We derived (\cref{sec:FC-RSI}) all the constraints between the band number and RSIs for Wannierizable phases in 2D. Their violation provides EFP criteria (\cref{tab:FC}.) equivalent to the ones found by the  affine monoid method \cite{song_fragile_2019} \cite{hwang_fragile_2019}. All EFPs have nonzero RSIs.  EOAP also have nonzero RSIs at atom-free Wyckoff position,



\section{Discussion and summary}
We have introduced a new bulk-boundary correspondence, the TBC, as a generic approach to experimentally detect a fragile topological state. As a parameter $\lambda$ is continuously tuned, symmetry-protected gap closings take place. To fully classify the spectral flow, we introduce the concept of RSI - a local topological invariant stabilized by PG symmetries, which in presence of transition symmetry, also classifies the eigenvalue  \uemph{stable and fragile} topological states. We explicitly work out all RSIs in all wallpaper groups, with and without SOC/TRS, and the corresponding TBCs to detect these RSIs.

Metamaterial systems \cite{prodan_topological_2009,kane2014meta,paulose_topological_2015,nash_topological_2015,susstrunk_observation_2015,wang_topological_2015,yang_topological_2015} are the ideal platforms where TBC can be  tuned by mechanical parameters. We imagine \cref{eq:Hk-main} is a Hamiltonian of a mechanical system consisting of mass points connected by rigid bonds or springs. 
The TBC of \cref{fig:TB-TBC}(d) can be realized by tuning the  springs connecting mass points in part I and II from their original values to zero, mimmicking $\lambda=1$ to $\lambda=0$. The gap between the fragile bands and the other bands must close during this process.


\textit{Note added.} Note the accompanying paper by Peri et al. ``Experimental characterization of spectral flow between fragile bands'' posted on the same day on the arXiv~\cite{Peri}.

\begin{acknowledgements}
We are grateful to Barry Bradlyn, Jennifer Cano, Emil Prodan, Nicolas Regnault for an earlier initial collaboration on the subject and to Biao Lian for helpful discussions.
Z. S. and B. A. B. are supported by the Department of Energy Grant No. DE-SC0016239, the National Science Foundation EAGER Grant No. DMR 1643312, Simons Investigator Grants No. 404513, ONR No. N00014-14-1-0330, and  NSF-MRSEC No. DMR-142051, the Packard Foundation, the Schmidt Fund for Innovative Research, and the  Guggenheim Fellowship. L. E. is supported by the Government of the Basque Country (project IT1301-19).
\end{acknowledgements}

\bibliography{ref}

\appendix
\onecolumngrid

\section{Group representation and topological quantum chemistry} \label{sec:representation}

\subsection{Induction and reduction of irreps of finite groups} \label{sec:induction}
\subsubsection{Without time-reversal symmetry}\label{sec:induction-NTR}
Given a group $G$ and one of its subgroups $H$, each irrep (shorthand for irreducible representation) of $H$ can induce a generally reducible representation of $G$.
The basic idea is that the coset elements $G/H$ generate a set of copies of the basis functions carrying the irrep of $H$; and these symmetric copies will form a representation of $G$.
We decompose $G$ into cosets of $H$
\begin{equation}
G = (\td{g}_1=e) H + \td{g}_2 H + \cdots \td{g}_n H,
\end{equation}
where $n=|G|/|H|$, and $|G|$ and $|H|$ are the numbers of group elements in $G$ and $H$, respectively.
We consider an $m$-dimensional irrep $\rho_H$ of $H$.
It is carried by the basis functions $\ket{\phi_\alpha}$ ($\alpha=1\cdots m$), \ie
\begin{equation}
\forall h \in H,\qquad h\ket{\phi_\alpha} = \sum_{\beta=1}^m [\rho_H(h)]_{\beta\alpha}\ket{\phi_\beta}.
\end{equation}
We define the $n$ copies of these basis functions as
\begin{equation}
\ket{\phi_{i\alpha}} = \td{g}_i\ket{\phi_\alpha},\qquad i=1\cdots n. \label{eq:sym-basis}
\end{equation}
Since $g\cdot \td{g}_i\in G$, there must exist such a coset $\td{g}_j H$ that $g\cdot \td{g}_i \in \td{g}_j H$.
Correspondingly, $g$ will transform the basis functions $\ket{\phi_{i\alpha}}$ in the $i$th copy of $\ket{\phi_\alpha}$ to the basis functions in the $j$th copy of $\ket{\phi_\alpha}$.
We introduce the matrix to 
\begin{equation}
[d(g)]_{ji} = \begin{cases}
1,\qquad & \text{if\ } g\cdot \td{g}_i \in \td{g}_j H \\
0,\qquad & \text{else}
\end{cases} \label{eq:small-d}
\end{equation}
denote the mapping between different copies.
For given $g$ and $\td{g}_i$, there is only one such $\td{g}_j$ that $g \td{g}_i \in \td{g}_j H$ and $d_{ji}(g)=1$.
Thus we obtain
\begin{equation}
g \ket{\phi_{i\alpha}} = g \td{g}_i \ket{\phi_{\alpha}} 
= \sum_j d_{ji}(g) \td{g}_j \cdot \td{g}_j^{-1} g \td{g}_i \ket{\phi_{\alpha}} .
\end{equation}
Since $ g \td{g}_i\in \td{g}_j H$, there must be $\td{g}_j^{-1}g \td{g}_i\in H$ and
\begin{equation}
g \ket{\phi_{i\alpha}} = \sum_{j\beta} d_{ji}(g) \brak{\rho_H (\td{g}_j^{-1} g \td{g}_i)}_{\beta\alpha} \td{g}_j \ket{\phi_{\beta}} = \sum_{j\beta} d_{ji}(g) \brak{\rho_H (\td{g}_j^{-1} g \td{g}_i)}_{\beta\alpha}\ket{\phi_{j\beta}} .
\end{equation}
Therefore, $\ket{\phi_{i\alpha}}$ form a representation of $G$.
We denote this induced representation as $\rho_H\up G$.
The induced representation decomposes into a set of irreps of $G$. 
We denote this decomposition as
\begin{equation}
\rho_H\up G = \bigoplus_{i} f(\rho_G^i | \rho_H\up G) \rho_{G}^i,\label{eq:induction1}
\end{equation}
where $\rho_{G}^i$ is the $i$th irrep of $G$ and $f(\rho_G^i | \rho_H\up G) $ is the multiplicity of $\rho_{G}^i$ in the decomposition.

For a given group $G$ and one of its subgroups $H$, each irrep $\rho_G$ of $G$ decomposes into a set of irreps of $H$.
We denote this reduced representation as $\rho_G \down H$, and its decomposition into irreps of $H$ as
\begin{equation}
\rho_G \down H = \bigoplus_{i} f(\rho_H^i | \rho_G \down H)  \rho_H^i, \label{eq:reduction1}
\end{equation}
where $\rho_H^i$ is the $i$th irrep of $H$ and $f(\rho_H^i | \rho_G \down H)$ is the corresponding multiplicity in the decomposition.
This multiplicity can be calculated as
\begin{equation}
f(\rho_H^i | \rho_G \down H) = \frac{1}{|H|} \sum_{h\in H} \Tr[\rho_H^i (h)]^* \Tr[\rho_G(h)]. \label{eq:reduction}
\end{equation}


The multiplicities in \cref{eq:induction1,eq:reduction1} are related through a particular case of the Frobenius reciprocity theorem.
For a \uemph{finite unitary} group $G$ and one of its subgroup $H$, the theorem states that
\begin{equation}
\boxed{    f(\rho_G | \rho_H\up G) = f(\rho_H | \rho_G \down H).} \label{eq:Frobenius}
\end{equation}
In practice, we will use \cref{eq:reduction,eq:Frobenius} to calculate the multiplicity $f(\rho_G | \rho_H\up G)$ in the induction.


\subsubsection{With time-reversal symmetry}\label{sec:TRS-irrep}

\begin{table}
\centering
\begin{tabular}{c|c|c}
\multicolumn{2}{c|}{PG 1} & 1  \\
\hline
$A$ & $A$ & 1  \\
\hline
$A_{1/2}$ & $\ovl{A}$ & 1  \\
\end{tabular}
~~~~~
\begin{tabular}{c|c|cc}
\multicolumn{2}{c|}{PG 2} & 1 & $2$ \\
\hline
$A$ & $A$ & 1 & 1 \\
$B$ & $B$ & 1 & -1 \\
\hline
$^2E_{1/2}$ & ${}^2\ovl{E}$ & 1 & $-i$ \\
$^1E_{1/2}$ & ${}^1\ovl{E}$ & 1 & $i$
\end{tabular}
~~~~~
\begin{tabular}{c|c|cc}
\multicolumn{2}{c|}{PG $m$} & 1 & $m$ \\
\hline
$A^\pr$ & $A^\pr$ & 1 & 1 \\
$A^\prpr$ & $A^\prpr$ & 1 & -1 \\
\hline
$^2E_{1/2}$ & ${}^2\ovl{E}$ & 1 & $-i$ \\
$^1E_{1/2}$ & ${}^1\ovl{E}$ & 1 & $i$
\end{tabular}
~~~~~
\begin{tabular}{c|c|cccc}
\multicolumn{2}{c|}{PG $2mm$} & 1 & $2$ & $m_{100}$ & $m_{010}$\\
\hline
$A_1$ & $A_1$ & 1 & 1 & 1 & 1\\
$A_2$ & $A_2$ & 1 & 1 & -1 & -1\\
$B_1$ & $B_1$ & 1 & -1 & -1 & 1\\
$B_2$ & $B_2$ & 1 & -1 & 1 & -1\\
\hline
$E_{1/2}$ & $\ovl{E}$ & 2 & 0 & 0 & 0
\end{tabular}
~~~~~
\begin{tabular}{c|c|cccc}
\multicolumn{2}{c|}{PG $4$} & 1 & $4^+$ & $2$ & $4^-$\\
\hline
$A$ & $A$ & 1 & 1 & 1 & 1\\
$B$ & $B$ & 1 & -1 & 1 & -1\\
$^1E$ & $^1E$ & 1 & $-i$ & -1 & $i$\\
$^2E$ & $^2E$ & 1 &  $i$ & -1 & $-i$\\
\hline
$^1E_{1/2}$ & $^2\ovl{E}_1$ & 1 & $e^{-i\frac{\pi}4}$ & $-i$ & $e^{i\frac{\pi}4}$\\
$^2E_{1/2}$ & $^1\ovl{E}_1$ & 1 & $e^{i\frac{\pi}4}$ & $i$ & $e^{-i\frac{\pi}4}$\\
$^1E_{3/2}$ & $^2\ovl{E}_2$ & 1 & $e^{i\frac{3\pi}4}$ & $-i$ & $e^{-i\frac{3\pi}4}$\\
$^2E_{3/2}$ & $^1\ovl{E}_2$ & 1 & $e^{-i\frac{3\pi}4}$ & $i$ & $e^{i\frac{3\pi}4}$
\end{tabular}
\vspace{0.5cm}

\begin{tabular}{c|c|ccccc}
\multicolumn{2}{c|}{PG $4mm$} & 1 & $\{4^+,4^-\}$ & $2$ & $\{m_{010},m_{100}\}$ & $\{m_{110},m_{1-10}\}$\\
\hline
$A_1$ & $A_1$ & 1 & 1 & 1 & 1 & 1\\
$A_2$ & $A_2$ & 1 & 1 & 1 &-1 &-1\\
$B_1$ & $B_1$ & 1 &-1 & 1 & 1 & -1\\
$B_2$ & $B_2$ & 1 &-1 & 1 &-1 & 1\\
$E$ & & 2 & 0 &-2 & 0 & 0 \\ 
\hline
$E_{1/2}$ & $\ovl{E}_1$ & 2 & $\sqrt2$ & 0 & 0 & 0\\
$E_{3/2}$ & $\ovl{E}_2$ & 2 & $-\sqrt2$ &0 & 0 & 0
\end{tabular}
~~~~~
\begin{tabular}{c|c|ccc}
\multicolumn{2}{c|}{PG 3} & 1 & $3^+$ & $3^-$ \\
\hline
$A$ & $A_1$ & 1 & 1 & 1 \\
$^2E$ & $^2E$ & 1 & $e^{i\frac{2\pi}3}$ & $e^{-i\frac{2\pi}3}$ \\
$^1E$ & $^1E$ & 1 & $e^{-i\frac{2\pi}3}$ & $e^{i\frac{2\pi}3}$ \\
\hline
$A_{3/2}$ & $\ovl{E}$ & 1 & -1 & -1 \\
$^2E_{1/2}$ & ${}^1\ovl{E}$ & 1 & $e^{-i\frac{\pi}3}$ & $e^{i\frac{\pi}3}$ \\
$^1E_{1/2}$ & ${}^2\ovl{E}$ & 1 & $e^{i\frac{\pi}3}$ & $e^{-i\frac{\pi}3}$
\end{tabular}
\vspace{0.5cm}

\begin{tabular}{c|c|ccc}
\multicolumn{2}{c|}{PG $3m$} & 1 & $\{3^+,3^-\}$ & $\{m_{120},m_{210},m_{1-10}\}$ \\
\hline
$A_1$ & $A_1$ & 1 & 1 & 1 \\
$A_2$ & $A_2$ & 1 & 1 &-1 \\
$E$   & $E$   & 2 &-1 & 0 \\
\hline
$E_{1/2}$ & $\ovl{E}_1$ & 2 & 1 & 0 \\
$^1E_{3/2}$ & ${}^1\ovl{E}$ & 1 &-1 & $i$ \\
$^2E_{3/2}$ & ${}^2\ovl{E}$ & 1 &-1 & $-i$
\end{tabular}
~~~~~
\begin{tabular}{c|c|cccccc}
\multicolumn{2}{c|}{PG 6} & 1 & $6^+$ & $3^+$ & $2$ & $3^-$ & $6^-$ \\
\hline
$A$ & $A$ & 1 & 1 & 1 & 1 & 1 & 1\\
$B$ & $B$ & 1 &-1 & 1 &-1 & 1 &-1\\
$^1E_1$ & $^1E_2$ & 1 & $e^{-i\frac{\pi}3}$ & $e^{-i\frac{2\pi}3}$ &-1 & $e^{i\frac{2\pi}3}$ & $e^{i\frac{\pi}3}$\\
$^2E_1$ & $^2E_2$ & 1 & $e^{i\frac{\pi}3}$ & $e^{i\frac{2\pi}3}$ &-1 & $e^{-i\frac{2\pi}3}$ & $e^{-i\frac{\pi}3}$\\
$^1E_2$ & $^1E_1$ & 1 & $e^{i\frac{2\pi}3}$ & $e^{i\frac{4\pi}3}$ & 1 & $e^{i\frac{2\pi}3}$ & $e^{i\frac{4\pi}3}$\\
$^2E_2$ & $^2E_1$ & 1 & $e^{-i\frac{2\pi}3}$ & $e^{-i\frac{4\pi}3}$ & 1 & $e^{-i\frac{2\pi}3}$ & $e^{-i\frac{4\pi}3}$\\
\hline
$^1E_{1/2}$ & ${}^2\ovl{E}_3$ & 1 & $e^{i\frac{\pi}6}$ & $e^{i\frac{\pi}3}$ & $i$ & $e^{-i\frac{\pi}3}$ & $e^{-i\frac{\pi}6}$\\
$^2E_{1/2}$ & ${}^1\ovl{E}_3$ & 1 & $e^{-i\frac{\pi}6}$ & $e^{-i\frac{\pi}3}$ & $-i$ & $e^{i\frac{\pi}3}$ & $e^{i\frac{\pi}6}$\\
$^1E_{3/2}$ & ${}^2\ovl{E}_1$ & 1 & $-i$ & $-1$ & $i$ & $-1$ & $i$\\
$^2E_{3/2}$ & ${}^1\ovl{E}_1$ & 1 & $i$ & $-1$ & $-i$ & $-1$ & $-i$\\
$^1E_{5/2}$ & ${}^2\ovl{E}_2$ & 1 & -$e^{i\frac{\pi}6}$ & $e^{i\frac{\pi}3}$ & $-i$ & $e^{-i\frac{\pi}3}$ & -$e^{-i\frac{\pi}6}$ \\
$^2E_{5/2}$ & ${}^1\ovl{E}_2$ & 1 & -$e^{-i\frac{\pi}6}$ & $e^{-i\frac{\pi}3}$ & $i$ & $e^{i\frac{\pi}3}$ & -$e^{i\frac{\pi}6}$
\end{tabular}
\vspace{0.5cm}

\begin{tabular}{c|c|cccccc}
\multicolumn{2}{c|}{PG $6mm$} & 1 & $\{6^+,6^-\}$ & $\{3^+,3^-\}$ & $2$ & $\{m_{100}, m_{010}, m_{110}\}$ & $\{m_{120},m_{210},m_{1-10}\}$ \\
\hline
$A_1$ & $A_1$ & 1 & 1 & 1 & 1 & 1 & 1\\
$A_2$ & $A_2$ & 1 & 1 & 1 & 1 &-1 &-1\\
$B_1$ & $B_1$ & 1 &-1 & 1 &-1 &-1 & 1\\
$B_2$ & $B_2$ & 1 &-1 & 1 &-1 & 1 &-1\\
$E_1$ & $E_1$ & 2 & 1 &-1 &-2 & 0 & 0\\
$E_2$ & $E_2$ & 2 &-1 &-1 & 2 & 0 & 0\\
\hline
$E_{1/2}$ & $\ovl{E}_1$ & 2 & $\sqrt3$ & 1 & 0 & 0 & 0 \\
$E_{3/2}$ & $\ovl{E}_3$ & 2 & 0 &-2 & 0 & 0 & 0 \\
$E_{5/2}$ & $\ovl{E}_2$ & 2 & $-\sqrt3$ & 1 & 0 & 0 & 0 \\
\end{tabular}
\caption{The character tables of 2D crystallographic PGs.
In each table, the symbols in the first and second columns are the Altmann-Herzig notations \cite{altmann_point-group_1994} and the \href{http://www.cryst.ehu.es/cgi-bin/cryst/programs/representations_point.pl?tipogrupo=dbg}{BCS notations} \cite{BCS1,*BCS2,*BCS3,Elcoro2017} of the irreps, respectively.
In each table, the linear irreps above the second horizontal line are the single-valued irreps (no-SOC), and the projective irreps below the second horizontal line are the double-valued irreps (SOC). \label{tab:2D-char}
}
\end{table}

Now we consider the time-reversal symmetry (TRS).
In presence of TRS, the group $G$ consists of a unitary part $G_U$ and an anti-unitary part $\TRS\cdot G_U$.
The basis functions of irreps of $G_U$, and their TRS-counterparts, if $\TRS$ transform them to other basis functions, form co-representation of $G$.
There are three cases of the irreps of $G_U$ \cite{Bradley2010}.
\begin{enumerate}[label=(\alph*)]
\item $\rho_{G_U}$ is equivalent to its complex conjugate $\rho_{G_U}^*$, \ie $\rho_{G_U} = N \rho_G^* N^{-1} $ for some unitary matrix $N$, and $N N^* = \TRS^2 $. Then $\rho_{G_U}$ itself form a co-representation of $\rho_{G}$.
\item $\rho_{G_U}$ is equivalent to $\rho_{G_U}^*$ and the transformation matrix satisfies $N N^* = -\TRS^2 $. Then the corresponding co-representation consists of the direct sum of two $\rho_{G_U}$ irreps, \ie $\rho_{G} = \rho_{G_U} \oplus \rho_{G_U}$ for unitary operations.
In the following we will denote such co-representation as $\rho_{G_U}\rho_{G_U}$.
\item $\rho_{G_U}$ is not equivalent to $\rho_{G_U}^*$, then the corresponding co-representation consists of $\rho_{G_U}$ and its complex conjugate, \ie  $\rho_{G} = \rho_{G_U} \oplus \rho_{G_U}^*$. In the following, we will denote such co-representation as $\rho_{G_U} \rho_{G_U}^*$.
\end{enumerate}
The three cases can be distinguished by the Frobenius-Schur indicator \cite{Bradley2010,Jennifer2018}.
In the following we will refer to irreducible co-representations of a group with TRS as co-irreps.
Here we take the PG (PG) $3$ ($C_3$) as an example to show the construction of co-representations. 
The character table of the irreps of PG $3$ is given in \cref{tab:2D-char}.
(Readers can find the character tables for all crystallographic PGs by the tool \href{http://www.cryst.ehu.es/cgi-bin/cryst/programs/representations_point.pl?tipogrupo=dbg}{Irreducible representations of the Double Point Groups} \cite{Bradlyn2017,Vergniory2017,Elcoro2017,Jennifer2018} on the \href{http://www.cryst.ehu.es}{Bilbao Crystallographic Server} (BCS) \cite{BCS1}.
Here we give the character tables of the 2D crystallographic PGs in \cref{tab:2D-char}.)
We first consider spinless case, where $\TRS^2=1$.
The PG $3$ has three single-valued irreps: $A$, ${}^1E$, and ${}^2E$. 
The first irrep $A$ is real and so belongs to case-(a). 
Thus the corresponding co-representation is just $A$.
The second and third irreps are complex conjugates of each other and hence they belong to case-(c). 
The corresponding co-prepresentation is ${}^1E\oplus{}^2E$ for the unitary operations (denoted as  ${}^1E{}^2E$).
Next we consider the spinfull case, where $\TRS^2=-1$.
The PG $3$ has three double-valued irreps: ${}^1\ovl{E}$, ${}^2\ovl{E}$, and $\ovl{E}$.
The first two are complex conjugates of each other and hence they belong to case-(c).
The corresponding co-representation is ${}^1\ovl{E} \oplus {}^2\ovl{E}$ (denoted as ${}^1\ovl{E}{}^2\ovl{E}$). 
The third irrep $\ovl{E}$ is real, \ie $\ovl{E} = N \cdot \ovl{E}^*\cdot N^\dagger$ where $N=1$.
Since $N\cdot N^*=1 =-\TRS^2$, the third irrep belongs to case-(b).
Thus the corresponding co-representation is $\ovl{E} \oplus \ovl{E}$ (denoted as $\ovl{E}\ovl{E}$).
Applying this algorithm to all the 2D crystallographic PGs, we obtain the co-irreps:
\begin{enumerate}[label=(\roman*)]
\item PG $1$. The no-SOC co-irrep: $A$. The SOC co-irrep: $\ovl{A}\ovl{A}$.
\item PG $2$. The no-SOC co-irreps: $A$, $B$. The SOC co-irrep: ${}^1\ovl{E}{}^2\ovl{E}$.
\item PG $m$. The no-SOC co-irreps: $A^\pr$, $A^\prpr$. The SOC co-irrep: ${}^1\ovl{E}{}^2\ovl{E}$.
\item PG $2mm$. The no-SOC co-irreps: $A_1$, $A_2$, $B_1$, $B_2$. The SOC co-irrep: $\ovl{E}\ovl{E}$.
\item PG $4$. The no-SOC co-irreps: $A$, $B$, ${}^1{E}{}^2{E}$. The SOC co-irreps: ${}^1\ovl{E}_1 {}^2\ovl{E}_1$, ${}^1\ovl{E}_2 {}^2\ovl{E}_2$.
\item PG $4mm$. The no-SOC co-irreps: $A_1$, $A_2$, $B_1$, $B_2$, $E$. The SOC co-irreps: $\ovl{E}_1$, $\ovl{E}_2$.
\item PG $3$. The no-SOC co-irreps: $A_1$, ${}^1E {}^2E$. The SOC co-irreps: $\ovl{E}\ovl{E}$, ${}^1\ovl{E} {}^2\ovl{E}$.
\item PG $3m$. The no-SOC co-irreps: $A_1$, $A_2$, $E$. The SOC co-irreps: $\ovl{E}_1$, ${}^1\ovl{E} {}^2\ovl{E}$.
\item PG $6$. The no-SOC co-irreps: $A$, $B$, ${}^1E_1 {}^2E_1$, ${}^1E_2 {}^2E_2$. The SOC co-irreps: ${}^1\ovl{E}_1 {}^2\ovl{E}_1$, ${}^1\ovl{E}_2 {}^2\ovl{E}_2$, ${}^1\ovl{E}_3 {}^2\ovl{E}_3$.
\item PG $6mm$. The no-SOC co-irreps: $A_1$, $A_2$, $B_1$, $B_2$, $E_1$, $E_2$. The SOC co-irreps: $\ovl{E}_1$, $\ovl{E}_2$, $\ovl{E}_3$.
\end{enumerate}

We consider the Frobenius reciprocity theorem in presence of TRS.
We assume that $H$ is a subgroup of $G$ that contains TRS and denote its maximal unitary subgroup as $H_U$, \ie $H=H_U + \TRS\cdot H_U$, $H\subset G$, $H_U\subset G_U$.
In order to find the relation between induction and subduction, we need to decompose the co-irreps of $G$ and $H$ into irreps of the unitary subgroups of $G$ and $H$, \ie $G_U$ and $H_U$.
If $\rho_{G}$ consists of a single case-(a) irrep $\rho_{G_U}$ of $G_U$, the multiplicity of $\rho_G$ in $\rho_H \up G$ coincides with the multiplicity of $\rho_{G_U}$ in $(\rho_H \up G)\down G_U$ because $\rho_G = \rho_{G_U}$.
If $\rho_{G}$ consists of case-(b) irreps of $G_U$, \ie $\rho_G = \rho_{G_U} \oplus \rho_{G_U}$, the multiplicity of $\rho_G$ in $\rho_H \up G$ is half the multiplicity of $\rho_{G_U}$ in $(\rho_H \up G)\down G_U$ because each $\rho_G$ decomposes into two $\rho_{G_U}$ in $G_U$.
If $\rho_{G}$ consists of case-(c) irreps of $G_U$, \ie $\rho_G = \rho_{G_U}\oplus \rho_{G_U}^*$,  the multiplicity of $\rho_G$ in $\rho_H \up G$ coincides with the multiplicity of $\rho_{G_U}$ in $(\rho_H \up G)\down G_U$ or the multiplicity of $\rho_{G_U}^*$ in $(\rho_H \up G)\down G_U$, thus it is half of the multiplicity of $\rho_{G_U}$ plus the multiplicity of $\rho_{G_U}^*$ in $(\rho_H \up G)\down G_U$.
Therefore, we have 
\begin{equation}
f(\rho_G | \rho_H \up G) =  \frac{1}{\zeta(\rho_G)} \sum_{i\in \rho_{G} \down G_U} f(\rho_{G_U}^i | (\rho_{H} \up G)\down G_U ) \label{eq:reciprocity-aux1}
\end{equation}
where $\zeta(\rho_G)=1,2,2$ for case-(a), (b), (c), respectively.
Notice that the irrep $\rho_{G_U}$ sums over each irrep in $\rho_G\down G_U$ only once.
For example, for $G$ is PG $3$ with TRS and SOC, if $\rho_G=\ovl{E}\ovl{E}$, which belongs to case-(b), there is
\begin{equation}
f(\ovl{E}\ovl{E} | \rho_H \up G) = \frac12 f(\ovl{E} | (\rho_H\up G)\down G_U);
\end{equation}
if $\rho_G={}^1\ovl{E}{}^2\ovl{E}$, which belongs to case-(c), there is
\begin{equation}
f({}^1\ovl{E}{}^2\ovl{E} | \rho_H \up G) = \frac12 f({}^1\ovl{E} | (\rho_H\up G)\down G_U) + \frac12 f({}^2\ovl{E} | (\rho_H\up G)\down G_U).
\end{equation}
The multiplicity of $\rho_{G_U}^i$ in $(\rho_{H} \up G)\down G_U $ can be calculated as
\begin{equation}
f(\rho_{G_U}^i | (\rho_{H} \up G)\down G_U ) =\frac{1}{|G_U|} \sum_{g\in G_U} \Tr[\rho_{G_U}^i(g)]^* \Tr[\rho_H(g)],
\end{equation}
The induced representation $\rho_{H} \up G$ is carried by the basis functions of $\rho_H$ and their copies in form of \cref{eq:sym-basis}, where $\td{g}_i \in G/H = G_U/H_U$.
The representation matrix of $g$ in $\rho_H\up G$ is $d(g)\otimes \rho_H(g)$, where $d(g)$ is defined in \cref{eq:small-d}.
Thus we obtain
\begin{equation}
f(\rho_{G_U}^i | (\rho_{H} \up G)\down G_U ) = \frac{1}{|G_U|} \sum_{\td{g}_k\in G_U/H_U} \sum_{g\in G_U} \Tr[\rho_{G_U}^i(g)]^* \Tr[ \rho_{H}(\td{g}_k^{-1} g \td{g}_k) ] d_{kk}(g).
\end{equation}
If $\rho_H$ consists of a single case-(a) irrep of $H_U$, there is $\Tr[ \rho_{H}(g^\pr) ] = \Tr[ \rho_{H_U}(g^\pr) ]$ because $\rho_H = \rho_{H_U}$, where $g^\pr =\td{g}_k^{-1} g \td{g}_k $.
If $\rho_H$ consists of case-(b) irreps of $H_U$, \ie $\rho_{H}=\rho_{H_U}\oplus \rho_{H_U}$, there is $\Tr[ \rho_{H}(g^\pr) ] = 2\Tr[ \rho_{H_U}(g^\pr) ]$ because $\rho_H$ decomposes into two $\rho_{H_U}$.
If $\rho_H$ consists of two case-(c) irreps of $H_U$, \ie $\rho_{H}=\rho_{H_U}\oplus \rho_{H_U}^*$, there is $\Tr[ \rho_{H}(g^\pr) ] = \Tr[ \rho_{H_U}(g^\pr) ] + \Tr[ \rho_{H_U}^*(g^\pr) ]$.
In summary there is
\begin{equation}
\Tr[ \rho_{H}(\td{g}_k^{-1} g \td{g}_k) ] = \sum_{j\in \rho_H\down H_U} \zeta^\pr(\rho_H) \Tr[\rho^j_{H_U}(\td{g}_k^{-1} g \td{g}_k) )],\label{eq:reciprocity-aux2}
\end{equation}
where $\zeta^\pr(\rho_H)=1,2,1$ for case-(a), (b), (c), respectively. 
Thus we obtain
\begin{align}
f(\rho_{G_U}^i | (\rho_{H} \up G)\down G_U ) &= \frac{1}{|G_U|} \sum_{j\in \rho_H\down H_U } \sum_{\td{g}_k\in G_U/H_U} \sum_{g\in G_U} \zeta^\pr(\rho_H) \Tr[\rho_{G_U}^i(g)]^* \Tr[ \rho_{H}^j(\td{g}_k^{-1} g \td{g}_k) ] d_{kk}(g) \nono \\
& = \zeta^\pr(\rho_H) \sum_{j\in \rho_H \down H_U} f(\rho_{G_U}^i | \rho_{H_U}^j \up G_U). 
\end{align}
Substituting this equation for \cref{eq:reciprocity-aux1}, we can write $f(\rho_G | \rho_H \up G)$ in terms of the induction coefficients of the irreps of unitary subgroups of $G$ and $H$
\begin{equation}
f(\rho_G | \rho_H \up G) = \frac{\zeta^\pr({\rho_H})}{\zeta(\rho_G)} \sum_{i \in \rho_{G} \down G_U} \sum_{j\in \rho_H \down H_U} f(\rho_{G_U}^i | \rho_{H_U}^j \up G_U).\label{eq:reciprocity-aux3}
\end{equation}

Now let us write $f(\rho_H | \rho_G \down H)$ in terms of the reduction coefficients of irreps of unitary subgroups of $G$ and $H$ and relate $f(\rho_H | \rho_G \down H)$ to $f(\rho_G | \rho_H \up G)$.
Following the same logic of \cref{eq:reciprocity-aux1}, we can calculate $f(\rho_H | \rho_G \down H)$ from $f(\rho_{H_U} | \rho_G \down H_U)$ as
\begin{equation}
f(\rho_H | \rho_G \down H) = \frac{1}{\zeta(\rho_H)} \sum_{j\in \rho_H\down H_U} f(\rho_{H_U}^j | \rho_G \down H_U ),
\end{equation}
where $\zeta(\rho_H)=1,2,2$ for case-(a), (b), (c), respectively.
Then, by decomposing $\rho_G$ into $\sum_{i\in \rho_G \down G_U} \zeta^\pr(\rho_{G_U}^i) \rho_{G}^i$, where $\zeta^\pr(\rho_H)=1,2,1$ for case-(a), (b), (c), respectively, we obtain
\begin{equation}
f(\rho_H | \rho_G \down H) = \frac{\zeta^\pr(\rho_G)}{\zeta(\rho_H)} \sum_{j\in \rho_H\down H_U} \sum_{i\in \rho_G \down G_U} f(\rho_{H_U}^j | \rho_{G_U}^i \down H_U ).
\end{equation}
Comparing this equation to \cref{eq:reciprocity-aux3} and making use of the Frobenius reciprocity of unitary groups $f(\rho_{H_U}^j | \rho_{G_U}^i \down H_U ) =f(\rho_{G_U}^i | \rho_{H_U}^j \up G_U) $, we find that
\begin{equation}\boxed{
\xi(\rho_G) f(\rho_G | \rho_H \up G) = \xi(\rho_H) f(\rho_H|\rho_G \down H), }\label{eq:Frobenius-TRS}
\end{equation}
where $\xi(\rho) = \zeta(\rho)\zeta^\pr(\rho) = 1,4,2$ if $\rho$ belongs to case-(a), (b), (c), respectively.
\cref{eq:Frobenius-TRS} is the Frobenius reciprocity for groups with TRS.

Here we take the group subgroup pair $G$= PG 4, $H=$ PG 1 with TRS $(\TRS^2=1)$ as an example to show the induction.
$A$, $B$ of PG 4 reduce the identity representation $A$ of PG 1, and $^1E^2E$ of PG4 reduce to two $A$ of PG 1, \ie
\begin{equation}
f( (A)_1 | (A)_4 \down 1) = f( (A)_1 | (B)_4 \down 1) = 1,\qquad
f( (A)_1 | ({}^1E{}^2E)_4 \down 1) = 2. 
\end{equation}
Here the subscripts of the irrep symbols represent the point groups to which the irreps belong.
Since $(A)_1$, $(A)_4$, $(B)_4$ belong to case-(a) co-irrep and $({}^1E{}^2E)_4$ belong to case-(c) co-irrep, we have $\xi((A)_1) = 1$, and $\xi((A)_4)=\xi((B)_4)=1$, $\xi(({}^1E{}^2E)_4)=2$.
According to \cref{eq:Frobenius-TRS}, we obtain
\begin{equation}
f( (A)_4 | (A)_1 \up 4) = f( (B)_4 | (A)_1 \up 4) =1,\qquad
f( ({}^1E{}^2E)_4 | (A)_1 \up 4)=2.
\end{equation}
In other words $(A)_1\up 4 = A \oplus B \oplus {}^1E{}^2E$.
One can also understand this by noting that the site-symmetry group of general position is identity; upon the representation of the identity group (identity) is induced to the higher group, we obtain the sum of all representation of the higher group, with the multiplicities equal to their dimensions, \ie $A\oplus B\oplus {}^1E \oplus {}^2E$, which with TRS becomes $A\oplus B\oplus {}^1E{}^2E$.

\subsection{Irrep of space group}\label{sec:irrep-SG}

\LE{Is it possible to change to the more standard $\{R|\tt\}$ for the symmetries in the Seitz notation? Moreover, the symbol $p$ is used for the coefficients of the EBRs. }
\SZD{I change $p_g$ to $R_g$.}

\LE{We define the group operation $\{R|\tt\}$ on momentum as $\kk \to R^T\kk $.}
\SZD{Fine. But I think the standard definition of group operation on wavefunctions is $\hat{R}\cdot \psi(\rr) = \psi(R^{-1}\rr)$. Then $R$ acts on momentum as $\kk \to R\kk$. I did notice that BCS used the definition $\hat{R}\cdot \psi(\rr) = \psi(R\rr)$, which caused some confusion when I applied the character tables on BCS to real materials. :)}

In a solid with translation symmetry, the Bloch wavefunctions at a fixed momentum form the irreps of the little group at the respective momentum.
The little group at momentum $\kk$ is defined as
\begin{equation}
G_\kk = \{ g=\{R_g|\tt_g\}\in G\; |\; R_g^T \kk \sim \kk \},
\end{equation}
where $G$ is the space group, $R_g$ is the rotational part of the group element $g\in G$, $\tt_g$ is the translation (in real space) part of $g$, $R_g^T \kk \sim \kk $ iff $R_g^T\kk -\kk$ is a vector of reciprocal lattice.
Notice that $G_\kk$ contains all the translation operations.
Suppose $\{R_1|\tt_1\}$, $\{R_2|\tt_2\}$, $\{R_3=R_1R_2|\tt_3\}$ are three operations in $G_\kk$ and they satisfy
\begin{equation}
\{R_1|\tt_1\} \{R_2|\tt_2\} =\{R_1R_2|\tt_1 + R_1\tt_2 \} = \{1|\tt_1 + R_1\tt_2 -\tt_3\} \{R_3|\tt_3\}
\end{equation}
Due to Bloch's theorem, the irreps of $G_\kk$ satisfy $\rho_{G_\kk}( \{1|\tt_1 + R_1\tt_2 -\tt_3\} ) = \exp(-i\kk\cdot(\tt_1 + R_1\tt_2 -\tt_3))$.
Thus we have
\begin{equation}
\rho_{G_\kk}(\{R_1|\tt_1\}) \rho_{G_\kk}(\{R_2|\tt_2\}) = \exp(-i\kk\cdot(\tt_1 + R_1\tt_2 -\tt_3))  \rho_{G_\kk}(\{R_3|\tt_3\}).
\end{equation}
We assign a $D(R_\nu)$ matrix to each operation $\{R_\nu|\tt_\nu\}$ in $G_\kk$.
Since $D(R_\nu)$ only depends on the rotational part $R_\nu$ of $\{R_\nu|\tt_\nu\}$, in principle we can only have $|G_\kk|/|T|$ different $D$ matrices, where $T$ is the infinite translation subgroup of $G_\kk$.
Making the transformation
\begin{equation}
\rho_{G_\kk}(\{R_\nu |\tt_\nu \}) = D(R_\nu) \exp(-i\kk\cdot \tt_\nu),   \label{eq:rho-Gk-D}
\end{equation}
and supposing $R_{\nu} R_{\nu^\pr}=R_\mu$
we find that 
\begin{align}
D(R_{\nu}) D(R_{\nu^\pr}) &= e^{i\kk\cdot(\tt_\nu + \tt_{\nu^\pr})} 
\rho_{G_\kk}(\{R_\nu | \tt_\nu \}) \rho_{G_\kk}(\{R_{\nu^\pr} | \tt_{\nu^\pr} \}) 
= e^{i\kk\cdot(\tt_{\nu^\pr}-R_\nu\tt_{\nu^\pr}+\tt_\mu)} \rho_{G_\kk}(\{R_\mu| \tt_\mu \}) \nono\\
&= \theta_{\nu,\nu^\pr} D(R_\mu), 
\end{align}
where
\begin{equation}
\theta_{\nu,\nu^\pr}= \exp(-i(R_{\nu}^T\kk - \kk)\cdot\tt_{\nu^\pr})
\end{equation}
The phase factor $\theta_{\nu,\nu^\pr}$ remains unchanged if we substitute $\tt_{\nu^\pr}$ by $\tt_{\nu^\pr} + \RR$, where $\RR$ is a lattice vector, because $R_{\nu}^T\kk - \kk$ is a reciprocal lattice and hence $(R_{\nu}^T\kk - \kk)\cdot\RR = 0 \mod 2\pi$.
$D$ is nothing but a projective representation of the PG \cite{Bradley2010}
\begin{equation}
P_\kk = \{ R_g \;|\; R_g^T \kk \sim \kk,\; g\in G \}.
\end{equation}
Due to \cref{eq:rho-Gk-D}, there is each irrep of $G_\kk$ corresponds to an projective irrep of the PG $P_\kk$.
(But in general not all projective irreps of $P_\kk$ correspond to irreps of $G_\kk$.)
\BAB{All all the projective irreps are realized? Or only some of them are realized?}
\SZD{Some of them.}
Tables of all the irreps of $G_\kk$ in all space groups are available at the \href{http://www.cryst.ehu.es/cgi-bin/cryst/programs/representations.pl?tipogrupo=dbg}{Irreducible representations of the Double Space Groups} tool \cite{Bradlyn2017,Elcoro2017,Vergniory2017,Jennifer2018} on the BCS \cite{BCS1}.

All the irreps of the space groups can be obtained as induced representations from irreps of little groups of some momenta, \ie $\rho_{G_\kk} \up G$.
The wavefunctions at equivalent momenta will be rotated to each other under the (finite) coset representatives in $G/G_\kk$ and hence will form the same irrep. 
Therefore, to obtain the complete irreps of $G$, one only needs to consider $\rho_{G_\kk} \up G$ for $\kk \in \mrm{IBZ}$, where IBZ stands for the irreducible Brillouin zone (BZ).
Interested readers might look at Ref. \cite{Bradley2010} for more details.

\subsection{A short review of topological quantum chemistry} \label{sec:TQC}
The symmetry property of a band structure is fully described by its decompositions into irreps of little groups at momenta in the Brillouin zone.
For gapped band structure, the multiplicities of the irreps at different $\kk$ are not independent: they satisfy the so-called compatibility relations \cite{Bradley2010}.
As a consequence, the irreps at all momenta are completely determined by the irreps at the so-called maximal momenta, the little groups of which are maximal subgroups of the space group \cite{Bradlyn2017,Po2017,Elcoro2017,Vergniory2017}.
We use symmetry data vector $B$ to denote the multiplicities in this decomposition
\begin{equation}
B = (m(\rho_{G_{K_1}}^1), m(\rho_{G_{K_1}}^2),\cdots,m(\rho_{G_{K_2}}^1),m(\rho_{G_{K_2}}^2),\cdots)^T,
\end{equation}
where $m(\rho_{G_{K_i}}^j)$ represents the multiplicity of the $j$th irrep of the little group at the maximal momentum $K_i$.
We denote the length, \ie the number of components, of symmetry data vector as $N_{S}$.
\BAB{The number of bands?}
\SZD{No, it's the number of components of $B$. 
I change the notation from $N_B$ to $N_S$ to avoid the misleading.}

Band representations (BR) are a special type of space group representations that are induced from irreps (orbitals) of the site-symmetry groups of the Wyckoff positions in real space \cite{Bradlyn2017,Elcoro2017,Vergniory2017,Jennifer2018}.
The site-symmetry group of a Wyckoff position is isomorphic to a PG and is defined as
\begin{equation}
    G_{w} = \{g = \{R_g|\tt_g\}\in G\ |\ R_g \tt_w + \tt_g = \tt_w\},
\end{equation}
where $\tt_w$ is the position of the Wyckoff position.
BR always decomposes into a multiple of elementary BRs (EBRs) \cite{Bradlyn2017,Elcoro2017,Vergniory2017}, which are defined as space group representations induced from irreps  the site-symmetry groups of the \uemph{maximal} Wyckoff positions.
The maximal Wyckoff positions those whose site-symmetry groups is not a subgroup of the site-symmetry group of another Wyckoff position connected to the first one.
The $B$-vector of the EBR induced from the $j$th irrep at the Wyckoff position $w$ is given as $B^{\rho_{G_{w}}^j}$.
We define the EBR matrix as the matrix where columns are $B$-vectors induced from irreps at all maximal Wyckoff positions
\begin{equation}
\EBR = ( B^{\rho_{G_{w_1}}^1}, B^{\rho_{G_{w_1}}^2},\cdots B^{\rho_{G_{w_2}}^1}, B^{\rho_{G_{w_2}}^2},\cdots ),\label{eq:EBR-def}
\end{equation}
where $B^{\rho_{G_{w_i}}^j}$ is the symmetry data vector of the band representation induced from the $j$th irrep at the $i$th maximal Wyckoff position.
The EBRs in all space groups are available at the \href{http://www.cryst.ehu.es/cgi-bin/cryst/programs/bandrep.pl}{Band representations of the Double Space Groups} section \cite{Bradlyn2017,Vergniory2017,Elcoro2017,Jennifer2018} on the BCS \cite{BCS1}.
We denote the number of EBRs as $N_{EBR}$. 
Thus $\EBR$ is an $N_{S}\times N_{\EBR}$ integer matrix.

The symmetry data vector of a gapped band structure \uemph{always} decomposes into a linear combination of EBRs, \ie  
\begin{equation}
B_i = \sum_j [\EBR]_{ij} p_j = [\EBR\cdot p]_i, \label{eq:B-EBR}
\end{equation}
where the $p$-vector
\begin{equation}
p =  ( p(\rho_{G_{w_1}}^1), p(\rho_{G_{w_1}}^2),\cdots p(\rho_{G_{w_2}}^1), p(\rho_{G_{w_2}}^2),\cdots )^T
\end{equation}
gives the coefficients on the EBRs.
There are four categories for the decomposition coefficients:

\noindent
\fbox{\parbox{\textwidth}{
\begin{enumerate}[label=(\Roman*)]
\item If some $p_i$ is fractional, then the band structure has a stable topology \cite{Po2017,Bradlyn2017}. Examples in this category include the inversion symmetric topological insulator, higher-order topological states protected by crystalline symmetries, and others \etc \cite{Khalaf2018,Song_NC_2018}.
\item If all $p_i$ are integers and some $p_i$ are \uemph{necessarily} negative, then the band structure at least has a fragile topology. 
In general, the coefficient vector $p$ is not unique because the EBRs may be linearly dependent.
Thus the condition for this category is that in \uemph{all} the possible integer decompositions some $p_i$ have to be negative.
Examples in this category include twisted bilayer graphene. We refer to such phase as the eigenvalue fragile phase (EFP). EFP may also has a stable topology which cannot be diagnosed from symmetry eigenvalues \cite{song_fragile_2019}. 
\item If a band structure does not belong to the first two categories, then we can choose $p$ as a nonnegative integer vector. In all these nonnegative integer decompositions, if some $p_i$ corresponding to unoccupied Wyckoff positions are \uemph{necessarily} nonzero, then the symmetry data of the band structure is consistent with an obstructed atomic insulator, which is Wannierizable but some WFs locate at unoccupied Wyckoff positions.
We refer to such phase as eigenvalue obstructed atomic phase (EOAP). EOAP may also has stable or fragile topology that cannot be diagnosed from symmetry eigenvalues.
\item If all $p_i$ are nonnegative integers and all nonzero $p_i$ correspond to occupied Wyckoff positions, then the symmetry data of the band structure is consistent with a trivial atomic insulator. We call such phase as eigenvalue trivial atomic phase (ETAP). ETAP may also have stable or fragile topology that cannot be diagnosed from symmetry eigenvalues, but can (always) be diagnosed from Wilson loops \cite{yu_WL_2011}.
\end{enumerate}}}


Here we briefly introduce an algorithm to determine the category of a given symmetry data vector.
Any integer matrix has a unique Smith normal form \cite{wiki:SNF}. 
\BAB{up to signs?}
\SZD{The diagonal elements of Smith normal form is usually chosen as positive.}
We write the Smith decomposition of the EBR matrix as
\begin{equation}
\EBR = L_\EBR \Lambda_\EBR R_\EBR, \label{eq:EBR-snf}
\end{equation}
where $L_{\EBR}$ is an $N_{S}\times N_{S}$ unimodular integer matrix, $R_{\EBR}$ is an $N_\EBR \times N_\EBR$ unimodular matrix, and $\Lambda_\EBR$ is the Smith normal form of $\EBR$, which is an $N_{S}\times N_\EBR$ integer matrix with nonzero elements only in the diagonal.
We denote the rank of $\EBR$ as $r$ and the Smith normal form $\Lambda_\EBR$ as
\begin{equation}
\Lambda_\EBR = \begin{pmatrix}
\lambda_1 & \cdots & 0     & 0 & \cdots & 0\\
\vdots & \ddots  & \vdots & \vdots & \ddots & \vdots\\
0   & \cdots & \lambda_{r} & 0 & \cdots & 0\\
0 & \cdots & 0 & 0 & \cdots & 0\\
\vdots & \ddots & \vdots & \vdots & \ddots & \vdots\\
0 & \cdots & 0 & 0 & \cdots & 0
\end{pmatrix}.
\end{equation}
Applying $L_\EBR^{-1}$ on both sides of \cref{eq:B-EBR}, we obtain
\begin{equation}
L_\EBR^{-1} B = \Lambda_\EBR R_\EBR\cdot p. \label{eq:B-p}
\end{equation}
In order for the solution $p$ to exist, only the first $r$ rows of both sides are nonzero.
We introduce the $y$-vector for the symmetry data vector $B$ as 
\begin{equation}
y_i = \frac{1}{\lambda_i}[L_\EBR^{-1} \cdot B]_i = [R_\EBR \cdot p]_i,\qquad i=1\cdots r. \label{eq:y-def}
\end{equation}
Notice that in general $y_i$ are rational numbers.
Then the $B$-vector can be expressed as 
\begin{equation}
B_j = \sum_{i=1}^{r} [L_\EBR]_{ji} y_i \lambda_i. \label{eq:B-y}
\end{equation}
The correspondence between the $B$-vector and the $y$-vector is one-to-one since $L_{EBR}$ is  unimodular.
However, in general, the correspondence between $y$ and $p$ is one-to-many.
For a given $B$ ($y$), the general solution of $p$ can be given as
\begin{equation}
p = R_\EBR^{-1} \cdot ( y_1, y_2, \cdots, y_r,
k_1,k_2,\cdots, k_{N_\EBR -r } )^T, \label{eq:p-y-k}
\end{equation}
where $k_j$ are free parameters. 
Which category $B$ belongs to can be diagnosed with the analysis the values of $y$.
Depending on character of $y$ we find four categories
\begin{enumerate}[label=(\Roman*)]
\item Since $R_\EBR$ is a unimodular matrix, $p$ is an integer vector iff all $y_i$ and $k_i$ are integers. As $k_i$ are free parameters, $p$ has an integer solution iff all $y_i$ are integers. One can define the symmetry-based-indicators \cite{Po2017} as
\begin{equation}\boxed{
z_i \equiv (y_i\cdot\lambda_i) \mod \lambda_i =(L_{EBR}^{-1} \cdot B)_i \mod \lambda_i,\qquad i=1\cdots r,}
\end{equation} 
which takes integer values from $0$ to $\lambda_i-1$.
$y_i$ is integer iff $z_i=0$.
Therefore, nonzero $z_i$ implies fractional $y$ and $p$ and hence a stable topology.
\item On the one hand, the entries of the symmetry data vector $B$ are multiplicities of irreps and thus are nonnegative, \ie $B\ge0$. This constraint defines a polyhedral cone in the $y$-space
\begin{equation}
Y=\brace{ y\in \mathbb{Q}^r\ \Big|\ \sum_{i=1}^{r} [L_\EBR]_{ji} y_i \lambda_i \ge 0\quad (j=1\cdots N_{S}) },
\end{equation}
where $\mathbb{Q}$ stands for rational number.
A band structure with zero symmetry-based indicators $z_i$ corresponds to an integer point, \ie $y\in \mbb{Z}^r$, in the polyhedral cone. The set formed by these integer points form the affine monoid $\ovl{Y} = \mathbb{Z}^r\cap Y$ \cite{song_fragile_2019}. 
On the other hand, all the $y$-vectors represent combinations of EBRs with nonnegative coefficients, which correspond to Wannierizable states.
They are given by another affine monoid \cite{song_fragile_2019}
\begin{equation}
\ovl{X} = \brace{ x\in \mathbb{Z}^r\ \Big|\ x_i = [R_\EBR\cdot p],\; p\in \mathbb{N}^{N_\EBR} },
\end{equation}
where $\mbb{N}$ stands for nonnegative integer. 
Therefore, the EFPs are given by integer points in $\ovl{Y}-\ovl{X}$, \ie
\begin{equation}\boxed{
    B\ \text{is EFP}\quad \Leftrightarrow\quad y \in \ovl{Y}-\ovl{X}.}
\end{equation}
As shown in Ref. \cite{song_fragile_2019}, the criterion of EFP is either an inequality or a $\mbb{Z}_2$ equation. 
\item If a symmetry data vector is neither diagnosed as stable topological phase nor an EFP, it might be an eigenvalue obstructed atomic phase (EOAP) or an eigenvalue trivial atomic phase (ETAP). 
For a given $B$ ($y$), there are finite different decompositions to EBRs with nonnegative coefficients. These decompositions are given as the integer points in the following polyhedral cone
\begin{equation}
\brace{ p \in \mbb{Q}^{N_\EBR}\ \Big|\ p\ge0,\quad \EBR \cdot p = B }.
\end{equation}
The number of different decompositions is finite because each component of $p$ is bounded by 
\begin{equation}
p (\rho_{G_w}^i) \times (\#\ \text{of bands of}\ \rho_{G_w}^i\up G) \le \#\ \text{of bands of}\ B.
\end{equation}
Here $p(\rho_{G_w}^i)$ is the component of $p$ corresponding to the $i$th irrep at the Wyckoff position $w$.
If in all the possible decompositions, the multiplicity of some irrep at an unoccupied Wyckoff position is always nonzero, then $B$ is an EOAP.
\item If $B$ does not belong to the first three categories, then $B$ belongs to ETAP.
\end{enumerate}


\section{Real space invariant}\label{sec:RSI}

\subsection{RSI as good quantum number in adiabatic process}\label{sec:RSI-quantum-number}

We consider a finite system that respects a PG symmetry.
The finite system can be a molecule, or a \BAB{parg ???} material with open boundary condition (OBC).
For the first case, we consider there is a finite gap between the highest occupied level and the lowest unoccupied level.
For the latter case, we only consider the trivial atomic insulator, obstructed atomic insulator, and fragile topological phase such that there is a finite gap above the occupied bulk energy levels.
The energy eigenstates, which can be many and which are finite-size orbitals, transform under irreps of the PG.
In an adiabatic process that mantains the PG symmetry, the Wannier centers of these eigenstates (in real space) can move.
A set of orbitals at far enough positions (outside the system) can move into the system and form an induced representation of the PG.
Conversely, a set of orbitals in the system can move outside the system if they form an induced representation.
In the following, we are going to find the invariants of the finite system in such adiabatic processes.

\subsection{Inversion example}
Before going to general case, we first take the PG $\bar1$ as an example.
$\bar1$ consists of the identity and the inversion symmetry.
There are only two irreps: $A_g$ (even) and $A_u$ (odd). 
We denote the multiplicities of the irreps formed by the occupied eigenstates \uemph{in} the system as $m(A_g)$ and $m(A_u)$, respectively.
\BAB{The eigenvalues are with ???}
Now we consider the displacements of two orbitals from outside the system into the system.
\BAB{Draw a figure please}
These two orbitals transform into each other under the inversion and so form the representation $A_g \oplus A_u$.
Once these two orbitals have been moved into the system, the multiplicities of the irreps change as $ m(A_g) \to m(A_g)+1 $,  $m(A_u) \to m(A_u)+1$.
The particle number is increased by 2.
However, the quantity
\begin{equation}
\dt  = m(A_u) - m(A_g) \label{eq:dt-inv}
\end{equation}
is unchanged in this process.
Therefore, \cref{eq:dt-inv} is a good quantum number in the adiabatic process.
We refer to \cref{eq:dt-inv} as the real space invariant (RSI) of the PG $\bar1$.

\subsection{Algorithm for RSI} \label{sec:algoritm-RSI}

Now we generalize the concept of RSI to a generic PG.
We denote the PG as $G$ and the symmetry data vector of the system as $p$, where its components give the multiplicities of the irreps of $G$,
\begin{equation}
p = (m(\rho_G^1), m(\rho_G^2),\cdots,m(\rho_G^{n_p}))^T.
\end{equation}
Here $m(\rho_G^i)$ is the multiplicity of the $i$th irrep of $G$, and $n_{p}$ is the number of irreps of $G$.
If we allow the components of $p$ to be negative, then all the $p$-vectors form an abelian group $\{p\}$.
We can induce (generically reducible) representations of $G$ from irreps of its subgroups.
This induction corresponds to adding orbitals from low-symmetry positions to the system.
We denote the abelian group formed by $p$-vectors induced from this trivial group as $\{q\}$.
Then the quotient group $\{p\}/\{q\}$ defines the good quantum numbers in the adiabatic process.
We call these good quantum numbers real space invariants (RSIs).
In principle we need to consider inductions from all the subgroups of $G$ that correspond to \uemph{lower symmetry} Wyckoff positions in real space.
However, in practice, we only need to consider the maximal subgroups of $G$ that correspond to lower symmetry Wyckoff positions, because the induced representation from non-maximal subgroup is always equivalent to a sum of induced representations from maximal subgroups.
To be specific, if $H_1\subset H_2 \subset G$, then $\rho_{H_1} \up G$ can be calculated by first inducing $\rho_{H_1}$ to ${H_2}$ and then inducing the induced irreps in $H_2$ to $G$, \ie $\rho_{H_1}\up G = (\rho_{H_1}\up H_2) \up G$.

Now we describe the algorithm to calculate the RSIs.
We take every irrep of every maximal subgroup of $G$ that correspond to lower symmetry Wyckoff position to induce a symmetry data vector of $G$.
We suppose there are $n_q$ such induced symmetry data vectors and denote them as $\aa_i$ ($i=1,2\cdots n_q$).
$\aa_i$ are the generators of $\{q\}$.
(Notice that each $\aa_i$ is a vector with length $n_p$ and $\aa_i$ may be linearly dependent on each other.)
We use $\aa_i$ to construct an $n_p\times n_q$ matrix $C=[a_1,a_2,\cdots a_{n_q}]$, which is similar with the EBR matrix (\cref{eq:EBR-def}) except that here both the supergroup and subgroup are PGs.
To obtain the quotient group $\{p\}/\{q\}$, we denote the generators of $\{p\}$ as $\bb_i$ ($i=1\cdots n_p$).
For simplicity, we can simply choose the generators as $[\bb_i]_j=\delta_{ij}$.
We can always expand $\aa_i$ as a linear combination of $\bb_j$ with integer coefficients
\begin{equation}
\aa_i = \sum_j \bb_j C_{ji}.\label{eq:abC}
\end{equation}
We apply the Smith decomposition of $C$ as
\begin{equation}
C = L_C \Lambda_C R_C,
\end{equation}
where $L_C$ is an $n_{p}\times n_p$ unimodular matrix, $R_C$ is an $n_q\times n_q$ unimodular matrix, and $\Lambda_C$ is an $n_p\times n_q$ integer matrix where the nonzero terms are only in the diagonal.
We write the explicit form of $\Lambda_C$ as
\begin{equation}
\Lambda_C = \begin{pmatrix}
\kappa_1 & \cdots & 0     & 0 & \cdots & 0\\
\vdots & \ddots  & \vdots & \vdots & \ddots & \vdots\\
0   & \cdots & \kappa_{r_q} & 0 & \cdots & 0\\
0 & \cdots & 0 & 0 & \cdots & 0\\
\vdots & \ddots & \vdots & \vdots & \ddots & \vdots\\
0 & \cdots & 0 & 0 & \cdots & 0
\end{pmatrix}.\label{eq:Lambda-C}
\end{equation}
Here $\kappa_i>0$, and $r_q$ is the number of linearly independent $\aa_i$ (or the rank of $\{q\}$).
The left top, right top, left bottom, and right bottom have the dimensions $r_q\times r_q$, $r_q\times (n_q-r_q)$, $(n_p-r_q)\times r_q$, and $(n_p-r_q)\times(n_q-r_q)$, respectively.
We define $ \bb_j^\pr = \sum_k \bb_{k} [L_C]_{k j}$ ($j=1\cdots n_p$) .
Since $L_C$ is unimodular, $\bb_j^\pr$ can be chosen as the generators of $\{p\}$.
Correspondingly, since $R_C$ is unimodular, $\aa_i^\pr= \sum_k \aa_k [R_C^{-1}]_{ki}$ ($i=1\cdots n_a$) can be chosen as generators of $\{q\}$.
Since $\aa_i^\pr = \sum_j \bb_j^\pr \Lambda_{ji}$, due to \cref{eq:Lambda-C}, we obtain
\begin{equation}
\aa_i^\pr = \begin{cases}
\bb_i^\pr \kappa_i,\qquad & i=1\cdots r_q\\
0,\qquad & i=r_q\cdots n_q
\end{cases}. \label{eq:ap-bp}
\end{equation}
We can decompose $\{p\}$ as a product of two parts: $\{p\}_1$ spanned by $\bb^\pr_i$ ($i=1\cdots r_q$) and $\{p\}_2$ spanned by $\bb^\pr_i$ ($i=r_q+1 \cdots n_p$).
Thus $\{p\}$ is isomorphic to $\{p\}_1\times \mbb{Z}^{n_p-r_q}$.
From \cref{eq:ap-bp}, it is clear that $\{q\} \subset \{p\}_1$ and $\{p\}_1/\{q\} = \mbb{Z}_{\kappa_1} \times \cdots \times \mbb{Z}_{\kappa_{r_q}}$.
Therefore we obtain
\begin{equation}\boxed{
\{p\}/\{q\} = \mbb{Z}_{\kappa_1} \times \cdots \times \mbb{Z}_{\kappa_{r_q}} \times \mbb{Z}^{n_p-r_q}. } \label{eq:RSI-group}
\end{equation}
Given a vector $p = \sum_i \bb_i p_i = \sum_i \bb_k^\pr [L_C^{-1}]_{ki} p_i$, we can determine its RSIs as
\begin{equation}\boxed{
\delta_i(p) = \begin{cases}
(L_C^{-1}\cdot p)_i \mod \kappa_i,& \qquad i\le r_q\\
(L_C^{-1}\cdot p)_i,& \qquad i> r_q\\
\end{cases} .} \label{eq:RSI}
\end{equation}

\subsection{RSI in 2D crystallographic PGs}\label{sec:RSI-PG}
In this section we calculate the RSIs in the 2D PGs: $1$, $2$, $m$, $2mm$, $4$, $4mm$, $3$, $3mm$, $6$, $6mm$.
We apply the method described in \cref{sec:RSI} to all the PGs and obtain the RSI groups (\cref{tab:RSI-group}) and explicit formula of RSIs (\cref{tab:RSI-formula}).
\begin{table}\scriptsize
\centering
\begin{tabular}{|c|L{4cm}|L{4cm}|L{4cm}|L{4cm}|}
\hline
 & NoSOC, NoTRS & NoSOC, TRS & SOC, NoTRS & SOC, TRS\\
\hline
\hline
\multirow{1}{*}{$2$}& $ \delta_{1} = - m(\mathrm{A}) + m(\mathrm{B}) $& $ \delta_{1} = - m(\mathrm{A}) + m(\mathrm{B}) $& $ \delta_{1} = - m(\mathrm{{}^{1}\overline{E}}) + m(\mathrm{{}^{2}\overline{E}}) $& $ \delta_{1} = m(\mathrm{{}^{1}\overline{E}{}^{2}\overline{E}}) \mod 2 $\\
\hline
\multirow{1}{*}{$m$}& $ \delta_{1} = - m(\mathrm{A'}) + m(\mathrm{A''}) $& $ \delta_{1} = - m(\mathrm{A'}) + m(\mathrm{A''}) $& $ \delta_{1} = - m(\mathrm{{}^{1}\overline{E}}) + m(\mathrm{{}^{2}\overline{E}}) $& $ \delta_{1} = m(\mathrm{{}^{1}\overline{E}{}^{2}\overline{E}}) \mod 2 $\\
\hline
\multirow{1}{*}{$2mm$}& $ \delta_{1} = - m(\mathrm{A_{1}}) - m(\mathrm{A_{2}}) + m(\mathrm{B_{1}}) + m(\mathrm{B_{2}}) $& $ \delta_{1} = - m(\mathrm{A_{1}}) - m(\mathrm{A_{2}}) + m(\mathrm{B_{1}}) + m(\mathrm{B_{2}}) $&& $ \delta_{1} = m(\mathrm{\overline{E}}) \mod 2 $\\
\hline
\multirow{3}{*}{$4$}& $ \delta_{1} = - m(\mathrm{A}) + m(\mathrm{{}^{1}E}) $& $ \delta_{1} = - m(\mathrm{A}) + m(\mathrm{{}^{1}E{}^{2}E}) $& $ \delta_{1} = m(\mathrm{{}^{2}\overline{E}_{1}}) - m(\mathrm{{}^{1}\overline{E}_{1}}) $& $ \delta_{1} = - m(\mathrm{{}^{1}\overline{E}_{1}{}^{2}\overline{E}_{1}}) + m(\mathrm{{}^{1}\overline{E}_{2}{}^{2}\overline{E}_{2}}) $\\
& $ \delta_{2} = - m(\mathrm{A}) + m(\mathrm{B}) $& $ \delta_{2} = - m(\mathrm{A}) + m(\mathrm{B}) $& $ \delta_{2} = - m(\mathrm{{}^{1}\overline{E}_{1}}) + m(\mathrm{{}^{1}\overline{E}_{2}}) $& $ \delta_{2} = m(\mathrm{{}^{1}\overline{E}_{1}{}^{2}\overline{E}_{1}}) \mod 2 $\\
& $ \delta_{3} = - m(\mathrm{A}) + m(\mathrm{{}^{2}E}) $&& $ \delta_{3} = - m(\mathrm{{}^{1}\overline{E}_{1}}) + m(\mathrm{{}^{2}\overline{E}_{2}}) $&\\
\hline
\multirow{2}{*}{$4mm$}& $ \delta_{1} = - m(\mathrm{A_{1}}) - m(\mathrm{A_{2}}) + m(\mathrm{E}) $& $ \delta_{1} = - m(\mathrm{A_{1}}) - m(\mathrm{A_{2}}) + m(\mathrm{E}) $& $ \delta_{1} = - m(\mathrm{\overline{E}_{1}}) + m(\mathrm{\overline{E}_{2}}) $& $ \delta_{1} = - m(\mathrm{\overline{E}_{1}}) + m(\mathrm{\overline{E}_{2}}) $\\
& $ \delta_{2} = - m(\mathrm{A_{1}}) - m(\mathrm{A_{2}}) + m(\mathrm{B_{1}}) + m(\mathrm{B_{2}}) $& $ \delta_{2} = - m(\mathrm{A_{1}}) - m(\mathrm{A_{2}}) + m(\mathrm{B_{1}}) + m(\mathrm{B_{2}}) $&& $ \delta_{2} = m(\mathrm{\overline{E}_{1}}) \mod 2 $\\
\hline
\multirow{2}{*}{$3$}& $ \delta_{1} = - m(\mathrm{A_{1}}) + m(\mathrm{{}^{1}E}) $& $ \delta_{1} = - m(\mathrm{A_{1}}) + m(\mathrm{{}^{1}E{}^{2}E}) $& $ \delta_{1} = - m(\mathrm{{}^{2}\overline{E}}) + m(\mathrm{{}^{1}\overline{E}}) $& $ \delta_{1} = - m(\mathrm{{}^{1}\overline{E}{}^{2}\overline{E}}) + 2 m(\mathrm{\overline{E}\overline{E}}) $\\
& $ \delta_{2} = - m(\mathrm{A_{1}}) + m(\mathrm{{}^{2}E}) $&& $ \delta_{2} = - m(\mathrm{{}^{2}\overline{E}}) + m(\mathrm{\overline{E}}) $&\\
\hline
\multirow{1}{*}{$3m$}& $ \delta_{1} = - m(\mathrm{A_{1}}) - m(\mathrm{A_{2}}) + m(\mathrm{E}) $& $ \delta_{1} = - m(\mathrm{A_{1}}) - m(\mathrm{A_{2}}) + m(\mathrm{E}) $& $ \delta_{1} = - m(\mathrm{\overline{E}_{1}}) + m(\mathrm{{}^{1}\overline{E}}) + m(\mathrm{{}^{2}\overline{E}}) $& $ \delta_{1} = - m(\mathrm{\overline{E}_{1}}) + 2 m(\mathrm{{}^{1}\overline{E}{}^{2}\overline{E}}) $\\
\hline
\multirow{5}{*}{$6$}& $ \delta_{1} = - m(\mathrm{A}) + m(\mathrm{{}^{1}E_{2}}) $& $ \delta_{1} = - m(\mathrm{A}) + m(\mathrm{{}^{1}E_{2}{}^{2}E_{2}}) $& $ \delta_{1} = - m(\mathrm{{}^{2}\overline{E}_{3}}) + m(\mathrm{{}^{1}\overline{E}_{3}}) $& $ \delta_{1} = - m(\mathrm{{}^{1}\overline{E}_{3}{}^{2}\overline{E}_{3}}) + m(\mathrm{{}^{1}\overline{E}_{1}{}^{2}\overline{E}_{1}}) $\\
& $ \delta_{2} = - m(\mathrm{A}) + m(\mathrm{{}^{2}E_{1}}) $& $ \delta_{2} = - m(\mathrm{A}) + m(\mathrm{{}^{1}E_{1}{}^{2}E_{1}}) $& $ \delta_{2} = - m(\mathrm{{}^{2}\overline{E}_{3}}) + m(\mathrm{{}^{2}\overline{E}_{1}}) $& $ \delta_{2} = - m(\mathrm{{}^{1}\overline{E}_{3}{}^{2}\overline{E}_{3}}) + m(\mathrm{{}^{1}\overline{E}_{2}{}^{2}\overline{E}_{2}}) $\\
& $ \delta_{3} = - m(\mathrm{A}) + m(\mathrm{B}) $& $ \delta_{3} = - m(\mathrm{A}) + m(\mathrm{B}) $& $ \delta_{3} = - m(\mathrm{{}^{2}\overline{E}_{3}}) + m(\mathrm{{}^{2}\overline{E}_{2}}) $& $ \delta_{3} = m(\mathrm{{}^{1}\overline{E}_{3}{}^{2}\overline{E}_{3}}) \mod 2 $\\
& $ \delta_{4} = - m(\mathrm{A}) + m(\mathrm{{}^{1}E_{1}}) $&& $ \delta_{4} = - m(\mathrm{{}^{2}\overline{E}_{3}}) + m(\mathrm{{}^{1}\overline{E}_{2}}) $&\\
& $ \delta_{5} = - m(\mathrm{A}) + m(\mathrm{{}^{2}E_{2}}) $&& $ \delta_{5} = - m(\mathrm{{}^{2}\overline{E}_{3}}) + m(\mathrm{{}^{1}\overline{E}_{1}}) $&\\
\hline
\multirow{3}{*}{$6mm$}& $ \delta_{1} = - m(\mathrm{A_{1}}) - m(\mathrm{A_{2}}) + m(\mathrm{E_{1}}) $& $ \delta_{1} = - m(\mathrm{A_{1}}) - m(\mathrm{A_{2}}) + m(\mathrm{E_{1}}) $& $ \delta_{1} = - m(\mathrm{\overline{E}_{1}}) + m(\mathrm{\overline{E}_{3}}) $& $ \delta_{1} = - m(\mathrm{\overline{E}_{1}}) + m(\mathrm{\overline{E}_{3}}) $\\
& $ \delta_{2} = - m(\mathrm{A_{1}}) - m(\mathrm{A_{2}}) + m(\mathrm{E_{2}}) $& $ \delta_{2} = - m(\mathrm{A_{1}}) - m(\mathrm{A_{2}}) + m(\mathrm{E_{2}}) $& $ \delta_{2} = - m(\mathrm{\overline{E}_{1}}) + m(\mathrm{\overline{E}_{2}}) $& $ \delta_{2} = - m(\mathrm{\overline{E}_{1}}) + m(\mathrm{\overline{E}_{2}}) $\\
& $ \delta_{3} = - m(\mathrm{A_{1}}) - m(\mathrm{A_{2}}) + m(\mathrm{B_{1}}) + m(\mathrm{B_{2}}) $& $ \delta_{3} = - m(\mathrm{A_{1}}) - m(\mathrm{A_{2}}) + m(\mathrm{B_{1}}) + m(\mathrm{B_{2}}) $&& $ \delta_{3} = m(\mathrm{\overline{E}_{1}}) \mod 2 $\\
\hline
\end{tabular}

\protect\caption{\label{tab:RSI-formula}The explicit expressions of RSIs in 2D crystallographic PGs. }
\end{table}

Here we take the PG $3$, which is generated from the $C_3$ rotation, as an example to show the calculation.
We first consider the case without SOC and TRS.
$3$ has only one subgroup, the identity group $1$; and all the three irreps $A$, $^1E$, $^2E$ (\cref{tab:2D-char}) reduce to the identity representation of $1$.
Then, according to the Frobenius reciprocity theorem (\cref{eq:Frobenius}), the induced representation from the identity group $1$ contains all the three irreps, where each irrep appears once, \ie
\begin{equation}
(A)_1 \up 3 = A \oplus {}^1E \oplus {}^2E.
\end{equation}
Here $(A)_1$ represents the identity representation of $1$, and $A$, $^1E$, $^2E$ are irreps of PG 3 (\cref{tab:2D-char}).
The $C$ matrix consists of $p$-vectors induced from irreps of the subgroups.
Since here the only subgroup is identity and the only irrep of the identity group is identity, $C$ is an one-column matrix.
We assume the orders of irreps of 3 in as $A$, $^1E$, $^2E$.
The $p$-vector induced from the identity group is $(1,1,1)^T$.
Thus $C=[1,1,1]^T$. 
The Smith decomposition of $C$ reads
\begin{equation}
C = L_C \Lambda_C R_C = 
\pare{\begin{array}{rrr}
1 & 0 & 0 \\
1 & 1 & 0 \\
1 & 0 & 1
\end{array}}
\begin{pmatrix}
1\\ 0 \\ 0
\end{pmatrix}
(1).
\end{equation}
According to \cref{eq:RSI}, we have two $\mbb{Z}$-type RSIs.
The inverse of the $L_C$ matrix is 
\begin{equation}
L_C^{-1} = 
\pare{\begin{array}{rrr}
1 & 0 & 0 \\
-1 & 1 & 0 \\
-1 & 0 & 1
\end{array}},
\end{equation}
and the formulae for the $\mbb{Z}$-type RSIs are
\begin{equation}
\dt_1^{\rm NSOC,NTR} = m(^1E) - m(A_1),\qquad \dt_2^{\rm NSOC,NTR} = m(^2E) - m(A_1). \label{eq:RSI-3-NSOC-NTR}
\end{equation}

We secondly consider the case with TRS but without SOC.
The RSIs can be obtained using the method introduced in \cref{sec:algoritm-RSI}, but here we will derive them from the RSIs of PG $3$ without TRS.
In presence of TRS, the irreps $^1E$ and $^2E$ of $3$ become degenerate (they are mutually conjugated) and form the co-irrep $^1E^2E$.
(See \cref{tab:2D-char} and \cref{sec:TRS-irrep}.)
Thus, if we forget the TRS and calculate the two no-TRS RSIs in \cref{eq:RSI-3-NSOC-NTR}, since $^1E^2E$ decomposes into $^1E \oplus {}^2E$, there will always be $\dt_1 = \dt_2 = m(^1E^2E)-m(A)$.
Thus we obtain the RSI with TRS as
\begin{equation}
\dt_1^{\rm NSOC,TR} = m(^1E^2E)-m(A).
\end{equation}
This explains the reduction of the RSI group from $\mbb{Z}\oplus \mbb{Z}$ to $\mbb{Z}$ due to the TRS.

We thirdly consider the case with SOC but without TRS.
There is a one-to-one correspondence between the no-SOC irreps and the SOC irreps: according to \cref{tab:2D-char}, the SOC irreps can be thought as the no-SOC irreps multiplied by the phase factor $e^{i\frac{\pi}3}$.
To be specific, $^2 \ovl{E}$ corresponds to $A$, $^1 \ovl{E}$ corresponds to $^1E$, and $\ovl{E}$ corresponds to $^2E$.
Therefore, the two RSIs can be obtained from \cref{eq:RSI-3-NSOC-NTR} by replacing the no-SOC irreps with SOC irreps
\begin{equation}
\dt_1^{\rm SOC,NTR} = m(^1\ovl{E}) - m(^2\ovl{E}),\qquad \dt_2^{\rm SOC,NTR} = m(\ovl{E}) - m(^2\ovl{E}). \label{eq:RSI-3-SOC-NTR}
\end{equation}
The same result is obtained using the algorithm in \cref{sec:algoritm-RSI}.

Finally, we consider the case with SOC and TRS.
$^1\ovl{E}$ and $^2\ovl{E}$ are complex conjugated irreps of each other, and thus they form the co-irrep $^1\ovl{E} {}^2\ovl{E}$.
On the other hand, $\ovl{E}$ is a 1D real representation and the matrix that transforms $\ovl{E}$ into its complex conjugate (itself) is just $1$.
According to the discussion in \cref{sec:TRS-irrep}, $\ovl{E}$ belongs to case-(b), where the transformation matrix squares to $-\TRS^2$ (=1), and hence the TRS co-irrep consists of two $\ovl{E}$.
We denote this co-irrep as $\ovl{E}\ovl{E}$.
We can also understand the doubling of $\ovl{E}$ from the Kramer's theorem: in presence of SOC, TRS always transforms an orbital to another orbital, which is orthogonal with respect to the original orbital, thus every co-irrep should be at least two-fold degenerate.
Now let us derive the RSI.
First we need to determine the induced representation from the subgroups of PG $3$.
PG $3$ has only one subgroup, the identity group $1$. 
Because of the Kramer's theorem, the identity irrep of PG $1$ is also doubled by the TRS.
We denote the doubled-valued co-irrep of PG $1$ as $\ovl{A}\ovl{A}$.
Both $^1\ovl{E} {}^2\ovl{E}$ and $\ovl{E}\ovl{E}$ reduce to $\ovl{A}\ovl{A}$ if we break the $C_3$ symmetry, \ie $f(\ovl{A}\ovl{A}\ |\ {}^1\ovl{E}{}^2\ovl{E}\down 1) = f(\ovl{A}\ovl{A}\ |\ {}\ovl{E}{}\ovl{E}\down 1) = 1$.
Then, according to \cref{eq:Frobenius-TRS}, we have
\begin{equation}
f(^1\ovl{E} {}^2\ovl{E}\ |\ \ovl{A}\ovl{A}\up 3) = \frac{\xi(\ovl{A}\ovl{A})}{\xi(^1\ovl{E} {}^2\ovl{E})} f(\ovl{A}\ovl{A}\ |\ ^1\ovl{E} {}^2\ovl{E} \down 1) = 2.
\end{equation}
\begin{equation}
f(\ovl{E}\ovl{E}\ |\ \ovl{A}\ovl{A}\up 3) = \frac{\xi(\ovl{A}\ovl{A})}{\xi(\ovl{E}\ovl{E})} f(\ovl{A}\ovl{A}\ |\ \ovl{E}\ovl{E} \down 1) = 1,
\end{equation}
where $\xi(\ovl{A}\ovl{A})=\xi(\ovl{E}\ovl{E})=4$ because $\ovl{A}\ovl{A}$ and $\ovl{E}\ovl{E}$ are case-(b) co-irreps, and $\xi(^1\ovl{E}{}^2\ovl{E})=2$ because $^1\ovl{E}{}^2\ovl{E}$ is a case-(c) co-irrep.
Thus the induced representation $\ovl{A}\ovl{A}\up 3$ is $2 ^1\ovl{E} {}^2\ovl{E} \oplus \ovl{E}\ovl{E}$.
We can also understand this result from a simple argument. 
First, since the orbitals forming $\ovl{A}\ovl{A}$ of PG 1 locate at some general Wyckoff position, the $C_3$ rotation must triple number of orbitals.  
Thus the induced representation of PG 3 must be six-fold.
There are four options: $3{}^1\ovl{E}{}^2\ovl{E}$,
$2{}^1\ovl{E}{}^2\ovl{E} \oplus \ovl{E}\ovl{E}$,
${}^1\ovl{E}{}^2\ovl{E} \oplus 2\ovl{E}\ovl{E}$,
$3\ovl{E}\ovl{E}$.
Secondly, since the orbitals transform into each other under $C_3$, the trace of the induced $C_3$ representation matrix must be zero.
According to \cref{tab:2D-char}, only $ 2{}^1\ovl{E}{}^2\ovl{E}\oplus \ovl{E}\ovl{E}$ satisfy this condition.
The quantity $2m(\ovl{E}\ovl{E}) -m({}^1\ovl{E} {}^2\ovl{E})$ is unchanged upon the induction, thus we have the $\mbb{Z}$-type RSI 
\begin{equation}
\dt_1^{\rm SOC,TR} = 2m(\ovl{E}\ovl{E}) -m({}^1\ovl{E} {}^2\ovl{E}). \label{eq:PG3-SOC-TRS-RSI}
\end{equation}
One can also derive this RSI by the method introduced in \cref{sec:algoritm-RSI}.
We assume the order of irreps in the symmetry data vector as ${}^1\ovl{E} {}^2\ovl{E}$, $\ovl{E}\ovl{E}$.
Then the induced representation has the symmetry data vector $p=(2,1)^T$.
Correspondingly, the $C$ matrix and its Smith decomposition $L_C\Lambda_C R_C$ are
\begin{equation}
C= \begin{pmatrix}
2 \\ 1
\end{pmatrix} = 
\begin{pmatrix}
2 & -1 \\
1 & 0
\end{pmatrix}
\begin{pmatrix}
1 & 0 \\
0 & 0
\end{pmatrix}
(1).
\end{equation}
The inverse of $L_C$ is
\begin{equation}
L_C^{-1} = \begin{pmatrix}
0 & 1\\
-1 & 2
\end{pmatrix}.
\end{equation}
Substituting $L_C$ and $\Lambda_C$ for \cref{eq:RSI}, we obtain \cref{eq:PG3-SOC-TRS-RSI}.

\subsection{Reduction of RSI}\label{sec:RSI-reduction}
Here we introduce the reduction of RSIs from a group to its subgroup. 
The results in this section will be used in deriving the twisted boundary conditions (TBC) in \cref{sec:TBC}.
Since the irreps of $G$ reduce to the irreps of $H$, we can calculate the $H$-RSIs of a representation of $G$ by decomposing the representation into irreps of $H$.
To be specific, for a given representation of $G$ where the multiplicities of irreps are $m(\rho_G)$, we can calculate the multiplicities of irreps of $H$ in the reduced representation as
\begin{equation}
    m^\pr(\rho_H) = \sum_j f(\rho_H\ |\ \rho_G^j \down H) \cdot m(\rho_G^j),
\end{equation}
where $j$ sums over all the irreps of $G$.
Then we can calculate the $H$-RSIs in terms of $m^\pr(\rho_H)$.
We call such $H$-RSIs as the reduced RSIs from RSIs of $G$.
If different RSIs of $G$ always reduce to different RSIs of $H$, we say that the RSIs of $G$ are \uemph{completely induced} from the RSIs of $H$.

For later use in \cref{sec:TBC}, we now explicitly work out the reductions of $\mbb{Z}$-type RSIs in 2D crystallographic groups. 
We find that the $\mbb{Z}$-type RSIs of $2mm$, $4mm$, $3m$, and $6mm$ are completely induced from the $\mbb{Z}$-type RSIs of $2$, $4$, $3$, and $6$, respectively.
The irreps used in the following discussion are defined in \cref{tab:2D-char}.

\noindent\textbf{PG $2mm$.}
\begin{enumerate}[label=(\arabic*)]
    \item Without SOC and TRS. The reductions from irreps of $2mm$ to $2$ are
    \begin{equation}
        {A_1} \down 2 = A,\quad {A_2} \down 2 = A,\quad {B_1} \down 2=B,\quad {B_2} \down 2=B.
    \end{equation}
    For a given representation of $G$ where the multiplicities of irreps are $m(\rho)$, the multiplicities of the $H$-irreps in the reduced representation are $m^\pr(B) = m(B_1) + m(B_2)$, $m^\pr(A) = m(A_1) + m(A_2)$.
    Thus the RSI $\delta_1 = m(B_1)+m(B_2) - m(A_1) - m(A_2)$ reduce to $\dt_1^\pr=m^\pr(B)-m^\pr(A)=\delta_1$.
    \item Without SOC, with TRS. The irreps and RSIs of $2mm$ and $2$ in this case are same with those of $2mm$ and $2$ without SOC and TRS. Thus $\dt_1$ of $2mm$ is also reduced to $\dt_1$ of $2$. 
    \item With SOC, without TRS. This case does not have RSI. The reduced RSI must be zero.
    \item With SOC and TRS. This case does not have $\mbb{Z}$-type RSI. The reduced $\mbb{Z}$-type RSI must be zero.
\end{enumerate}
\textbf{PG $4mm$.}
\begin{enumerate}[label=(\arabic*)]
    \item Without SOC and TRS. The reductions from irreps of $4mm$ to $4$ are
    \begin{equation}
        A_1\down 4 = A,\quad A_2\down 4 = A,\quad B_1\down 4 = B, B_2\down 4 = B,\quad E\down 4 = {}^1E \oplus {}^2E.
    \end{equation}
    Thus we have $m^\pr(A) = m(A_1)+m(A_2)$, $m^\pr(B) = m(B_1) + m(B_2)$, $m^\pr(^1E) = m^\pr(^2E) = m(E)$.
    According to the formulae in \cref{tab:RSI-formula}, the RSIs $(\dt_1,\dt_2)$ reduce to $(\dt_1^\pr,\dt_2^\pr,\dt_3^\pr) = (\dt_1,\dt_2,\dt_1)$.
    \item Without SOC, with TRS. Each co-irrep of $4mm$ with TRS consists of a single irrep of  $4mm$ without TRS. (See \cref{sec:TRS-irrep}.) Correspondingly, the RSIs with TRS are the same as those without TRS. On the other hand, the irreps $^1E$ and $^2E$ of $4$ form the co-irrep $^1E {}^2E$ in presence of TRS. Therefore, we have $m^\pr(A) = m(A_1)+m(A_2)$, $m^\pr(B) = m(B_1) + m(B_2)$, $m^\pr(^1E^2E) = m(E)$.
    According to formulae in \cref{tab:RSI-formula}, the RSIs $(\dt_1,\dt_2)$ reduce to $(\dt_1^\pr,\dt_2^\pr) = (\dt_1,\dt_2)$.
    \item With SOC, without TRS. The reduction from irreps of $4mm$ to irreps of $4$ are
    \begin{equation}
        \ovl{E}_1\down 4 = {}^1\ovl{E}_1\oplus {}^2\ovl{E}_1,\quad
        \ovl{E}_2\down 4 = {}^1\ovl{E}_2\oplus {}^2\ovl{E}_2.
    \end{equation} 
    We have $m^\pr(^1\ovl{E}_1) = m^\pr(^2\ovl{E}_1) = m(\ovl{E}_1)$ and $m^\pr(^1\ovl{E}_2) = m^\pr(^2\ovl{E}_2) = m(\ovl{E}_2)$.
    According to formulae in \cref{tab:RSI-formula}, the RSI $\dt_1$ reduces to $(\dt_1^\pr,\dt_2^\pr,\dt_3^\pr) = (0,\dt_1,\dt_1)$.
    \item With SOC and TRS. The irreps ${}^1\ovl{E}_i$ and ${}^2\ovl{E}_i$ ($i=1,2$) of $4$ are mutually conjugated and hence form the co-irrep ${}^1\ovl{E}_i{}^2\ovl{E}_i$ in presence of TRS. 
    While the irreps $\ovl{E}_1$ and $\ovl{E}_2$ of $4mm$ themselves form the co-irreps in presence of TRS. 
    \BAB{Always refer to the table where we define characters.}
    \SZD{All the characters are defined in \cref{tab:2D-char} and I add a reference to it at the beginning this reduction discussion.}
    According to formulae in \cref{tab:RSI-formula}, the RSI $\dt_1$ reduces to $\dt_1^\pr = \dt_1$.
\end{enumerate}
\textbf{PG $3m$.}
\begin{enumerate}[label=(\arabic*)]
    \item Without SOC and TRS. The reductions from irreps of $3m$ to $3$ are
    \begin{equation}
        {A_1} \down 3 = A_1,\quad {A_2} \down 3 = A_1,\quad {E} \down 3= {}^1E \oplus {}^2E.
    \end{equation}
    Thus $m^\pr(A_1)=m(A_1)+m(A_2)$, $m^\pr(^1E)=m^\pr(^2E)=m(E)$. According to formulae in \cref{tab:RSI-formula}, the RSI $\dt_1$ reduces to $(\dt_1^\pr,\dt_2^\pr,\dt_3^\pr) = (0,\dt_1,\dt_1)$. \BAB{I would put this table, a central result, on one page!} \SZD{Do you mean that we put \cref{tab:RSI-formula} on a page where we do not put anything else?}
    \item Without SOC, with TRS. Each co-irrep of $3m$ with TRS consists of a single irrep $3m$ without TRS. Correspondingly, the RSIs with TRS are the same as those without TRS. On the other hand, the irreps $^1E$ and $^2E$ of $3$ are complex conjugate of each other and hence form the co-irrep $^1E^2E$ in presence of TRS.
    According to formulae in \cref{tab:RSI-formula}, the RSI $\dt_1$ reduces to $\dt_1^\pr= \dt_1$.
    \item With SOC, without TRS. The reductions from irreps of $3m$ to $3$ are
    \begin{equation}
        \ovl{E}_1 \down 3 = {}^1\ovl{E} + {}^{2}\ovl{E},\quad
        {}^1\ovl{E} \down 3 = \ovl{E},\quad
        {}^2\ovl{E} \down 3 = \ovl{E}.
    \end{equation}
    Thus $m^\pr(^1\ovl{E}) = m^\pr(^2\ovl{E}) = m(\ovl{E}_1)$, $m^\pr(\ovl{E}) = m(^1\ovl{E}) + m(^2\ovl{E})$.
    According to formulae in \cref{tab:RSI-formula}, the RSI $\dt_1$ reduces to $\dt_1^\pr = \dt_1$.
    \item With SOC and TRS. The reduction from irreps of $3m$ to irreps of $3$ are 
    \begin{equation}
        {}^1\ovl{E}{}^2\ovl{E} \down 3 = \ovl{E}\ovl{E}, \quad
        \ovl{E}_1 \down 3 = {}^1\ovl{E}{}^2\ovl{E}.
    \end{equation}
    Thus $m^\pr(\ovl{E}\ovl{E}) = m(^1\ovl{E}{}^2\ovl{E})$, $m^\pr({}^1\ovl{E}{}^2\ovl{E}) = m(\ovl{E}\ovl{E})$. 
    According to formulae in \cref{tab:RSI-formula}, the RSI $\dt_1$ reduces to $\dt_1^\pr = \dt_1$.
\end{enumerate}
\textbf{PG $6mm$.}
\begin{enumerate}[label=(\arabic*)]
    \item Without SOC and TRS. The reductions from irreps of $6mm$ to $6$ are
    \begin{equation}
        A_1 \down 6 = A,\quad
        A_2 \down 6 = A,\quad
        B_1 \down 6 = B,\quad
        B_2 \down 6 = B,\quad
    \end{equation}
    \begin{equation}
        E_1 \down 6 = {}^1E_2 \oplus {}^2E_2,\quad
        E_2 \down 6 = {}^1E_1 \oplus {}^2E_1.\quad
    \end{equation}
    We obtain $m^\pr(A) = m(A_1) + m(A_2)$, $m^\pr(B)=m(B_1)+m(B_2)$, $m^\pr(^1E_1)=m^\pr(^2E_1) = m(E_2)$, $m^\pr(^1E_2)=m^\pr(^2E_2) = m(E_1)$.
    According to formulae in \cref{tab:RSI-formula}, the RSIs $(\dt_1,\dt_2,\dt_3)$ reduce to $(\dt_1^\pr,\dt_2^\pr,\dt_3^\pr,\dt_4^\pr,\dt_5^\pr) = (\dt_1,\dt_2,\dt_3,\dt_2,\dt_1)$.
    \item Without SOC, with TRS. Each co-irrep of $6mm$ with TRS consists of a single irrep of  $6mm$ without TRS. Correspondingly, the RSIs with TRS are the same as those without TRS. On the other hand, the irreps $^1E_i$ and $^2E_i$ ($i=1,2$) of $6$ form the co-irrep $^1E_i {}^2E_i$ in presence of TRS. Therefore, we have $m^\pr(A) = m(A_1)+m(A_2)$, $m^\pr(B) = m(B_1) + m(B_2)$, $m^\pr(^1E_1 {}^2E_1) = m(E_2)$, $m^\pr(^1E_2 {}^2E_2) = m(E_1)$. According to formulae in \cref{tab:RSI-formula}, the RSIs $(\dt_1,\dt_2,\dt_3)$ reduce to $(\dt_1^\pr,\dt_2^\pr,\dt_3^\pr) = (\dt_1,\dt_2,\dt_3)$.
    \item With SOC, without TRS. The reductions from irreps of $6mm$ to $6$ are
    \begin{equation}
        \ovl{E}_1 \down 6 = {}^1\ovl{E}_3 \oplus {}^2\ovl{E}_3,\quad
        \ovl{E}_2 \down 6 = {}^1\ovl{E}_2 \oplus {}^2\ovl{E}_2,\quad
        \ovl{E}_3 \down 6 = {}^1\ovl{E}_1 \oplus {}^2\ovl{E}_1.
    \end{equation}
    We obtain $m^\pr({}^1\ovl{E}_3) = m^\pr({}^2\ovl{E}_3) = m(\ovl{E}_1)$, $m^\pr({}^1\ovl{E}_2) = m^\pr({}^2\ovl{E}_2) = m(\ovl{E}_2)$, $m^\pr({}^1\ovl{E}_1) = m^\pr({}^2\ovl{E}_1) = m(\ovl{E}_3)$.
    According to formulae in \cref{tab:RSI-formula}, the RSIs $(\dt_1,\dt_2)$ reduce to $(\dt_1^\pr,\dt_2^\pr,\dt_3^\pr,\dt_4^\pr,\dt_5^\pr) = (0,\dt_1,\dt_2,\dt_2,\dt_1)$.
    \item With SOC and TRS.  Each co-irrep of $6mm$ with TRS consists of a single irrep of  $6mm$ without TRS. Correspondingly, the RSIs with TRS are the same as those without TRS. 
    On the other hand, the irreps $^1\ovl{E}_i$ and $^2\ovl{E}_i$ ($i=1,2,3$) of $6$ form the co-irrep $^1\ovl{E}_i {}^2\ovl{E}_i$ in presence of TRS. 
    Therefore, we have $m^\pr(^1\ovl{E}_3 {}^2\ovl{E}_3) = m(\ovl{E}_1)$, $m^\pr(^1\ovl{E}_2 {}^2\ovl{E}_2) = m(\ovl{E}_2)$, $m^\pr(^1\ovl{E}_1 {}^2\ovl{E}_1) = m(\ovl{E}_3)$.
    According to formulae in \cref{tab:RSI-formula}, the RSIs $(\dt_1,\dt_2)$ reduce to $(\dt_1^\pr,\dt_2^\pr) = (\dt_1,\dt_2)$.
\end{enumerate}

\subsection{Constraints about orbital number}\label{sec:constraint-nobt}
In \cref{sec:algoritm-RSI,sec:RSI-PG}, we allow the multiplicity of an irrep to be any integer.
We have shown that the number of orbitals in the finite open system is not a conserved quantity in adiabatic process as we can bring orbitals into the system from infinity.
However, as will be shown below, if the multiplicity of an irrep is restricted to be nonnegative, as will be in any real system, the number of orbitals has a nonzero lower bound given if the RSIs are nonzero.

\begin{table}\scriptsize
\centering
\begin{tabular}{|c|l|l|l|l|}
\hline
 & NoSOC, NoTRS & NoSOC, TRS & SOC, NoTRS & SOC, TRS \\
\hline
\multirow{1}{*}{$1$}&&&& $ N_\mathrm{orb} = 0 \mod 2 $\\
\hline
\multirow{3}{*}{$2$}& $ N_\mathrm{orb} = \delta_{1}\mod 2 $& $ N_\mathrm{orb} = \delta_{1}\mod 2 $& $ N_\mathrm{orb} = \delta_{1}\mod 2 $& $ N_\mathrm{orb} = 2\delta_1 \mod 4 $\\
& $ N_{\mathrm{orb}} \ge - \delta_{1} $& $ N_{\mathrm{orb}} \ge - \delta_{1} $& $ N_{\mathrm{orb}} \ge - \delta_{1} $&\\
& $ N_{\mathrm{orb}} \ge \delta_{1} $& $ N_{\mathrm{orb}} \ge \delta_{1} $& $ N_{\mathrm{orb}} \ge \delta_{1} $&\\
\hline
\multirow{3}{*}{$m$}& $ N_\mathrm{orb} = \delta_{1}\mod 2 $& $ N_\mathrm{orb} = \delta_{1}\mod 2 $& $ N_\mathrm{orb} = \delta_{1}\mod 2 $& $ N_\mathrm{orb} = 2\delta_1 \mod 4 $\\
& $ N_{\mathrm{orb}} \ge - \delta_{1} $& $ N_{\mathrm{orb}} \ge - \delta_{1} $& $ N_{\mathrm{orb}} \ge - \delta_{1} $&\\
& $ N_{\mathrm{orb}} \ge \delta_{1} $& $ N_{\mathrm{orb}} \ge \delta_{1} $& $ N_{\mathrm{orb}} \ge \delta_{1} $&\\
\hline
\multirow{3}{*}{$2mm$}& $ N_\mathrm{orb} = \delta_{1}\mod 2 $& $ N_\mathrm{orb} = \delta_{1}\mod 2 $& $ N_\mathrm{orb} = 0 \mod 2 $& $ N_\mathrm{orb} = 2\delta_1 \mod 4 $\\
& $ N_{\mathrm{orb}} \ge \delta_{1} $& $ N_{\mathrm{orb}} \ge \delta_{1} $&&\\
& $ N_{\mathrm{orb}} \ge - \delta_{1} $& $ N_{\mathrm{orb}} \ge - \delta_{1} $&&\\
\hline
\multirow{5}{*}{$4$}& $ N_\mathrm{orb} = \delta_{1} + \delta_{2} + \delta_{3}\mod 4 $& $ N_\mathrm{orb} = 2 \delta_{1} + \delta_{2}\mod 4 $& $ N_\mathrm{orb} = \delta_{1} + \delta_{2} + \delta_{3}\mod 4 $& $ N_\mathrm{orb} = 2 \delta_{1}\mod 4 $\\
& $ N_{\mathrm{orb}} \ge -3 \delta_{1} + \delta_{2} + \delta_{3} $& $ N_{\mathrm{orb}} \ge -2 \delta_{1} + \delta_{2} $& $ N_{\mathrm{orb}} \ge -3 \delta_{1} + \delta_{2} + \delta_{3} $& $ N_{\mathrm{orb}} \ge -2 \delta_{1} $\\
& $ N_{\mathrm{orb}} \ge \delta_{1} -3 \delta_{2} + \delta_{3} $& $ N_{\mathrm{orb}} \ge 2 \delta_{1} -3 \delta_{2} $& $ N_{\mathrm{orb}} \ge \delta_{1} -3 \delta_{2} + \delta_{3} $& $ N_{\mathrm{orb}} \ge 2 \delta_{1} $\\
& $ N_{\mathrm{orb}} \ge \delta_{1} + \delta_{2} -3 \delta_{3} $& $ N_{\mathrm{orb}} \ge 2 \delta_{1} + \delta_{2} $& $ N_{\mathrm{orb}} \ge \delta_{1} + \delta_{2} -3 \delta_{3} $&\\
& $ N_{\mathrm{orb}} \ge \delta_{1} + \delta_{2} + \delta_{3} $&& $ N_{\mathrm{orb}} \ge \delta_{1} + \delta_{2} + \delta_{3} $&\\
\hline
\multirow{4}{*}{$4mm$}& $ N_\mathrm{orb} = 2 \delta_{1} + \delta_{2}\mod 4 $& $ N_\mathrm{orb} = 2 \delta_{1} + \delta_{2}\mod 4 $& $ N_\mathrm{orb} = 2 \delta_{1}\mod 4 $& $ N_\mathrm{orb} = 2 \delta_{1}\mod 4 $\\
& $ N_{\mathrm{orb}} \ge -2 \delta_{1} + \delta_{2} $& $ N_{\mathrm{orb}} \ge -2 \delta_{1} + \delta_{2} $& $ N_{\mathrm{orb}} \ge -2 \delta_{1} $& $ N_{\mathrm{orb}} \ge -2 \delta_{1} $\\
& $ N_{\mathrm{orb}} \ge 2 \delta_{1} + \delta_{2} $& $ N_{\mathrm{orb}} \ge 2 \delta_{1} + \delta_{2} $& $ N_{\mathrm{orb}} \ge 2 \delta_{1} $& $ N_{\mathrm{orb}} \ge 2 \delta_{1} $\\
& $ N_{\mathrm{orb}} \ge 2 \delta_{1} -3 \delta_{2} $& $ N_{\mathrm{orb}} \ge 2 \delta_{1} -3 \delta_{2} $&&\\
\hline
\multirow{4}{*}{$3$}& $ N_\mathrm{orb} = \delta_{1} + \delta_{2}\mod 3 $& $ N_\mathrm{orb} = 2 \delta_{1}\mod 3 $& $ N_\mathrm{orb} = \delta_{1} + \delta_{2}\mod 3 $& $ N_\mathrm{orb} = 4 \delta_{1}\mod 6 $\\
& $ N_{\mathrm{orb}} \ge -2 \delta_{1} + \delta_{2} $& $ N_{\mathrm{orb}} \ge - \delta_{1} $& $ N_{\mathrm{orb}} \ge -2 \delta_{1} + \delta_{2} $& $ N_{\mathrm{orb}} \ge \delta_{1} $\\
& $ N_{\mathrm{orb}} \ge \delta_{1} -2 \delta_{2} $& $ N_{\mathrm{orb}} \ge 2 \delta_{1} $& $ N_{\mathrm{orb}} \ge \delta_{1} -2 \delta_{2} $& $ N_{\mathrm{orb}} \ge -2 \delta_{1} $\\
& $ N_{\mathrm{orb}} \ge \delta_{1} + \delta_{2} $&& $ N_{\mathrm{orb}} \ge \delta_{1} + \delta_{2} $&\\
\hline
\multirow{3}{*}{$3m$}& $ N_\mathrm{orb} = 2 \delta_{1}\mod 3 $& $ N_\mathrm{orb} = 2 \delta_{1}\mod 3 $& $ N_\mathrm{orb} = \delta_{1}\mod 3 $& $ N_\mathrm{orb} = 4 \delta_{1}\mod 6 $\\
& $ N_{\mathrm{orb}} \ge - \delta_{1} $& $ N_{\mathrm{orb}} \ge - \delta_{1} $& $ N_{\mathrm{orb}} \ge \delta_{1} $& $ N_{\mathrm{orb}} \ge \delta_{1} $\\
& $ N_{\mathrm{orb}} \ge 2 \delta_{1} $& $ N_{\mathrm{orb}} \ge 2 \delta_{1} $& $ N_{\mathrm{orb}} \ge -2 \delta_{1} $& $ N_{\mathrm{orb}} \ge -2 \delta_{1} $\\
\hline
\multirow{7}{*}{$6$}& $ N_\mathrm{orb} = \delta_{1} + \delta_{2} + \delta_{3} + \delta_{4} + \delta_{5}\mod 6 $& $ N_\mathrm{orb} = 2 \delta_{1} + 2 \delta_{2} + \delta_{3}\mod 6 $& $ N_\mathrm{orb} = \delta_{1} + \delta_{2} + \delta_{3} + \delta_{4} + \delta_{5}\mod 6 $& $ N_\mathrm{orb} = 2 \delta_{1} + 2 \delta_{2}\mod 6 $\\
& $ N_{\mathrm{orb}} \ge -5 \delta_{1} + \delta_{2} + \delta_{3} + \delta_{4} + \delta_{5} $& $ N_{\mathrm{orb}} \ge -4 \delta_{1} + 2 \delta_{2} + \delta_{3} $& $ N_{\mathrm{orb}} \ge -5 \delta_{1} + \delta_{2} + \delta_{3} + \delta_{4} + \delta_{5} $& \\
& $ N_{\mathrm{orb}} \ge \delta_{1} -5 \delta_{2} + \delta_{3} + \delta_{4} + \delta_{5} $& $ N_{\mathrm{orb}} \ge 2 \delta_{1} -4 \delta_{2} + \delta_{3} $& $ N_{\mathrm{orb}} \ge \delta_{1} -5 \delta_{2} + \delta_{3} + \delta_{4} + \delta_{5} $& $ N_{\mathrm{orb}} \ge -4 \delta_{1} + 2 \delta_{2} $\\
& $ N_{\mathrm{orb}} \ge \delta_{1} + \delta_{2} -5 \delta_{3} + \delta_{4} + \delta_{5} $& $ N_{\mathrm{orb}} \ge 2 \delta_{1} + 2 \delta_{2} -5 \delta_{3} $& $ N_{\mathrm{orb}} \ge \delta_{1} + \delta_{2} -5 \delta_{3} + \delta_{4} + \delta_{5} $& $ N_{\mathrm{orb}} \ge 2 \delta_{1} -4 \delta_{2} $\\
& $ N_{\mathrm{orb}} \ge \delta_{1} + \delta_{2} + \delta_{3} -5 \delta_{4} + \delta_{5} $& $ N_{\mathrm{orb}} \ge 2 \delta_{1} + 2 \delta_{2} + \delta_{3} $& $ N_{\mathrm{orb}} \ge \delta_{1} + \delta_{2} + \delta_{3} -5 \delta_{4} + \delta_{5} $& $ N_{\mathrm{orb}} \ge 2 \delta_{1} + 2 \delta_{2} $\\
& $ N_{\mathrm{orb}} \ge \delta_{1} + \delta_{2} + \delta_{3} + \delta_{4} -5 \delta_{5} $&& $ N_{\mathrm{orb}} \ge \delta_{1} + \delta_{2} + \delta_{3} + \delta_{4} -5 \delta_{5} $&\\
& $ N_{\mathrm{orb}} \ge \delta_{1} + \delta_{2} + \delta_{3} + \delta_{4} + \delta_{5} $&& $ N_{\mathrm{orb}} \ge \delta_{1} + \delta_{2} + \delta_{3} + \delta_{4} + \delta_{5} $&\\
\hline
\multirow{5}{*}{$6mm$}& $ N_\mathrm{orb} = 2 \delta_{1} + 2 \delta_{2} + \delta_{3}\mod 6 $& $ N_\mathrm{orb} = 2 \delta_{1} + 2 \delta_{2} + \delta_{3}\mod 6 $& $ N_\mathrm{orb} = 2 \delta_{1} + 2 \delta_{2}\mod 6 $& $ N_\mathrm{orb} = 2 \delta_{1} + 2 \delta_{2}\mod 6 $\\
& $ N_{\mathrm{orb}} \ge -4 \delta_{1} + 2 \delta_{2} + \delta_{3} $& $ N_{\mathrm{orb}} \ge -4 \delta_{1} + 2 \delta_{2} + \delta_{3} $& $ N_{\mathrm{orb}} \ge -4 \delta_{1} + 2 \delta_{2} $& \\
& $ N_{\mathrm{orb}} \ge 2 \delta_{1} -4 \delta_{2} + \delta_{3} $& $ N_{\mathrm{orb}} \ge 2 \delta_{1} -4 \delta_{2} + \delta_{3} $& $ N_{\mathrm{orb}} \ge 2 \delta_{1} -4 \delta_{2} $& $ N_{\mathrm{orb}} \ge -4 \delta_{1} + 2 \delta_{2} $\\
& $ N_{\mathrm{orb}} \ge 2 \delta_{1} + 2 \delta_{2} -5 \delta_{3} $& $ N_{\mathrm{orb}} \ge 2 \delta_{1} + 2 \delta_{2} -5 \delta_{3} $& $ N_{\mathrm{orb}} \ge 2 \delta_{1} + 2 \delta_{2} $& $ N_{\mathrm{orb}} \ge 2 \delta_{1} -4 \delta_{2} $\\
& $ N_{\mathrm{orb}} \ge 2 \delta_{1} + 2 \delta_{2} + \delta_{3} $& $ N_{\mathrm{orb}} \ge 2 \delta_{1} + 2 \delta_{2} + \delta_{3} $& & $ N_{\mathrm{orb}} \ge 2 \delta_{1} + 2 \delta_{2} $\\
\hline
\end{tabular}

\protect\caption{\label{tab:Nocc-RSI} The constraints about orbital number and RSIs in 2D crystallographic PGs. }
\end{table}

We first take the PG $\bar1$ as an example. 
The number of orbitals is given by $N_\mrm{orb} = m(A_u) + m(A_g)$, and the RSI is given by $\delta = m(A_u) - m(A_g)$ (\cref{eq:dt-inv}).
If $m(A_u)$ and $m(A_g)$ are nonnegative integers, there must be
\begin{equation}
    N_\mrm{orb} \ge m(A_u) - m(A_g) = \delta,
\end{equation}
and
\begin{equation}
    N_\mrm{orb} \ge -m(A_u) + m(A_g) =-\delta.
\end{equation}
Therefore, we always have $N_\mrm{orb}\ge |\delta|$.
This is easy to understand: $\delta$ is nonzero means that there is some orbital in the system (otherwise $m(A_g)=m(A_u)=0\; \Rightarrow\; \delta=0$) and hence the number of orbitals cannot be zero.

Now we generalize this analysis to general PGs.
First, we look at the $\mbb{Z}_2$ RSIs.
Only the PGs $2$, $m$, $2mm$, $4$, $4mm$, $6$, $6mm$ with TRS and SOC have $\mbb{Z}_2$ RSIs; and all the $\mbb{Z}_2$ RSIs are nothing but the parity of number of Kramer's pairs (\cref{tab:RSI-formula}).
We denote these RSIs as $\delta_{\mbb{Z}_2}$
Since each Kramer pair contribute two orbitals, the orbital number $N_{\rm orb}$ must satisfy
\begin{equation}
N_{\rm orb} = 2\delta_{\mbb{Z}_2} \mod 4.
\end{equation}

Second, we first consider the $\mbb{Z}$-type RSIs.
We denote the linear mapping from $p=(m(\rho_G^1),m(\rho_G^2)\cdots)^T$ to these RSIs as the matrix $F_1$, \ie
\begin{equation}
\delta = F_1 \cdot p.
\end{equation}
The explicit form of $F_1$ is determined by the second case in \cref{eq:RSI}.
$F_1$ consists of rows of $L_C^{-1}$ that correspond to $\kappa_i=0$.
\BAB{Sorry, why are we introducing new notations? We clearly had $\delta=L_C^{-1} \cdot p$}
\SZD{Because $F_1$ only consists of the rows of $L_C^{-1}$ that correspond to $\kappa=0$ and hence is \uemph{not} identical to $L_C^{-1}$.}
We denote the linear mapping from $p$ to number of orbitals as the row-vector $F_0$, \ie
\begin{equation}
N_\mrm{orb} = F_0 \cdot p, \qquad F_0 = (|\rho_G^1|,\ |\rho_G^2)|\cdots),
\end{equation}
where $|\rho_G|$ is the dimension of the irrep $\rho_G$.
For convenience, we define the matrix $F = \begin{pmatrix} F_0\\ F_1 \end{pmatrix}$; and we denote the number of rows of $F$ as $n_F$, which is the number of $\mathbb{Z}$-type RSIs plus 1.
We denote the length of $p$, or the number of irreps, as $n_p$, hence $F$ is an $n_F\times n_p$ integer matrix.
Since all the $\mbb{Z}$-type RSIs and the orbital number are independent numbers, $F$ is a rank-$n_F$ matrix, and there is always $n_F\le n_p$ because matrix rank cannot be larger than the number of columns.
Then, all the combinations of $N_\mrm{orb}$ and $\delta$ that are realizable in real systems form the affine monoid
\begin{equation}
\overline{\Delta} = \brace{ \begin{pmatrix} N_\mrm{orb}\\ \delta \end{pmatrix} = F \cdot p \;\bigg|\; p\in \mbb{N}^{n_p} },
\end{equation}
where $\mbb{N}$ stands for nonnegative integer.
In the following, by applying the affine monoid method introduced in Ref. \cite{song_fragile_2019}, we will figure out the conditions for given $N_\mrm{orb}$ and $\delta$ to belong to $\ovl{\Delta}$ or not.
The affine monoid $\ovl{\Delta}$ is the intersection of the lattice
\begin{equation}
\mcl{L} =  \brace{ F \cdot p \;|\; p\in \mbb{Z}^{n_p} }, \label{eq:L-lattice}
\end{equation}
and the polyhedral cone
\begin{equation}
\Delta =  \brace{ F \cdot p \;|\; p\in \mbb{R}_+^{n_p} },\label{eq:Dt-polyhedron}
\end{equation}
where $\mbb{R}_+$ stands for nonnegative real number.
We first look at the conditions for $(N_\mrm{orb},\delta)^T$ to belong to the lattice $\mcl{L}$.
We write the Smith decomposition of $F$ as
\begin{equation}
F= L_F \Lambda_F R_F,
\end{equation}
where $L_F$ is an $n_F \times n_F$ unimodular matrix, $R_F$ is an $n_p\times n_p$ unimodular matrix, and $\Lambda_F$ is an $n_F\times n_p$ integer matrix, where all the off-diagonal entries are zero.
Notice that $[\Lambda_F]_{ii}\neq0$ for $i=1\cdots n_F$ because $F$ is a rank-$n_F$ matrix.
Therefore, $f=(N_\mrm{orb},\delta)^T$ belongs to $\mcl{L}$ (\cref{eq:L-lattice}) iff exists some $p\in \mathbb{Z}^{n_p}$ such that $L_F^{-1} \cdot f = \Lambda_F R_F \cdot p$.
Since $R_F$ is a unimodular matrix, such $p$ vector exists iff $ [L_F^{-1} \cdot (N_\mrm{orb},\delta)^T ]_i / [\Lambda_F]_{ii} \in \mathbb{Z}$ for $i=1\cdots n_F$.
We can rewrite this condition as
\begin{equation}
\begin{pmatrix} N_\mrm{orb}\\ \delta \end{pmatrix} \in \mcl{L}\quad\Leftrightarrow\quad
\brak{ L_F^{-1} \cdot \begin{pmatrix} N_\mrm{orb}\\ \delta \end{pmatrix} }_i \mod [\Lambda_F]_{ii} = 0\quad (i=1\cdots n_F). \label{eq:lattice-cond}
\end{equation}
Then we look at the conditions for $(N_\mrm{orb},\delta)^T$ to belong to the polyhedral cone $\Delta$.
\cref{eq:Dt-polyhedron} is the so-called V-representation of a polyhedral cone, where $F$ consisting of the defining rays of the cone \cite{song_fragile_2019}.
We can use a standard mathematics package, such as \href{http://www.sagemath.org/}{\it SageMath} \cite{Sage}, to obtain the inequalities defining $\Delta$ (the so-called H-representation of polyhedron).
The fulfillment of these inequalities are the conditions for $(N_\mrm{orb},\delta)^T$ to belong to the polyhedral cone $\Delta$.
We will not discuss the general algorithm to obtain the H-representation from the V-representation.
Interested readers can find more applications of this algorithm in Ref. \cite{song_fragile_2019}.

Combining the orbital number constraints from $\mbb{Z}$-type RSIs and $\mbb{Z}_2$-type RSIs, we obtain \cref{tab:Nocc-RSI}.

Here we use PG $3$ with SOC and TRS as an example to show the constraints on the orbital number.
The PG $3$ with SOC and TRS has two two-fold co-irreps ${}^1\ovl{E} {}^2\ovl{E}$ and $\ovl{E}\ovl{E}$.
According to \cref{tab:RSI-formula}, $3$ has only one RSI $\dt_1 = 2m(\ovl{E}\ovl{E}) - m({}^1\ovl{E} {}^2\ovl{E})$. 
Thus, we can write the mapping from multiplicities to $N_\mrm{orb}$ and $\dt_1$ as
\begin{equation}
\begin{pmatrix}
N_\mrm{orb}\\ \dt_1 
\end{pmatrix}
= \pare{\begin{array}{rr}
2 & 2\\ -1 & 2
\end{array}}
\begin{pmatrix}
m({}^1\ovl{E}{}^2\ovl{E})\\ m(\ovl{E}\ovl{E})
\end{pmatrix}.
\end{equation}
The 2D polyhedral cone $\Delta$ (\cref{eq:Dt-polyhedron}) is spanned by the two columns of $F$, \ie $(N_{\rm orb}, \delta)^T = (2,-1)^T$ and $(N_{\rm orb}, \delta)^T = (2,2)^T$.
The affine monoid $\ovl{\Delta}$ is the intersection of $\Delta$ and the lattice \cref{eq:L-lattice}.
The equation satisfied by the two rays are $ N_\mrm{orb} =-2 \dt_1$ and $N_{\rm orb} = \dt_1$, respectively.
Therefore, the 2D polyhedral cone is
\begin{equation}
\Delta = \{(N_{\rm orb}, \dt_1)\in \mbb{R}^2 \ |\ N_{\rm orb}\ge -2\dt_1,\; N_{\rm orb}\ge \dt_1\}.\label{eq:PG3-Delta}
\end{equation}
The Smith normal form of $F$ is 
\begin{equation}
F = 
\begin{pmatrix} 4 & 1 \\ 1 & 0 \end{pmatrix}
\begin{pmatrix} 1 & 0 \\ 0 & 6 \end{pmatrix}
\begin{pmatrix} -1 & 2 \\ 1 &-1 \end{pmatrix}.
\end{equation}
The inverse of left matrix is $L^{-1}_F = \begin{pmatrix}0 & 1\\ 1 & -4 \end{pmatrix}$.
Due to \cref{eq:lattice-cond}, the equivalent condition for $(N_{\rm orb}, \dt_1)^T$ to be in the lattice $\mcl{L}$ is $N_{\rm orb} - 4\dt_1 = 0 \mod 6$.
Therefore, there are in total three different constraints on $N_{\rm orb}$ and $\dt_1$
\begin{equation}
N_{\rm orb}\ge -2\dt_1,\qquad N_{\rm orb}\ge \dt_1,\qquad N_{\rm orb} = 4\dt_1\mod 6.
\end{equation}
\BAB{These two are unclear now, you obtain them in general, where did you show the ??? (???)}
\SZD{The two inequalities come from \cref{eq:PG3-Delta}, which we derive explicitly. Why they are not clear?}

\section{RSI in wallpaper groups} \label{sec:RSI-WG}
In this section, we will calculate the RSIs of Bloch wavefunctions in terms of the momentum space irreps.
We will only consider the EFP (eigenvalue fragile phase), EOAP (eigenvalue obstructed atomic phase), and ETAP (eigenvalue trivial atomic phase) cases.

\subsection{Wannierization of band structure without stable topology}\label{sec:wann}

As introduced in \cref{sec:TQC}, the representation formed by band structure with trivial symmetry indicators decomposes into a set of induced representation from irreps at \uemph{maximal} Wyckoff positions, \ie
\begin{equation}
\text{representation of bands} = \bigoplus_{w i} p(\rho_{G_w}^i) (\rho_{G_w}^i \up G), \label{eq:rep-of-band}
\end{equation}
where $\rho_{G_w}^i$ represents the $i$th irrep of the site-symmetry group $G_w$ at the maximal Wyckoff position $w$.
For EOAP and ETAP, $p(\rho_{G_w}^i)$ can be chosen as nonnegative integers; whereas for EFP, some components of $p$ must be negative.
In all these cases, \cref{eq:rep-of-band} suggests that the single-particle density matrix 
\begin{equation}
\sum_{\kk n}^\pr \ket{\psi_{\kk n}}\bra{\psi_{\kk n}}, \label{eq:density-psi}
\end{equation}
where $n$ sums over the occupied bands or a group of bands under consideration, has the same transformation matrix under the space group operations as the following trial density matrix
\begin{equation}
\sum_{\RR w} \sum_{ij} \sum_{l=1}^{|p(\rho_{G_w}^i)|} \mrm{sgn}\pare{ p(\rho_{G_w}^i) }\ket{W_{l i j}(\RR+\tt_w) } \bra{W_{l i j}(\RR+\tt_w)}  \label{eq:density-W}
\end{equation}
\BAB{The density matrix should have nonnegative eigenvalues.}
\SZD{I completely agree.}
\BAB{For fragile, \cref{eq:density-W} has negative eigenvalues. So it is not correct to call it density matrix.}
\SZD{I don't agree. Negative coefficients in \cref{eq:density-W} does \uemph{not} imply negative eigenvalue because the Wannier functions here are not linearly independent. \cref{eq:density-W} is an ansatz for the density matrix of fragile phase, which has all nonnegative eigenvalues.}
Here $\RR$ sums over all the lattice vectors, $w$ sums over all the maximal Wyckoff positions, $\tt_w$ is the position vector of $w$ in the chosen unit cell, $i$ sums over all the irreps of $G_w$ and $j$ sums over all the basis functions of $\rho_{G_w}^i$.
The localized Wannier function $\ket{W_{l i j}(\tt_w)}$ sites on  $\tt_{w}$ and belongs to the $j$th basis function of the irrep $\rho_{G_w}^i$; $\ket{W_{l i j}(\RR+\tt_w)}$ is obtained from $\ket{W_{l i j}(\tt_w)}$ by a translation.
Notice that we allow the Wannier functions to be linearly dependent.
For obstructed atomic phases and trivial atomic phases, the density matrix \cref{eq:density-psi} can always be written in the form of \cref{eq:density-W}, with all $p$ components being nonnegative.
For fragile phases, the density matrix \cref{eq:density-psi} can also be written in the form of \cref{eq:density-psi}, with some $p$ components negative.
\BAB{I would not call \cref{eq:density-psi} density matrix.}
\SZD{Why? All eigenvalues of \cref{eq:density-psi} are 1.}
It should be emphasized that all the eigenvalues of the density matrix must be nonnegative, which is possible even with negative $p$ components because the Wannier functions are allowed to be linearly dependent.
By definition, the fragile phase will become Wannierizable after being coupled to a particular group of Wannierizable bands.
We assume the density matrix of this particular group of Wannierizable bands, which need to be added to \cref{eq:density-psi} to make it Wannierizable, and the density matrix of the trivialized total state formed by \cref{eq:density-psi} and the added bands to be
\begin{equation}
\sum_{\RR w} \sum_{ij} \sum_{l=1}^{q(\rho_{G_w}^i)} \ket{W^\pr_{l i j}(\RR+\tt_w) } \bra{W^\pr_{l i j}(\RR+\tt_w)},\qquad
\sum_{\RR w} \sum_{ij} \sum_{l=1}^{p^\pr(\rho_{G_w}^i)} \ket{W^\prpr_{l i j}(\RR+\tt_w) } \bra{W^\prpr_{l i j}(\RR+\tt_w)},
\end{equation}
respectively, where $q$ and $p^\pr$ are vectors with nonnegative components.
Then the density matrix of the fragile phase is
\begin{equation}
\sum_{\kk n}^\pr \ket{\psi_{\kk n}}\bra{\psi_{\kk n}} = \sum_{\RR w} \sum_{ij} \sum_{l=1}^{p^\pr(\rho_{G_w}^i)} \ket{W^\prpr_{l i j}(\RR+\tt_w) } \bra{W^\prpr_{l i j}(\RR+\tt_w)} - \sum_{\RR w} \sum_{ij} \sum_{l=1}^{q(\rho_{G_w}^i)} \ket{W^\pr_{l i j}(\RR+\tt_w) } \bra{W^\pr_{l i j}(\RR+\tt_w)},
\end{equation}
which has the form of \cref{eq:density-W}.
However, if the band structure has a stable topology which may or may not be diagnosed through eigenvalues but can always be diagnosed through Wilson loops, the density matrix \uemph{cannot} be written in the form of \cref{eq:density-W}.

We emphasize that although different $p$ may correspond to the same induced representation (in terms of symmetry eigenvalues), they can represent topologically distinct phases. 
To be specific, we consider a 1D system with only $P\TR$ symmetry, \ie time-reversal followed by inversion, where $(P\TR)^2=1$.
The (magnetic) PG of this magnetic space group is $\bar{1}^\pr$.
All the momenta in the BZ are $P\TR$ symmetric and their little group is the whole space group. 
As the unitary subgroup of the little group is just the set of translations, at every $\kk$, the little group generated by has only one irrep.
Therefore, all the band structures with the same number of bands have the same symmetry data vector.
There are two maximal Wyckoff positions: $a\ (x=0)$ and $b\ (x=1/2)$, and both have a site-symmetry group isomorphic to $\bar{1}^\pr$.
We consider two atomic phases: one is formed by one local orbital at $a$, another is formed by one local orbital at $b$.
These two atomic phases have the same symmetry data vector, but they have different $p$-vectors - the first one has $p(\rho_{G_a})=1$, $p(\rho_{G_b})=0$ and the second one has $p(\rho_{G_a})=0$, $p(\rho_{G_b})=1$.
They are topologically distinct: a single orbital at $a$ cannot be moved to $b$ without breaking the $P\TR$ symmetry.
The two phases can be distinguished by Wilson loop.

If two band structures have the same $p$-vector, \ie their density matrices can be written in the form of \cref{eq:density-W} with the same $p$, then the two band structures must be topologically equivalent.
\BAB{This is in contradict to the first sentence of the previous paragraph!}
\SZD{No, it does not. The sentence in last paragraph is basically ``different $p$ corresponding to the same eigenvalues can represent distinct phases''.}
\BAB{The $p$ vector meaning bands at high symmetry points do not fully quantify the band representation.}
\SZD{Notice that $p$ vector is in real space; it is not the $B$-vector; it contains information beyond eigenvalues, such as the exmaple in last paragraph. I think such $p$ fully determine the band representation. No? If it is not clear, we can delete this whole paragraph.}
We denote the Wannier functions of the first band structure as $\ket{W(1)_{lij}}$ and the Wannier functions of the second band structure as $\ket{W^{(2)}_{lij}}$.
Since $\ket{W(1)_{lij}}$ and $\ket{W^{(2)}_{lij}}$ belong to the same basis of the same irrep at the same Wyckoff position, we can continuously deform $\ket{W(1)_{lij}}$ to $\ket{W^{(2)}_{lij}}$.
Thus the two band structures are related by an adiabatic process.
We conclude that the topology of a band structure without stable topology is uniquely determined by the $p$-vector.

\subsection{RSIs in terms of momentum space irreps}\label{sec:RSI-k-irrep}

For a given band structure without stable topology, the strategy to calculate RSIs is (i) find the $p$-vector that induces the correct representation and corresponds to the correct topology, (ii) substitute $p$ for \cref{eq:RSI} to obtain the RSIs. 
The determination of the $p$-vector that gives the correct topology is in general not direct because it involves calculation based on wavefunctions, such as Wilson loops.
Here we focus on a relatively easier problem: can we determine the RSIs only through the symmetry data vector?
The answer is yes if the RSIs do not depend on how we choose $p$, because then we can pick any $p$-vector corresponding to the correct symmetry data to calculate the RSIs.
In the following we will first give the algorithm to calculate the RSIs in terms of $p$ and then find which RSIs do not depend on how we choose $p$.

As all the $p$ components correspond to irreps at maximal Wyckoff positions, it is direct to calculate the RSIs at maximal Wyckoff positions - we only need to substitute the corresponding $p$ in \cref{eq:RSI}.
Since the maximal Wyckoff positions are special points of non-maximal Wyckoff position, these $p$-vectors can also be used to calculate RSIs at non-maximal Wyckoff positions.
We denote the non-maximal Wyckoff positions as $w$, and a maximal Wyckoff position that belongs to $w$ as $w^\pr$.
The site-symmetry group of $w$ is a subgroup of the site-symmetry group of $w^\pr$, \ie $G_w \subset G_{w^\pr}$.
The representation of $G_w$ assigned to the WFs at $w^\pr$ that form the irrep $\rho_{G_{w^\pr}}$ is the reduced representation $\rho_{G_{w^\pr}}\down G_w$.
Therefore, the multiplicity of the $i$th irrep of $G_w$ can be calculated as
\begin{equation}
    p(\rho_{G_w}^i) = \sum_{\substack{ w^\pr\\ G_w \subset G_{w^\pr}}} \sum_j f(\rho_{G_w}^i | \rho_{G_{w^\pr}}^j \down G_w ) p(\rho_{G_{w^\pr}}^j),\label{eq:p-NMWP}
\end{equation}
where $w^\pr$ only sums over maximal Wyckoff positions.
Substituting the $p$-vector at non-maximal Wyckoff positions into \cref{eq:RSI}, we can obtain the corresponding RSIs.

As discussed in the first paragraph in this subsection, we can determine the RSIs in terms of momentum space irreps only if the RSIs do not depend on how we choose $p$.
Now we check which RSIs are independent of the choice of $p$.
According to \cref{eq:B-p,eq:p-y-k}, $p$ and $p + \sum_{j=1}^{N_{\EBR}-r} k_j [R_{\EBR}^{-1}]_{r+j}$ correspond to the same symmetry data vector $B$.
Here $R_\EBR$ is the right transformation matrix in the Smith decomposition of the EBR matrix (\cref{eq:EBR-snf}), $r$ is the rank of the EBR matrix, $N_\EBR$ is the number of EBRs, $[R_{\EBR}^{-1}]_{r+j}$ is the $(r+j)$th column of $R_\EBR^{-1}$, and $k_j$ are free (integer) parameters.
If a RSI calculated from $[R_{\EBR}^{-1}]_{r+j}$ ($j=1\cdots N_\EBR -r$) is always zero, then this RSI is independent of the choice of $p$ and hence it can be determined from symmetry eigenvalues.
Applying this criteria to all the wallpaper groups, we find that the $\mbb{Z}$-type RSIs are always independent of choice of $p$, whereas $\mbb{Z}_2$-type RSIs are either redundant or depend on the choice of $p$.
Here ``redundant'' means that the $\mbb{Z}_2$-type RSI is determined by the other RSIs.
In \cref{tab:WGRSI}, we tabulate all the formulae of RSIs in wallpaper groups.

Here we take the wallpaper group $p4$ with SOC and TRS as an example to show how to derive the RSI formulae in terms of momentum space irreps.
$p4$ has three momenta with maximal symmetry $\Gamma\ (0,0)$, $X\ (0,\pi)$, $M\ (\pi,\pi)$ and three maximal Wyckoff position $1a\ (0,0)$, $1b\ (\frac12,\frac12)$, $2c\ (0,\frac12)\ (\frac12,0)$.
The irreps at $\Gamma$, $X$, $M$ are $\ovl{\Gamma}_5\ovl{\Gamma}_7$, $\ovl{\Gamma}_6\ovl{\Gamma}_8$, $\ovl{M}_5\ovl{M}_7$, $\ovl{M}_6\ovl{M}_8$, $\ovl{X}_3\ovl{X}_4$.
The definitions of these irreps can be found at the \href{http://www.cryst.ehu.es/cgi-bin/cryst/programs/representations.pl?tipogrupo=dbg}{Irreducible Representations of the Double Space Groups} tool \cite{Bradlyn2017,Vergniory2017,Elcoro2017,Jennifer2018} on the BCS \cite{BCS1}.
The site-symmetry group of $a$ and $b$ is $4$, which has two different irreps ${}^1\ovl{E}_1 {}^2\ovl{E}_1$, ${}^1\ovl{E}_2 {}^2\ovl{E}_2$.
The site-symmetry of $c$ is $2$, which has only one irrep ${}^1\ovl{E} {}^2\ovl{E}$.
One can refer to \cref{tab:2D-char} for the definitions of these irreps.
Thus we have five EBRs, two induced from $a$, two induced from $b$, and one induced from $c$.
From the \href{http://www.cryst.ehu.es/cgi-bin/cryst/programs/bandrep.pl}{Band Representations of the Double Space Groups} tool \cite{Bradlyn2017,Vergniory2017,Elcoro2017,Jennifer2018} on the BCS \cite{BCS1} we obtain these EBRs
\begin{equation}
({}^1\ovl{E}_1 {}^2\ovl{E}_1)_a \up G = \ovl{\Gamma}_6\ovl{\Gamma}_8 + \ovl{M}_6\ovl{M}_8 + \ovl{X}_3\ovl{X}_4,
\end{equation}
\begin{equation}
({}^1\ovl{E}_2 {}^2\ovl{E}_2)_a \up G = \ovl{\Gamma}_5\ovl{\Gamma}_7 + \ovl{M}_5\ovl{M}_7 + \ovl{X}_3\ovl{X}_4,
\end{equation}
\begin{equation}
({}^1\ovl{E}_1 {}^2\ovl{E}_1)_b \up G = \ovl{\Gamma}_6\ovl{\Gamma}_8 + \ovl{M}_5\ovl{M}_7 + \ovl{X}_3\ovl{X}_4,
\end{equation}
\begin{equation}
({}^1\ovl{E}_2 {}^2\ovl{E}_2)_b \up G = \ovl{\Gamma}_5\ovl{\Gamma}_7 + \ovl{M}_6\ovl{M}_8 + \ovl{X}_3\ovl{X}_4,
\end{equation}
\begin{equation}
({}^1\ovl{E} {}^2\ovl{E})_c \up G = \ovl{\Gamma}_5\ovl{\Gamma}_7 + \ovl{\Gamma}_6\ovl{\Gamma}_8 + \ovl{M}_5\ovl{M}_7 + \ovl{M}_6\ovl{M}_8 + 2 \ovl{X}_3\ovl{X}_4.
\end{equation}
Here we use $(\rho)_w$ ($w=a,b,c$) to represent the irreps of $G_w$. 
We can write the EBR matrix as
\begin{equation}
EBR = \pare{\begin{array}{rrrrr}
0 & 1 & 0 & 1 & 1\\
1 & 0 & 1 & 0 & 1\\
0 & 1 & 1 & 0 & 1\\
1 & 0 & 0 & 1 & 1\\
1 & 1 & 1 & 1 & 2
\end{array}}, \label{eq:EBR-4-SOC-TR}
\end{equation}
where the five columns correspond to the five EBRs, and the five rows represent the momentum space irreps $\ovl{\Gamma}_5\ovl{\Gamma}_7$, $\ovl{\Gamma}_6\ovl{\Gamma}_8$, $\ovl{\Gamma}_5\ovl{\Gamma}_7$, $\ovl{M}_5\ovl{M}_7$, $\ovl{M}_6\ovl{M}_8$, $\ovl{X}_3\ovl{X}_4$.
The symmetry data vector of an EFP, EOAP, or ETAP can be written as $B = EBR \cdot p$ with $p$ an integer vector. 
According to \cref{eq:y-def,eq:p-y-k}, we can write the $p$ vector in terms of $B$ as 
\begin{equation}
p_i = \sum_{j=1}^{r} [R_\EBR^{-1}]_{ij} \frac{1}{[\Lambda_\EBR]{jj}} (L^{-1}_\EBR B)_j + \sum_{j=1}^{N_{\EBR} - 1} [R_\EBR^{-1}]_{i,j+r} k_j, 
\end{equation}
where $k_j$ are free integer parameters.
The Smith decomposition of \cref{eq:EBR-4-SOC-TR} is
\begin{equation}
\EBR = \left(\begin{array}{rrrrr}
1 & 0 & -1 & 0 & 0 \\
0 & 0 & 1 & 0 & 0 \\
1 & -1 & 0 & 0 & 0 \\
0 & 1 & 0 & 0 & 1 \\
1 & 0 & 0 & 1 & 0
\end{array}\right) 
\left(\begin{array}{rrrrr}
1 & 0 & 0 & 0 & 0 \\
0 & 1 & 0 & 0 & 0 \\
0 & 0 & 1 & 0 & 0 \\
0 & 0 & 0 & 0 & 0 \\
0 & 0 & 0 & 0 & 0
\end{array}\right)
\left(\begin{array}{rrrrr}
1 & 1 & 1 & 1 & 2 \\
1 & 0 & 0 & 1 & 1 \\
1 & 0 & 1 & 0 & 1 \\
0 & 0 & 0 & 0 & 1 \\
0 & 0 & 0 & 1 & 0
\end{array}\right),
\end{equation}
and the inverses of the left and right transformation matrices are
\begin{equation}
L_\EBR^{-1} = \left(\begin{array}{rrrrr}
1 & 1 & 0 & 0 & 0 \\
1 & 1 & -1 & 0 & 0 \\
0 & 1 & 0 & 0 & 0 \\
-1 & -1 & 0 & 0 & 1 \\
-1 & -1 & 1 & 1 & 0
\end{array}\right),\qquad
R_\EBR^{-1} = 
\left(\begin{array}{rrrrr}
0 & 1 & 0 & -1 & -1 \\
1 & 0 & -1 & -1 & -1 \\
0 & -1 & 1 & 0 & 1 \\
0 & 0 & 0 & 0 & 1 \\
0 & 0 & 0 & 1 & 0
\end{array}\right).
\end{equation}
Therefore, the $p$ vector can be written as 
\begin{equation}
p = \left(\begin{array}{rrrrr}
1 & 1 & -1 & 0 & 0 \\
1 & 0 & 0 & 0 & 0 \\
-1 & 0 & 1 & 0 & 0 \\
0 & 0 & 0 & 0 & 0 \\
0 & 0 & 0 & 0 & 0
\end{array}\right) B + 
\left(\begin{array}{r}
-1 \\-1 \\0 \\0 \\1 \end{array}\right)k_1 +
\left(\begin{array}{r}
-1 \\ -1 \\ 1 \\ 1 \\ 0 \end{array}\right) k_2.
\end{equation}
Notice that the order of real space irreps is $({}^1\ovl{E}_1 {}^2\ovl{E}_1)_a$, $({}^1\ovl{E}_2 {}^2\ovl{E}_2)_a$, $({}^1\ovl{E}_1 {}^2\ovl{E}_1)_b$, $({}^1\ovl{E}_2 {}^2\ovl{E}_2)_b$, $({}^1\ovl{E} {}^2\ovl{E})_c$.
We can write the $p$ vector in terms of multiplicities of momentum space irreps  explicitly as
\begin{equation}
p(({}^1\ovl{E}_1 {}^2\ovl{E}_1)_a) = m(\ovl{\Gamma}_5\ovl{\Gamma}_7) + m(\ovl{\Gamma}_6\ovl{\Gamma}_8) - m(\ovl{M}_5\ovl{M}_7) - k_1 - k_2.
\end{equation}
\begin{equation}
p( ( {}^1\ovl{E}_2 {}^2\ovl{E}_2)_a ) = m(\ovl{\Gamma}_5\ovl{\Gamma}_7)  - k_1 - k_2.
\end{equation}
\begin{equation}
p( ( {}^1\ovl{E}_1 {}^2\ovl{E}_1 )_b ) =-m(\ovl{\Gamma}_5\ovl{\Gamma}_7) + m(\ovl{M}_5\ovl{M}_7) + k_2.
\end{equation}
\begin{equation}
p( ( {}^1\ovl{E}_2 {}^2\ovl{E}_2)_b ) = k_2.
\end{equation}
\begin{equation}
p( ({}^1\ovl{E} {}^2\ovl{E})_c ) = k_1.
\end{equation}
Then according to the RSIs defined in \cref{tab:RSI-formula}, we obtain
\begin{equation}
\dt_{a1} = p(( {}^1\ovl{E}_2 {}^2\ovl{E}_2)_a) - p(({}^1\ovl{E}_1 {}^2\ovl{E}_1)_a) =  - m(\ovl{\Gamma}_6\ovl{\Gamma}_8) + m(\ovl{M}_5\ovl{M}_7),
\end{equation}
\begin{equation}
\dt_{a2} = p(( {}^1\ovl{E}_2 {}^2\ovl{E}_2)_a) + p(({}^1\ovl{E}_1 {}^2\ovl{E}_1)_a) \mod 2=  \dt_{a1} \mod 2,
\end{equation}
\begin{equation}
\dt_{b1} = p(( {}^1\ovl{E}_2 {}^2\ovl{E}_2)_b) - p(({}^1\ovl{E}_1 {}^2\ovl{E}_1)_b)=  m(\ovl{\Gamma}_5\ovl{\Gamma}_7) - m(\ovl{M}_5\ovl{M}_7),
\end{equation}
\begin{equation}
\dt_{b2} = p(( {}^1\ovl{E}_2 {}^2\ovl{E}_2)_b) + p(({}^1\ovl{E}_1 {}^2\ovl{E}_1)_b) \mod 2= \dt_{b1} \mod2,
\end{equation}
\begin{equation}
\dt_{c1} = p( ({}^1\ovl{E} {}^2\ovl{E})_c ) \mod 2= k_1 \mod2.
\end{equation}
In last paragraph we show that an RSI can be determined from momentum space irreps if it does not depend on how we choose $p$ for a given $B$ vector.
Here different parameters $k_{1,2}$ represent different choices of $p$ for a given $B$.
Therefore, we conclude that $\dt_{a1,2}$ and $\dt_{b1,2}$ can be determined from the momentum space irreps, whereas $\dt_{c}$ cannot be determined from the momentum space irreps.


\BAB{This is confusing. In \cref{sec:wann,sec:RSI-k-irrep} you only used the high symmetry points multiplicities in obtaining the RSIs.}
\SZD{Yes, we only need the momentum space irrep multiplicities to calculate the RSIs. But the underlying logic is that, the band can be Wannierized in form of \cref{eq:density-W} and we use the multiplicities of Wannier functions, $p$, to calculate the RSI.
If the band is not Wannierizable, then this method is not justified.}

\BAB{I don't fully understand how you compute the global-RSI and why they are not just related to the RSI.}
\SZD{Perhaps this logic flaw does not worth such a section to explain and I  wasted at least one week on it. I admit that this section is very confusing and seems not necessary. Now I delete the whole section. I will just say: applying the method in the above two sections to stable topologicla band we will get fractional RSI.}

\subsection{Diagnosis for topological states} \label{sec:FC-RSI}

We find that, in all the space groups, the number of $\mbb{Z}$-type RSIs plus 1 always equals to the rank of the EBR matrix (\cref{eq:EBR-def}), and there is a one-to-one correspondence between the $y$-vector (\cref{eq:y-def}), which is a faithfull and non-redundant representation of the symmetry eigenvalues, and the $\mbb{Z}$-type RSIs plus the band number.
Since the $y$-vector is a faithfull representation of the symmetry data vector, the set of $\mbb{Z}$-type RSIs plus the band number is another faithfull representation of the symmetry data vector.
We write the band number and $\mbb{Z}$-type RSIs in terms of $B$ as
\begin{equation}
(N, \delta_{a1},\delta_{a2},\cdots,\delta_{b1},\cdots)^T = \mcl{F}\cdot B,
\end{equation}
where $N$ is the band number, $\delta_{w i}$ is the $i$th $\mbb{Z}$-type RSI of the site-symmetry group $G_{w}$ of the Wyckoff position $w$, the first row of $\mcl{F}$ consists of the dimensions of the momentum space irreps at the first high symmetry momentum and zeros at other positions, and the remaining rows in $\mcl{F}$ can be obtained by the method in \cref{sec:RSI-k-irrep}.
\BAB{This is the RSI in terms of momentum space irreps, you mean \cref{sec:algoritm-RSI}?}
\SZD{No, I mean exactly the RSI in terms of momentum space irreps. Before I wrote $(N,\delta)^T = F\cdot B$, which is confusing because in \cref{sec:constraint-nobt} I wrote $(N_w,\delta)^T = F\cdot p$ at a single Wyckoff position. Now I change the notation here from $F$ to $\mcl{F}$ to avoid this confusion.}
According to \cref{eq:B-y}, the $(N,\delta)$-vector can be calculated from $y$ as
\begin{equation}
(N, \delta_{a1},\delta_{a2},\cdots,\delta_{b1},\cdots)^T = \mcl{F} \cdot \pare{\sum_{i=1}^r [L_\EBR]_i \lambda_i y_i} =  \mcl{F} \cdot L_{\EBR}\cdot [\Lambda_\EBR]_{:,1:r} \cdot y,
\end{equation}
where $[L_\EBR]_i$ is the $i$th column of $L_\EBR$ and $[\Lambda_\EBR]_{:,1:r}$ represents the submatrix of $\Lambda_\EBR$ consisting of the first $r$ columns of $\Lambda_\EBR$.
\LE{Is this true?}
We find that in all space groups $\mcl{F}\cdot L_{\EBR}\cdot [\Lambda_\EBR]_{:,1:r}$ is an \uemph{invertable} integer matrix.
Thus the correspondence between $(N,\delta)$-vectors and $y$-vectors are one-to-one. 
For bands with zero symmetry indicators, where $y$ is integer vector, the RSIs are always integer.
Furthermore, in all the wallpaper groups we find that integer RSIs always imply zero symmetry indicators.
Thus the sufficient and necessary condition for band to have nonzero indicators is that $\delta$ is fractional.
Therefore, RSIs diagnose all the stable topological phases implied by symmetry eigenvalues.
\SZD{Luis: do you agree with thses statements?}

Ref.~\cite{song_fragile_2019} shows that the criteria for a band structure to be EFP are inequalities or $\mbb{Z}_2$ equations of the $y$-vector.
Since the $\mbb{Z}$-type RSIs are equivalent to the $y$-vectors, all these criteria can be expressed in terms of the RSIs.
In \cref{sec:constraint-nobt} we derived some constraints between the orbital number at Wyckoff positions and the RSIs at those Wyckoff positions. 
The orbital number at a Wyckoff position must be larger than some lower bounds determined by the RSIs at the Wyckoff position, and satisfies some mod equation (\cref{tab:Nocc-RSI}). 
Here we present a method to derive the fragile criteria based on these orbital number constraints.
We will show that all the criteria in Ref.~\cite{song_fragile_2019} can be obtained by this new method.
First let us derive the eigenvalue condition for a band structure to be Wannierizable (EOAP or ETAP).
We assume a band structure is Wannierizable and write the band number as a sum of orbital numbers at \uemph{maximal} Wyckoff positions
\begin{equation}
N = \sum_{w}^\pr N_{w},
\end{equation}
where $w$ sums over all maximal Wyckoff positions, and $N_{w}$ ($\ge 0$) is the orbital numbers at $w$.
To be specific, for the state \cref{eq:density-W}, $N_{w}$ is
\begin{equation}
N_w = \sum_i |\rho_{G_w}^i| \cdot p(\rho_{G_w}^i),
\end{equation}
where $|\rho_{G_w}^i|$ is the dimension of $\rho_{G_w}^i$ and $p(\rho_{G_w}^i)$ is the multiplicity on the irrep $\rho_{G_w}^i$.
In general the symmetry data vector does not uniquely determine the $p$-vector.
For a given symmetry data vector $B$, we have many choices of $p$, which correspond to the same $B$ but can represent topologically distinct phases (\cref{sec:wann}).
\BAB{Sorry, this is just wrong/confusing. $B$ is the symmetry data vector. The only way that different phases can have the same $B$ is that they have different Wilson loops! Your method so far relies on eigenvalues.}
\SZD{Sorry, I'm 100 percent sure this sentence is correct. 
I'm talking about different $p$ with same $B$, and I never said that $p$ can be determined from symmetry eigenvalues! And I have given an example (in \cref{sec:wann}) where different $p$'s corresponding to the same $B$ (symmetry eigenvalues) represent distinct phases.}
Since $N_w$ is determined by the $p$-vector, $B$ cannot determine $N_w$.
However, in the case that all the $p$ components at the position $w$ are nonnegative, we have some constraints on the $N_w$, as we discuss in \cref{sec:constraint-nobt}.
Therefore, if $B$ belongs to EOAP or ETAP, where among all the choices there exists at least a nonnegative $p$, then the constraints between $N_w$ and $\{\delta_w\}$, which by definition are equivalent to $p\in \mbb{N}$, are satisfied.
Thus we conclude that
\begin{equation}
B\in \mrm{EOAP}\ \mrm{or}\ \mrm{ETAP}\quad \Rightarrow\quad
\exists \{N_{w}\} \  \text{subject to constraints}(\{\delta_{wi}\})\quad s.t.\quad N = \sum_{w} N_w, \label{eq:EFP-C-tmp}
\end{equation}
where $\mrm{constraints}(\{\delta_{wi}\})$ are lower bounds of $N_w$ and the mod $n$ ($n=2,4,6$) restrictions of $N_w$ in \cref{tab:Nocc-RSI}.
Since we have excluded the eigenvalue stable topological phases in discussion, then, for a given set of RSIs, the violation of the r.h.s. of \cref{eq:EFP-C-tmp} implies EFP.
Mathematically, the \uemph{sufficient} condition for a symmetry data vector with zero symmetry indicators to be EFP can be written as
\begin{equation}\boxed{
\not\exists\  \{N_w\} \ \text{subject to constraints}(\{\delta_{wi}\})\quad s.t.\quad N = \sum_{w} N_w\quad\Rightarrow\quad B\in \mrm{EFP}.} \label{eq:EFP-C}
\end{equation}

Here we use the wallpaper group $p3$ with SOC and TRS as an example to show how to derive the fragile criteria.
We find that, in $p3$, \cref{eq:EFP-C} is not only sufficient but also \uemph{necessary} for a band to be EFP.
Later we will prove that \cref{eq:EFP-C} is necessary in all wallpaper groups.
$p3$ has three maximal Wyckoff positions $a$, $b$, $c$, the site-symmetry groups of which are PG 3.
According to \cref{tab:RSI-formula}, PG 3 has a single $\mbb{Z}$-type RSI.
We denote the $\mbb{Z}$-type RSIs on the three positions as $\dt_{a}$, $\dt_{b}$, $\dt_{c}$, respectively.
We first consider the inequality constraints.
As shown in Ref. \cite{song_fragile_2019} and tabulated in \cref{tab:FC}, there are three inequality criteria for $p3$: 
\begin{equation}
N< \dt_{a1} -2 \dt_{b1} - 2\dt_{c1},\qquad N< -2\dt_{a1} + \dt_{b1} - 2\dt_{c1},\qquad N< -2\dt_{a1} -2 \dt_{b1} + \dt_{c1}. \label{eq:p3-FC-1}
\end{equation}
Now we show that all of them can be obtained from \cref{eq:EFP-C}.
We denote the numbers of WFs at the three positions as $N_a$, $N_b$, $N_c$, respectively.
If the $p$-vector on a maximal Wyckoff position $w$ can be chosen as nonnegative, then, according to \cref{tab:Nocc-RSI}, we have the constraints 
\begin{equation}
N_w \ge \dt_w,\qquad N_w\ge -2\dt_w. \label{eq:p3-Nw-dw}
\end{equation} 
(The constraint  $N_w = 4\dt_w \mod 6$ will lead to $\mbb{Z}_2$ criteria and hence will be considered later.)
Since the number of bands is the sum of $N_{a,b,c}$, \ie $N=N_a+N_b+N_c$, if the system is an EOAP or ETAP, where the $p$-vector is nonnegative, there must be 
\begin{equation}
N\ge \max(\dt_{a}, -2\dt_{a}) + \max(\dt_{b}, -2\dt_{b}) + \max(\dt_{c}, -2\dt_{c}). \label{eq:p3-trivial-1}
\end{equation}
Therefore, the violation of this constraint implies negative $p$ components and hence an EFP phase.
\cref{eq:p3-FC-1} are three cases for \cref{eq:p3-trivial-1} to be violated.
Accutually \cref{eq:p3-trivial-1} is equivalent to eight linear inequalities.
We can obtain these inequalities by replacing $\max(\dt_{w},-2\dt_w)$ with $\dt_w$ or $-2\dt_w$ as the lower bound of $N_w$. 
The violation of each of these eight inequalities implies an EFP phase.
Readers may ask where the other five inequalities, such as $N<\dt_a+\dt_b+\dt_c$, go?
The answer is that the other five inequalities can never be violated in gapped band structure satisfying compatibility relations.
Here we only show by example that $N<\dt_a + \dt_b + \dt_c$ is forbidden by compatibility relations. 
$p3$ has three maximal Wyckoff positions $\Gamma$, $K$, $KA$. 
$\Gamma$ has two two-dimensional irreps $\ovl{\Gamma}_4\ovl{\Gamma}_4$, $\ovl{\Gamma}_5\ovl{\Gamma}_6$; $K$ has three one-dimensional irreps $\ovl{K}_4$, $\ovl{K}_5$, $\ovl{K}_6$; $KA$ also has three one-dimensional irreps $\ovl{KA}_4$, $\ovl{KA}_5$, $\ovl{KA}_6$.
The compatibility relations enforced by band number and TRS are
\begin{equation}
2m(\ovl{\Gamma}_4\ovl{\Gamma}_4) + 2 m(\ovl{\Gamma}_5\ovl{\Gamma}_6) = m(\ovl{K}_4) + m(\ovl{K}_5) + m(\ovl{K}_6),
\end{equation}
and 
\begin{equation}
m(\ovl{K}_4) = m(\ovl{KA}_4),\qquad
m(\ovl{K}_5) = m(\ovl{KA}_5),\qquad
m(\ovl{K}_6) = m(\ovl{KA}_6).
\end{equation}
On the one hand, according to \cref{tab:WGRSI}, $ \dt_a + \dt_b + \dt_c$ is given by
\begin{equation}
\dt_a + \dt_b + \dt_c = -2m(\ovl{\Gamma}_4\ovl{\Gamma}_4)-5m(\ovl{\Gamma}_5\ovl{\Gamma}_5) + 2m(\ovl{K}_5) + 2m(\ovl{K}_6) + 2m(\ovl{KA}_4) = 2m(\ovl{\Gamma}_4\ovl{\Gamma}_4) - m(\ovl{\Gamma}_5\ovl{\Gamma}_5).
\end{equation}
On the other hand, the band number is $N = 2m(\ovl{\Gamma}_4\ovl{\Gamma}_4) + 2m(\ovl{\Gamma}_5\ovl{\Gamma}_5)$.
Thus the relation $N\ge \dt_a + \dt_b+\dt_c$ is always satisfied because $m(\ovl{\Gamma}_5\ovl{\Gamma}_5)$ is a nonnegative number.
All the inequalities implied by the positivity of the multiplicities of momentum space irreps can be derived in a similar way; they are
\begin{equation}
N\ge \dt_a + \dt_b + \dt_c,\qquad 
N\ge \dt_a + \dt_b - 2\dt_c,\qquad
N\ge \dt_a - 2\dt_b + \dt_c,\qquad
N\ge - 2\dt_a + \dt_b + \dt_c,\qquad
N\ge - 2\dt_a -2 \dt_b -2 \dt_c.
\end{equation}
They forbid the other five cases for \cref{eq:p3-trivial-1} to be violated in gapped band structures satisfying compatibility relations.

Now we consider the $\mbb{Z}_2$ criteria of $p3$.
\BAB{I thought only the $\mbb{Z}$ criteria matter! You kept saying that the $\mbb{Z}_2$ don't matter; some of them are even not invariant under the $p$-vector re-choice.}
\SZD{I kept saying that $\mbb{Z}_2$-RSIs don't matter. Here I'm talking about $\mbb{Z}_2$ criteria, which are also given by $\mbb{Z}$-RSIs.}
One should not confuse $\mbb{Z}_2$ criteria with $\mbb{Z}_2$-type RSIs, which are introduced in \cref{sec:RSI} and shown to be either redundant or dependent on the choices of $p$-vector in \cref{sec:RSI-WG}.
The $\mbb{Z}_2$ criteria are $\mbb{Z}_2$ equations of band number and $\mbb{Z}$-type RSIs.
Making use of the fact $N$ is even (due to spinfull TRS), we can rewrite the five $\mbb{Z}_2$ criteria in Ref.~\cite{song_fragile_2019}  (also tabulated in \cref{tab:FC}) as
\begin{equation}
\dt_{b}=1\mmod2,\qquad N = \dt_a + \dt_b -2\dt_c, \label{eq:p3-FC-2}
\end{equation}
\begin{equation}
\{\dt_{b}\mmod2,\;\dt_c\mmod2\} =\{0,1;\; 1,0;\; 1,1\},\qquad N = \dt_a + \dt_b + \dt_c, \label{eq:p3-FC-3}
\end{equation}
\begin{equation}
\dt_{b}\mmod2 = \dt_c\mmod2=1,\qquad N = \dt_a + \dt_b + \dt_c +3, \label{eq:p3-FC-4}
\end{equation}
\begin{equation}
\dt_{c}\mmod2 =1,\qquad N = \dt_a  -2 \dt_b + \dt_c, \label{eq:p3-FC-5}
\end{equation}
\begin{equation}
\dt_{b}\mmod2 =1,\qquad N = -2 \dt_a  + \dt_b + \dt_c. \label{eq:p3-FC-6}
\end{equation}
We will show that all of these criteria can be derived from the mod-equation constraints $N_w = 4 \dt_w \mod 6$.
We write $N_w = 4 \dt_w + 6 n_w$ ($n_w\in \mbb{Z}$) such that the mod-equation constraints are automatically satisfied and the two inequalities $N_w\ge \delta_w$, $N_w\ge -2\delta_w$ (\cref{eq:p3-Nw-dw}) become $ n_w \ge -\frac12 \dt_w$ and $ n_w \ge -\dt_w$.
When $\dt_w \in odd$, the lower bound $-\frac12 \dt_w$ of $n_w$ cannot be saturated because $-\frac12\dt_w$ is not an integer.
Instead, the lower bound of $n_w$ becomes $\max(-\frac12 \dt_w + \frac12,-\dt_w)$.
Correspondingly, in a Wannierizable state where the $p$ vector is nonnegative integer, the number of Wannier states $N_w=4\dt_w + 6n_w$ at the Wyckoff position $w$ should satisfy 
\begin{align}
N_w\ge & \max(\dt_w + 3,-2\dt_w) \qquad (\dt_w\in odd),\nono\\
N_w\ge & \max(\dt_w,-2\dt_w) \qquad (\dt_w\in even). \label{eq:p3-Nw-tighter}
\end{align}
Therefore, even if $(N,\dt_w)$ satisfies \cref{eq:p3-trivial-1}, $(N,\dt_w)$ may also represent an EFP because $\dt_{a,b,v}$ may be odd and $N$ may be smaller than the tighter lower bound.
We first consider \cref{eq:p3-FC-2}.
When $\dt_b\mod2=1$, the lower bound of $N_b$ becomes $\max(\dt_b+3,-2\dt_b)$, then, in a Wannierizable phase the relation $N=N_{a}+N_b +N_c \ge \dt_a + \dt_b - 2\dt_c +3$ is fulfilled according to \cref{eq:p3-Nw-tighter}.
Thus $N =\dt_a + \dt_b - 2\dt_c$ does not satisfy this constraint and hence represents an EFP.
We then look at \cref{eq:p3-FC-3,eq:p3-FC-4}.
If $\dt_b$ ($\dt_c$) is odd, then in a Wannierizable phase there is $N_b\ge \max(\dt_b+3,-2\dt_b)$ ($N_c\ge \max(\dt_c+3,-2\dt_c)$) and hence $N=N_a+N_b+N_c\ge \dt_a+\dt_b+\dt_c+3$ according to \cref{eq:p3-Nw-tighter}.
Thus $N = \dt_a+\dt_b+\dt_c$ represents EFP.
If both $\dt_b$ and $\dt_c$ are odd, then in a Wannierizable phase the relation $N\ge \dt_a+\dt_b+\dt_c+6$ is true.
Thus $N = \dt_a+\dt_b+\dt_c$ and $N = \dt_a+\dt_b+\dt_c+3 $ represent EFP. 
At last we consider \cref{eq:p3-FC-5,eq:p3-FC-6}.
When $\dt_b$ ($\dt_c$) is odd, then in a Wannierizable phase
$N_b\ge \max(\dt_b+3,-2\dt_b)$ ($N_c\ge \max(\dt_c+3,-2\dt_c)$) is true and hence $N=N_a+N_b+N_c\ge \dt_a-2\dt_b+\dt_c+3$ ($N\ge -2\dt_a+\dt_b+\dt_c+3$) is true according to \cref{eq:p3-Nw-tighter}.
Thus $N= \dt_a-2\dt_b+\dt_c$ ($N= -2\dt_a+\dt_b+\dt_c$ ) represents EFP.

As shown in the example in last paragraph, for Wannierizable phase we can obtain a set of inequality constraints between the band number and the RSIs from the inequality constraints between orbital numbers and RSIs at all maximal Wyckoff positions.
The inequality type of fragile criteria can be obtained by violating these constraints.
We have exhaustively compared the inequality criteria obtained in this way with those given by the polyhedron method in Ref. \cite{song_fragile_2019} in all wallpaper groups and find that they are \uemph{always} equivalent.

As for the $\mbb{Z}_2$ type criteria, we only need to consider $p3$, $p3m1$, $p31m$, $p6$, $p6mm$ with SOC and TRS, which are the only five wallpaper groups having $\mbb{Z}_2$ criteria.
It should be noticed that all these groups are supergroups of $p3$; and it is direct to verify that the $\mbb{Z}_2$ criteria in all these groups are induced from those of $p3$.
Here we take $p3m1$ as an example.
In \cref{sec:RSI-reduction} we introduced the reduction of RSIs from supergroup to subgroup.
Now we use these reduction relation to show that all $\mbb{Z}_2$ criteria in $p31m$ are induced from thoese in $p3$.
$p31m$ has two maximal Wyckoff positions $a\ (0,0)$, $b\ (1/3,2/3),\ (2/3,1/3)$; $a$ has site-symmetry group $3m$ and $b$ has site-symmetry group $3$.
As shown in \cref{tab:RSI-formula}, both PG $3m$ and PG $3$ with SOC and TRS have a single $\mbb{Z}$-type RSI. 
We denote the RSIs at the two positions as $\dt_a$, $\dt_b$, respectively.
$p3$ has three maximal Wyckoff positions $a\ (0,0)$, $b\ (1/3,2/3)$, $c\ (2/3,1/3)$; all of them have the site-symmetry PG $3$.
We denote the RSIs at the three positions as $\dt_{a}^\pr$, $\dt_{b}^\pr$, $\dt_c^\pr$ .
According to the reduction relations in \cref{sec:RSI-reduction}, the $\dt_1$ of PG $3m$ reduces to $\dt_1^\pr$ of PG $3$, \ie $\dt_1^\pr=\dt_1$, if we do not take the mirror symmetry into account.
Therefore, the reduction from RSIs of $p31m$ to RSIs of $p3$ is
\begin{equation}
\dt_{a}^\pr = \dt_a,\qquad \dt_b^\pr=\dt_c^\pr = \dt_b.\label{eq:RSI-p31m-p3}
\end{equation}
$p31m$ with SOC and TRS has three $\mbb{Z}_2$ criteria (\cref{tab:FC}):
\begin{equation}
\dt_b\mmod2=1, \qquad N = \dt_a + 2\dt_b,
\end{equation}
\begin{equation}
\dt_b\mmod2=1, \qquad N = \dt_a + 2\dt_b + 3,
\end{equation}
\begin{equation}
\dt_b\mmod2=1, \qquad N = -2\dt_a + 2\dt_b.    
\end{equation}
The three criteria can be written in terms of the reduced RSIs (\cref{eq:RSI-p31m-p3}) as
\begin{equation}
\dt_b^\pr \mmod2 = \dt_c^\pr \mmod2=1, \qquad N=\dt_a^\pr + \dt_b^\pr + \dt_c^\pr,
\end{equation}
\begin{equation}
\dt_b^\pr \mmod2 = \dt_c^\pr \mmod2=1, \qquad N=\dt_a^\pr + \dt_b^\pr + \dt_c^\pr + 3,
\end{equation}
\begin{equation}
\dt_b^\pr \mmod2 = \dt_c^\pr \mmod2=1, \qquad N=-2\dt_a^\pr + \dt_b^\pr + \dt_c^\pr,
\end{equation}
which are implied by \cref{eq:p3-FC-3,eq:p3-FC-4,eq:p3-FC-6}, respectively.
Thus the $\mbb{Z}_2$ criteria of $p31m$ are induced from those of $p3$.
Similarly, one can verify that all the $\mbb{Z}_2$ criteria are induced from those of $p3$.
Since we have shown that the $\mbb{Z}_2$ criteria of $p3$ can be derived from \cref{eq:EFP-C}, we conclude that all the $\mbb{Z}_2$ criteria can be derived from \cref{eq:EFP-C}.
In summary, by exhaustive calculation, we show that \cref{eq:EFP-C} is a \uemph{necessary and sufficient} for a band to be EFP.

\section{Detecting RSIs using twisted boundary conditions}\label{sec:TBC}

In this section we introduce a systematical method, the twisted boundary condition (TBC), to detect the nonzero RSIs at empty Wyckoff positions.
Due to the symmetry of the system, we divide the system into several parts (\cref{fig:TBC-2D-open}) and  multiply the hoppings between orbitals in different parts by some factors.
All these factors are functions of a single parameter $\lambda$.
As we tune $\lambda$ along some specific path, which depends on the symmetry group and the RSI, the spectrum of the system must close gap during the process.

\subsection{Centrosymmetric 1D example: SSH chain}\label{sec:1D-SSH}
\subsubsection{Solution under PBC}
We consider the SSH-chain model with inversion symmetry.
We put one $s$ orbital and one $p$ orbital at the $a$ position.
And we set the onsite energies and hopping as
\begin{equation}
\bra{x,s} \hat{H} \ket{x,s} = - \bra{x,p} \hat{H} \ket{x,p} = \Delta. \label{eq:SSH-h1}
\end{equation}
\begin{equation}
\bra{x,s} \hat{H} \ket{x+1,s} = - \bra{x,p} \hat{H} \ket{x+1,p} = \frac12 t_1.\label{eq:SSH-h2}
\end{equation}
\begin{equation}
\bra{x,s} \hat{H} \ket{x+1,p} = - \bra{x,p} \hat{H} \ket{x+1,s} =-\frac12 t_2.\label{eq:SSH-h3}
\end{equation}
Here we choose $\Delta,t_1,t_2,$ to be real numbers. 
\BAB{These are not the most general hoppings consistent with inversion.}
\SZD{They are not. As I said in next sentence, this model has accidental symmetries. But we will see these accidental symmetrys won't change our conclusion.}
This model has accidental time-reversal symmetry (TRS) ($\mcl{T}=\mcl{K}$) and chiral symmetry ($S=\sigma_x$).
But as be shown below, the inversion symmetry is enough to protect the boundary states.
We define a basis of Bloch wavefunctions as
\begin{equation}
\ket{\phi_{k\alpha}} = \sum_x e^{ikx} \ket{x,\alpha},\qquad \alpha=s,p. \label{eq:SSH-phi}
\end{equation}
Notice that $\ket{\phi_{k+2\pi,\alpha}}=\ket{\phi_{k\alpha}}$.
In the bases $\ket{\phi_{k\alpha}}$ the Hamiltonian is given as
\begin{equation}
H(k) = \begin{pmatrix}
\Delta + \frac12 t_1 (e^{ik} + e^{-ik}) & -\frac12 t_2 (e^{ik}-e^{-ik}) \\
-\frac12 t_2 (e^{-ik}-e^{ik}) &  -\Delta - \frac12 t_1 (e^{ik} + e^{-ik})
\end{pmatrix}
= (\Delta + t_1 \cos(k) ) \sigma_z + t_2 \sin(k) \sigma_y.
\end{equation}
The inversion symmetry operator in momentum space is given by $P=\sigma_z$.
According to the discussion in \cref{sec:RSI-k-irrep}, an 1D centrosymmetric system has two $\mbb{Z}$-type RSIs, $\delta_{a}$ and $\delta_b$ (one at each of the two Wyckoff positions), which represents the number of odd states minus the numbers of even states at $x=0,\frac12$, respectively.
It is direct to verify that
\begin{equation}
\delta_a, \delta_b = \begin{cases}
0,1,\qquad & t_1>|\Delta|\\
0,-1,\qquad      & t_1<-|\Delta|\\
1,0,\qquad       & |t_1|<|\Delta|,\ \Delta>0\\
-1,0,\qquad       & |t_1|<|\Delta|,\ \Delta<0
\end{cases}
\end{equation}
The $\delta_b\neq0$ phases are EOAP.
With an open boundary condition (OBC) that preserves the inversion symmetry centered at $x=\frac12$, the system has two zero modes at the two ends of the SSH-chain.
The two zero modes can be pushed away from zero energy by surface potential (or chiral symmetry breaking term in the bulk) without breaking inversion symmetry.
However, we emphasize that the half-filling many-body ground state has to be double-degenerate, because if one raise the energy of the mode at one end, then due to inversion symmetry the energy of the mode at another end will also be raised. 
Then the ground state remains double-degenerate.

For latter reference, here we solve the model in the bonding limit. 
We set $\Delta=0$, $t_1=t_2$.
Then the occupied Bloch wavefunctions and their eigenenergies are
\begin{equation}
u_{1}(k) = e^{-ik/2}( i \sin\frac{k}2,\;  \cos\frac{k}2)^T,\qquad E_1(k) = -t_1,
\end{equation}
\begin{equation}
u_{2}(k) = e^{-ik/2}( \cos\frac{k}2,\;  i\sin\frac{k}2)^T,\qquad E_2(k) = t_1.
\end{equation}
We choose the gauge $u_n(k+2\pi)=u_n(k)$ to make the Bloch wavefunction $\ket{\psi_{k n}} = \sum_\alpha \ket{\phi_{k\alpha}} u_{\alpha n}(k)$ to be periodic in momentum space.
The corresponding Wannier functions are
\begin{equation}
\ket{w_n(x)} = \frac{1}{\sqrt{N}} \sum_k e^{-i x k}\ket{\phi_{k\alpha}} u_{\alpha n}(k),
\end{equation}
where $x=0,\pm1\cdots$ represents the unit cell.
We obtain the Wannier states at the $x=0$ unit cell as
\begin{align}
\ket{w_1(0)} &=  \frac{1}{N} \sum_k\sum_x e^{-i\frac12 k}\pare{ \ket{x,s}  e^{ikx} i\sin\frac{k}2 + \ket{x,p} e^{ikx} \ket{x,p} \cos\frac{k}2 } \nono\\
&= \frac1{N}\sum_{k}\sum_x e^{-i\frac12 k} \pare{ \ket{x,s} \frac{e^{ikx+ik/2}-e^{ikx-ik/2}}{2} + \ket{x,p} \frac{e^{ikx+ik/2}+e^{ikx-ik/2}}{2} } \nono\\
& = \frac12 \pare{ \pare{\ket{0,s}+ \ket{0,p}} - \pare{\ket{1,s}-\ket{1,p}} } \label{eq:wan1-SSH}
\end{align}
\begin{align}
\ket{w_2(0)} = \frac12 \pare{ \pare{\ket{0,s}+ \ket{0,p}} + \pare{\ket{1,s}-\ket{1,p}} } \label{eq:wan2-SSH}
\end{align}
\BAB{What is $\ket{1,s}$?}
\SZD{Same with $\ket{x,s/p}$ in \cref{eq:SSH-h1,eq:SSH-h2,eq:SSH-h3,eq:SSH-phi}.}
$\ket{w_1}$ and $\ket{w_2}$ can be thought as the anti-bonding and the bonding states of the $sp^1$-hybridized orbitals, respectively.
We will use these two Wannier functions to study the edge states under TBC.

\subsubsection{Boundary states under twisted boundary condition}\label{sec:response-SSH}

\begin{figure}
\begin{centering}
\includegraphics[width=0.9\linewidth]{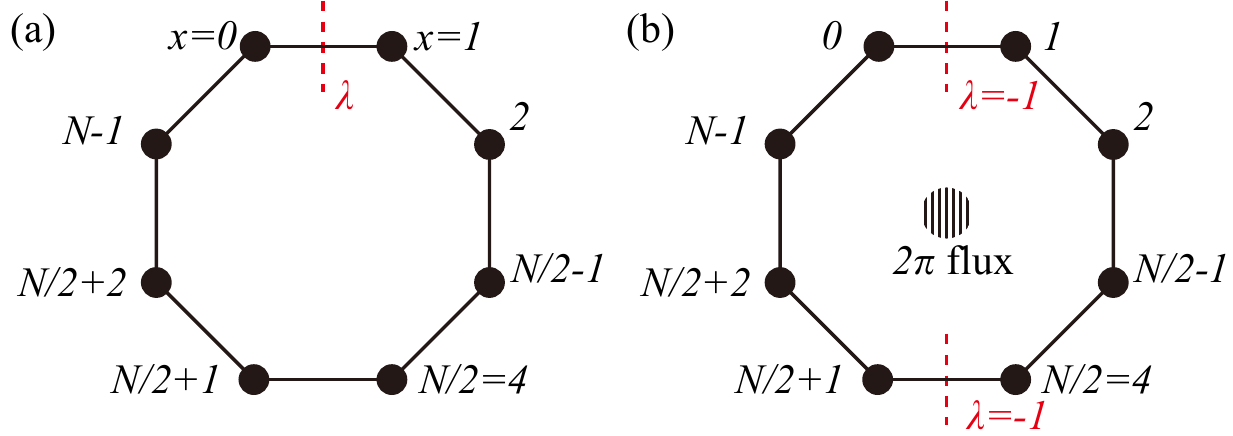}
\par\end{centering}
\protect\caption{(a) SSH-chain under TBC. Hopping between the $N$th and $1$th sites is multiplied by a real number $\lambda$.
The number of even states minus the number of odd states centered at the middle point between $1$th and $N$th sites is a good quantum number.
(b) Two bonds in the chain are multiplied by $\lambda$. When $\lambda=-1$, the Hamiltonian is equivalent to the original Hamiltonian up to a gauge transformation.
\label{fig:SSH-TBC}}
\end{figure}

\begin{figure}
\begin{centering}
\includegraphics[width=1\linewidth]{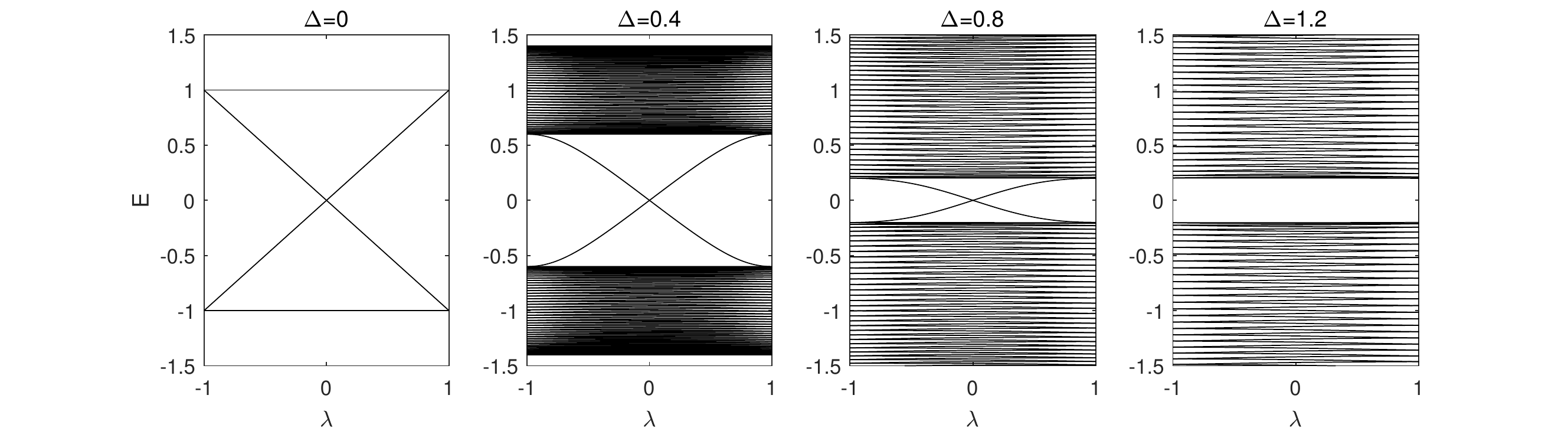}
\par\end{centering}
\protect\caption{Spectrum of SSH-chain under TBC, where $t_1=t_2=1$, and $\Delta=0, 0.4, 0.8, 1.2$ in the four plots, respectively.
\label{fig:SSG-spectrum}}
\end{figure}

We consider the SSH-chain under TBC (\cref{fig:SSH-TBC}a).
We multiply the hopping between from the $0$th site to the $1$st site by a number $\lambda$; and the hopping from the $N$th site to the $1$st site by $\lambda^*$.
We denote the Hamiltonian under TBC as $\hat{H}{(\lambda)}$.
We assume $|a\rangle$ is an orbital at $1$, then $P|a\rangle$ is a orbital at $N$, where $P$ is the inversion operation at $x=\frac12$.
Then the inversion symmetry requires 
\begin{equation}
\langle P a | \hat{H}(\lambda) |a \rangle = \langle a| \hat{H}(\lambda) P |a\rangle \quad \Rightarrow \quad  
\lambda \langle P a | \hat{H}(1) |a \rangle = \lambda^* \langle a| \hat{H}(1) P |a\rangle.
\end{equation}
Since $ \langle P a | \hat{H}(1) |a \rangle = \langle a| \hat{H}(1) P |a\rangle$, to preserve the inversion symmetry, $\lambda$ must be a real number. 
First we consider the model in the bonding limit, where the energy eigenstates are given as the localized Wannier functions in \cref{eq:wan1-SSH,eq:wan2-SSH}.
As $\lambda$ does not introduce hoppings between the two Wannier functions because they have different inversion eigenvalues, \cref{eq:wan1-SSH,eq:wan2-SSH} remain to be the energy eigenstates under the TBC.
Since the hopping is just the on-site energy of the Wannier functions, the energies of the two Wannier functions at $x=\frac12$ are modified to be $E_1 = -\lambda t_1$, $E_2 = \lambda t_1$, respectively.
As we change $\lambda$ from $1$ to $-1$ slowly, the energies of these two states integerchange; and the gap closes at the point $\lambda=0$.
Since these two Wannier functions have opposite inversion eigenvalues, the level crossing is stable against perturbations respecting inversion symmetry.
To verify this, we numerically calculate the spectrum, as shown in \cref{fig:SSG-spectrum}.
The gap closing disappears after the phase transition point $|\Delta|>|t_1|$.

Now we prove that as long as $\delta_2\neq 0$, there must be a gap closing of the spectrum as we change $\lambda$ from 1 to -1.
We make use of the property that $\lambda$ can only change the states near the modified bond (in a gapped state). 
We introduce two far-separated modified bonds in the chain, one between $0$ and 1 and another one between $\frac{N}{2}$ and $\frac{N}{2}+1$ (\cref{fig:SSH-TBC}b).
The Hamiltonian in presence of $\lambda$ is given as
\begin{equation}
\bra{x,\alpha} \hH(\lambda) \ket{y,\beta} = 
\begin{cases}
\lambda\bra{x,\alpha} \hH(1) \ket{y,\beta},\qquad & 1\le x \le \frac{N}2,\; \frac{N}2+1 \le y \le N,\\
\lambda\bra{x,\alpha} \hH(1) \ket{y,\beta},\qquad & \frac{N}2+1 \le x \le N,\; 1\le y \le \frac{N}2, \\
\bra{x,\alpha} \hH(1) \ket{y,\beta},\qquad & \text{else}. \label{eq:TBC-SSH}
\end{cases}
\end{equation}
Since the modified bond only change the states locally, we expect that \cref{fig:SSH-TBC}b will have the identical spectrum with \cref{fig:SSH-TBC}a, except that the level participating the crossing become double degenerate.
\BAB{Why can't you do \cref{fig:SSH-TBC}a? Why do you have to take \cref{fig:SSH-TBC}b?}
\SZD{Because I need to define the gauge transformation.}
We denote the inversion operation centered at $x=\frac12$ as $P$; and the corresponding RSI as $\delta_2^{\rm TBC}$.
Notice that $P$ leaves two positions, \ie $x=\frac12, \frac{N+1}{2}$, invariant.
Thus $\delta_2$ is given as the number of local even states centered at both $x=\frac12,\frac{N+1}{2}$ minus the number of local odd states centered at both $x=\frac12,\frac{N+1}{2}$.
Under the PBC ($\lambda=1$), due to the translation symmetry, this number is determined by $\delta_2^{\rm TBC} = 2\delta_2$.
At the end of the adiabatic process ($\lambda=-1$), the signs of hoppings passing through $x=\frac12$ and hoppings passing through $x=\frac{N+1}2$, \ie the first two cases in \cref{eq:TBC-SSH}, are flipped, and the system can be thought as the original system but with a $2\pi$-flux, as shown in \cref{fig:SSH-TBC}b.
Thus the wavefunctions at $\lambda=-1$ differ from the wavefunctions at $\lambda=1$ only by a gauge transformation.
To be specific, we introduce a new set of local orbitals
\begin{equation}
\ket{x,\alpha}^\pr = \begin{cases}
-\ket{x,\alpha},\qquad & 1\le x \le \frac{N}{2} \\
\ket{x,\alpha},\qquad & \frac{N}{2}+1\le x\le N
\end{cases}
\end{equation}
such that the Hamiltonian matrix under the $\ket{x,\alpha}^\pr$ basis remains the same as the original Hamiltonian matrix
\begin{equation}
\bra{x,\alpha}^\pr \hat{H}(\lambda=-1) \ket{y,\beta}^\pr = \bra{x,\alpha} \hat{H}(\lambda=1) \ket{y,\beta}. \label{eq:Hnew-SSH}
\end{equation}
And the inversion operator in the new basis is given by
\begin{equation}
P \ket{x,\alpha}^\pr = - \sum_{\alpha^\pr} \eta_\alpha \ket{1-x,\alpha^\pr}, \label{eq:Pnew-SSH}
\end{equation}
where $\eta_s=1$ and $\eta_p=-1$.
Suppose 
\begin{equation}
\ket{\psi} = \sum_{x \alpha} \ket{x,\alpha} c_{x,\alpha}
\end{equation} 
is an eigenstate of $\hat{H}(\lambda=1)$ with energy $E$ and $P$ eigenvalue $\xi$, where $c_{x,\alpha}$ is the amplitude on the local bases.
Then due to \cref{eq:Hnew-SSH,eq:Pnew-SSH}
\begin{equation}
\ket{\psi}^\pr = \sum_{x \alpha} \ket{x,\alpha}^\pr c_{x,\alpha}
\end{equation}
is an eigenstate of $\hat{H}(\lambda=-1)$ with energy $E$ and $P$ eigenvalue $-\xi$.
In other words the $P$ eigenvalues of all the occupied states are flipped.
Therefore, there must be $\delta_2^{\rm TBC} (-1)=-\delta_2^{\rm TBC}(1)$.
Since $\delta(\lambda)$ is a good quantum number in the whole process, there must be gap closing.

\subsection{Detecting 2D RSIs}\label{sec:TBC-2D}
In this subsection, we generalize the detecting method in \cref{sec:1D-SSH} to 2D systems.
The basic idea is to divide the 2D system into several parts, which transform to each other in turn under the 2D PG operations, as exemplified in \cref{fig:TBC-2D-open}.
Then we obtain the TBC $\hH(\lambda)$ by multiplying the hoppings between different parts by factors, which depend on a \uemph{single} parameter $\lambda$. 
For $\lambda=1$, the TBC reduces to the original Hamiltonian.
For some other value $\lambda=\lambda_0$, $\hH(\lambda_0)$ is equivalent to $\hH(1)$ up to a gauge transformation.
Thus the RSIs of $\hH(\lambda_0)$ can be determined from those of $\hH(1)$.
If the RSIs of $\hH(\lambda_0)$ and $\hH(1)$ are distinct, as we continuously tune $\lambda$ from 1 to $\lambda_0$, the gap of the system must close during this process.
This is because the RSIs are good quantum numbers in adiabatic processes, and change of RSIs must imply gap closing.

For simplicity, we put the unperturbed Hamiltonian $\hH(1)$ on an OBC geometry (\cref{fig:TBC-2D-open}).
Only the WFs at the center of the system contribute the RSIs of the total system.
For example for PG 2 in \cref{fig:TBC-2D-open}a, only the WFs at the $C_2$ invariant point at the center of the system will contribute the RSIs.
If the sample is a crystal and large enough, the RSIs of the total system can be calculated in terms of momentum space irreps, using the method introduced in \cref{sec:RSI-k-irrep}.
As we discuss in \cref{sec:RSI-reduction}, the RSIs of PGs $2mm$, $4mm$, $3m$, $6mm$ are completely induced from the RSIs of PGs $2$, $4$, $3$, $6$, respectively.
That means, a state having nonzero RSIs of these groups also have nonzero RSIs of the corresponding cyclic subgroup if the extra mirror symmetries are broken. 
Therefore, to detect the state with nonzero RSIs of PGs $2mm$, $4mm$, $3m$, $6mm$, we can just use the TBCs designed for $2$, $4$, $3$, $6$.
In the following, we only discuss the TBC of PGs $2$, $m$, $4$, $3$, $6$.



\subsubsection{Without SOC and TRS}
\textbf{PG $2$.}
PG $2$ has a single $\mbb{Z}$-type RSI $\delta$ (\cref{tab:RSI-formula}), which is the difference of the number of $C_2$-odd states and the number of $C_2$-even states.
\BAB{at every Wyckoff position.}
\SZD{PG 2 is a PG, there is only one $C_2$ invariant point.}
As shown in \cref{fig:TBC-2D-open}a, we divide the system into two parts: I and II, which transform to each other under the $C_2$-rotation.
Notice that I and II should be properly chosen such that no orbital in the system sits on the boundary between I and II.
We introduce the TBC by multiplying the hoppings from I to II by $\lambda$ and hoppings from II to I by $\lambda^*$ (for hermicity)
\begin{equation}
\bra{\RR,\alpha} \hat{H}(\lambda) \ket{\RR^\pr,\beta} = \begin{cases}
\bra{\RR,\alpha} \hat{H}(1) \ket{\RR^\pr,\beta},\qquad & \RR+\tt_\alpha\in {\rm I},\ \RR^\pr+\tt_\beta \in {\rm I} \\
\bra{\RR,\alpha} \hat{H}(1) \ket{\RR^\pr,\beta},\qquad & \RR+\tt_\alpha\in {\rm II},\ \RR^\pr+\tt_\beta \in {\rm II} \\
\lambda \bra{\RR,\alpha} \hat{H}(1) \ket{\RR^\pr,\beta},\qquad & \RR+\tt_\alpha\in {\rm I},\ \RR^\pr+\tt_\beta \in {\rm II} \\
\lambda^* \bra{\RR,\alpha} \hat{H}(1) \ket{\RR^\pr,\beta},\qquad & \RR+\tt_\alpha\in {\rm II},\ \RR^\pr+\tt_\beta \in {\rm I} \\
\end{cases}, \label{eq:TBC-2-NSOC-NTRS}
\end{equation}
where $\ket{\RR,\alpha}$ is the $\alpha$th orbital in the lattice $\RR$, and $\tt_\alpha$ is the position of the $\alpha$th orbital in each unit cell.
We assume $|a\rangle$ is an orbital in I, then $C_2|a\rangle$ is an orbital in II. Then the $C_2$ symmetry requires 
\begin{equation}
\langle a | \hat{H}(\lambda) |  C_2  a \rangle 
= \langle a | C_2 \hat{H}(\lambda) | a \rangle 
= \langle  C_2  a| \hat{H}(\lambda) |a\rangle \quad \Rightarrow \quad  
\lambda \langle a | \hat{H}(1) | C_2 a \rangle = \lambda^* \langle  C_2 a| \hat{H}(1) |a\rangle.\label{eq:C2-real-lambda}
\end{equation}
Since the $C_2$ symmetry enforces the relation $ \langle C_2 a | \hat{H}(1) |a \rangle  =  \langle  a | \hat{H}(1) |C_2 a \rangle $, $\lambda$ must be a real number.
For $\lambda=1$, the Hamiltonian reduces to the unperturbed Hamiltonian and thus $\delta^{\rm}(1)$ can be calculated from the momentum space irreps.
Now we consider $\lambda=-1$.
As in the example of the SSH chain in \cref{sec:1D-SSH}, for $\lambda=-1$, we can transfer the minus signs of the modified hoppings to the orbitals in II, \ie $\ket{a}\to -\ket{a}$ for $a$ in II.
Thus $\hat{H}{(-1)} $ is equivalent to the original Hamiltonian up to a gauge transformation.
We write this transformation as
\begin{equation}
\hat{H}{(-1)} = \hat{V} \hat{H}(1) \hat{V}^\dagger,
\end{equation}
with 
\begin{equation}
\hat{V} \ket{\RR,\alpha} = \begin{cases}
\ket{\RR,\alpha}, \qquad &\RR+\tt_\alpha \in {\rm I}\\
-\ket{\RR,\alpha}, \qquad &\RR+\tt_\alpha \in {\rm II}\\
\end{cases}.\label{eq:V-2-NSOC-NTRS}
\end{equation}
Now we show that $\hat{V}$ anti-commutes with the $C_2$-rotation.
Suppose $C_2$ transform an orbital in I $\ket{a}$ to another orbital $\ket{b}$ in II.
Then we have
\begin{equation}
\hat{C}_2 \hat{V} \ket{a}=  \hat{C}_2 \ket{a}=  \ket{b},
\end{equation}
and
\begin{equation}
\hat{V} \hat{C}_2 \ket{a}=  \hat{V} \ket{b}=  -\ket{b}.
\end{equation}
Thus $\{\hat{C}_2,\hat{V}\}=0$.
If $\ket{\psi}$ is an eigenstate of $\hat{H}(1)$ with the $C_2$ eigenvalue $\xi$, then $\hat{V}\ket{\psi}$ is an eigenstate of $\hat{H}{(-1)}$ with the  $C_2$ eigenvalue $-\xi$.
The $C_2$ RSI at $\hat{H}{(-1)}$, which is defined as the difference of the number of odd states and the number of even states, is opposite to the $C_2$ RSI for $\hat{H}(1)$, \ie $\delta_1(-1) = - \delta_1(1)$.
Therefore, for a state with $\delta_1(1)\neq 0$, as we continuously change $\lambda$ from $1$ to $-1$, $\delta_1(1)$ odd states will move from occupied bands to unoccupied bands and $\delta_1(1)$ even states will move from unoccupied bands to occupied bands, leading to $|\delta_1(1)|$ times of gap closing.
In conclusion, the change of multiplicities of irreps in this processes are
\begin{equation} 2:\quad
\begin{tabular}{c|c}
$\lambda\in\mbb{R}$ & $1\to -1$\\\hline\hline
$\Delta m(A)$ & $\dt_1$ \\
$\Delta m(B)$ & $-\dt_1$\\
\end{tabular}, \label{eq:path-2-NSOC-NTRS}
\end{equation}

\textbf{PG $m$.}
$m$ has a single $\mbb{Z}$-type RSI, $\delta_1$, which is defined as the number of mirror-odd states minus the number of mirror-even states (\cref{tab:RSI-formula}).
As shown in \cref{fig:TBC-2D-open}(b), we divide the system under PBC into two parts, I and II, which transform to each other under the mirror plane.
We introduce the TBC multiplying the hoppings from I to II by a factor $\lambda$, and multiplying the hoppings from II to I by $\lambda^*$. 
The TBC has the form \cref{eq:TBC-2-NSOC-NTRS}, with I and II defined in \cref{fig:TBC-2D-open}(b).
For the TBC to be invariant under the mirror plane, $\lambda$ must be real number as discussed above \cref{eq:C2-real-lambda}.
$H{(-1)}$ is equivalent to the $H(1)$ up to a gauge transformation $\hat{V}$, and the transformation has the form \cref{eq:V-2-NSOC-NTRS}, except that I and II are defined in \cref{fig:TBC-2D-open}(b).
It is direct to verify that the mirror operation anti-commutes with $\hat{V}$. 
Thus $H{(-1)}$ and $H(1)$ have the same gapped energy spectrum but opposite mirror eigenvalues.
Therefore, if we continuously tune $\lambda$ from 1 to -1, $\delta_1(1)$ odd states will move from occupied bands to unoccupied bands, and $\delta_1(1)$ even states will move from unoccupied bands to occupied bands, leading to $|\dt_1(1)|$ times of gap closing, \ie
\begin{equation} m:\quad
\begin{tabular}{c|c}
$\lambda\in\mbb{R}$ & $1\to -1$\\\hline\hline
$\Delta m(A^\pr)$ & $\dt_1$ \\
$\Delta m(A^\prpr)$ & $-\dt_1$\\
\end{tabular}, \label{eq:path-m-NSOC-NTRS}
\end{equation}

\textbf{PGs $4,3,6$.}
We divide a $C_n$-symmetric ($n=4,3,6$) system into $n$ parts, which transform to each other in turn under the $C_n$ rotation (\cref{fig:TBC-2D-open}(c-e)).
We define the TBC for $n=4,3,6$ as
\begin{equation} 4:\quad
\bra{\RR,\alpha} \hat{H}(\lambda) \ket{\RR^\pr,\beta} = 
\begin{cases}
\bra{\RR,\alpha} \hat{H}(1) \ket{\RR^\pr,\beta},\qquad & \mrm{idx}(\RR^\pr+\tt_\beta) = \mrm{idx}(\RR + \tt_\alpha)\\
\lambda \bra{\RR,\alpha} \hat{H}(1) \ket{\RR^\pr,\beta},\qquad & \mrm{idx}(\RR^\pr+\tt_\beta) = \mrm{idx}(\RR + \tt_\alpha)+1\\
\lambda^* \bra{\RR,\alpha} \hat{H}(1) \ket{\RR^\pr,\beta},\qquad & \mrm{idx}(\RR^\pr+\tt_\beta) = \mrm{idx}(\RR + \tt_\alpha)-1\\
\Re(\lambda^2) \bra{\RR,\alpha} \hat{H}(1) \ket{\RR^\pr,\beta},\qquad & \mrm{idx}(\RR^\pr+\tt_\beta) = \mrm{idx}(\RR + \tt_\alpha)+2\\
\end{cases}, \label{eq:TBC-4-NSOC-NTR}
\end{equation}
\begin{equation} 3:\quad
\bra{\RR,\alpha} \hat{H}(\lambda) \ket{\RR^\pr,\beta} = 
\begin{cases}
\bra{\RR,\alpha} \hat{H}(1) \ket{\RR^\pr,\beta},\qquad & \mrm{idx}(\RR^\pr+\tt_\beta) = \mrm{idx}(\RR + \tt_\alpha)\\
\lambda \bra{\RR,\alpha} \hat{H}(1) \ket{\RR^\pr,\beta},\qquad & \mrm{idx}(\RR^\pr+\tt_\beta) = \mrm{idx}(\RR + \tt_\alpha)+1\\
\lambda^* \bra{\RR,\alpha} \hat{H}(1) \ket{\RR^\pr,\beta},\qquad & \mrm{idx}(\RR^\pr+\tt_\beta) = \mrm{idx}(\RR + \tt_\alpha)-1\\
\end{cases}, \label{eq:TBC-3-NSOC-NTR}
\end{equation}
\begin{equation} 6:\quad
\bra{\RR,\alpha} \hat{H}(\lambda) \ket{\RR^\pr,\beta} = 
\begin{cases}
\bra{\RR,\alpha} \hat{H}(1) \ket{\RR^\pr,\beta},\qquad & \mrm{idx}(\RR^\pr+\tt_\beta) = \mrm{idx}(\RR + \tt_\alpha)\\
\lambda \bra{\RR,\alpha} \hat{H}(1) \ket{\RR^\pr,\beta},\qquad & \mrm{idx}(\RR^\pr+\tt_\beta) = \mrm{idx}(\RR + \tt_\alpha)+1\\
\lambda^* \bra{\RR,\alpha} \hat{H}(1) \ket{\RR^\pr,\beta},\qquad & \mrm{idx}(\RR^\pr+\tt_\beta) = \mrm{idx}(\RR + \tt_\alpha)-1\\
\lambda^2 \bra{\RR,\alpha} \hat{H}(1) \ket{\RR^\pr,\beta},\qquad & \mrm{idx}(\RR^\pr+\tt_\beta) = \mrm{idx}(\RR + \tt_\alpha)+2\\
\lambda^{*2} \bra{\RR,\alpha} \hat{H}(1) \ket{\RR^\pr,\beta},\qquad & \mrm{idx}(\RR^\pr+\tt_\beta) = \mrm{idx}(\RR + \tt_\alpha)-2\\
\Re(\lambda^3) \bra{\RR,\alpha} \hat{H}(1) \ket{\RR^\pr,\beta},\qquad & \mrm{idx}(\RR^\pr+\tt_\beta) = \mrm{idx}(\RR + \tt_\alpha)+3
\end{cases}, \label{eq:TBC-6-NSOC-NTR}
\end{equation}
respectively.
Here $\mrm{idx}(\rr)=$I, II, III, IV, V, VI represents the region where $\rr$ is located.
As discussed below \cref{eq:C2-real-lambda}, in presence of the $C_2$ symmetry, to preserve the $C_2$ symmetry the hoppings between $C_2$ partners should be real.
Thus the factor of the hoppings between the $j$th part and the $(j+2)$th ($(j+3)$th) part in PG $4$ ($6$) should be real.
For this purpose, we choose this real factor as $\Re(\lambda^2)$ and $\Re(\lambda^3)$ for PG 4 and PG 6, respectively.
We emphasize that the parameter $\lambda$ itself can be chosen as \uemph{any} complex number.
It is direct to verify that \cref{eq:TBC-4-NSOC-NTR,eq:TBC-3-NSOC-NTR,eq:TBC-6-NSOC-NTR} preserve the $C_n$ symmetries. 
For $\lambda= \exp(-i\frac{2\pi}{n} p)$ ($p=0,1\cdots n-1$), the TBC is equivalent to the unperturbed Hamiltonian up to a gauge transformation, \ie
\begin{equation}
\hat{H}(\lambda) = \hat{V}(\lambda)\hat{H}(1) \hat{V}^\dagger{(\lambda)},
\end{equation}
and the gauge transformation is
\begin{equation}
\hat{V}{(\lambda)} \ket{\RR,\alpha} = (\lambda^*)^{ \mrm{idx}(\RR+\tt_\alpha)-1} \ket{\RR,\alpha}. \label{eq:V-general}
\end{equation}
For example, for PG 4 and $\lambda=-i$, the explicit TBC and gauge transformation are 
\begin{equation}
\bra{\RR,\alpha} \hat{H}(-i) \ket{\RR^\pr,\beta} = 
\begin{cases}
\bra{\RR,\alpha} \hat{H}(1) \ket{\RR^\pr,\beta},\qquad & \mrm{idx}(\RR^\pr+\tt_\beta) = \mrm{idx}(\RR + \tt_\alpha)\\
-i \bra{\RR,\alpha} \hat{H}(1) \ket{\RR^\pr,\beta},\qquad & \mrm{idx}(\RR^\pr+\tt_\beta) = \mrm{idx}(\RR + \tt_\alpha)+1\\
i \bra{\RR,\alpha} \hat{H}(1) \ket{\RR^\pr,\beta},\qquad & \mrm{idx}(\RR^\pr+\tt_\beta) = \mrm{idx}(\RR + \tt_\alpha)-1\\
- \bra{\RR,\alpha} \hat{H}(1) \ket{\RR^\pr,\beta},\qquad & \mrm{idx}(\RR^\pr+\tt_\beta) = \mrm{idx}(\RR + \tt_\alpha)+2\\
\end{cases}
\end{equation}
and 
\begin{equation}
\hat{V}(-i) \ket{\RR,\alpha} = \begin{cases}
\ket{\RR,\alpha}, \qquad & \mrm{idx}(\RR,\alpha)=\mrm{I},\\ 
i\ket{\RR,\alpha}, \qquad & \mrm{idx}(\RR,\alpha)=\mrm{II},\\
-\ket{\RR,\alpha}, \qquad & \mrm{idx}(\RR,\alpha)=\mrm{III},\\  
-i\ket{\RR,\alpha}, \qquad & \mrm{idx}(\RR,\alpha)=\mrm{IV}.
\end{cases}
\end{equation}
respectively.
The gauge transformation does \uemph{not} commutes with the rotation symmetry.
Instead, a phase factor, $\lambda$, appears after the rotation commutes with $\hV$, \ie $\hat{C}_n \hV(\lambda) = \lambda \hV(\lambda) \hat{C}_n $.
Therefore, if $\ket{\psi}$ is an eigenstate of $H(1)$ with the $C_n$ eigenvalue $\xi$, then $\hV{(\lambda)} \ket{\psi}$ is an eigenstate of $H(\lambda)$ with the $C_n$ eigenvalue $\lambda \xi$.
The mappings from the irreps of $\ket{\psi}$ to the irreps of $\hV\ket{\psi}$ are
\begin{equation} 4:\quad 
\begin{tabular}{c|c|c|c|c}
$\lambda$ & 1 & $-i$ & -1 & $i$ \\\hline\hline
$A$ & $A$ & $^1E$ & $B$ & $^2E$\\
$^1E$ & $^1E$ & $B$ & $^2E$ & $A$ \\
$B$ & $B$ & $^2E$ & $A$ & $^1E$\\ 
$^2E$ & $^2E$ & $A$ & $^1E$ & $B$
\end{tabular},
\end{equation}
\begin{equation}3:\quad 
\begin{tabular}{c|c|c|c}
$\lambda$ & 1 & $ e^{-i\frac{2\pi}3}$ & $e^{i\frac{2\pi}3}$ \\\hline\hline
$A_1$ & $A_1$ & $^1E$ & $^2E$ \\
$^1E$ & $^1E$ & $^2E$ & $A_1$ \\
$^2E$ & $^2E$ & $A_1$ & $^1E$
\end{tabular},
\end{equation}
\begin{equation}6:\quad 
\begin{tabular}{c|c|c|c|c|c|c}
$\lambda$ & 1 & $ e^{-i\frac{\pi}3}$ & $e^{-i\frac{2\pi}3}$ & -1 & $e^{i\frac{2\pi}3}$ & $e^{i\frac{\pi}3}$\\\hline\hline
$A$ & $A$ & $^1E_2$ & $^2E_1$ & $B$ & $^1E_1$ & $^2E_2$ \\
$^1E_2$ & $^1E_2$ & $^2E_1$ & $B$ & $^1E_1$ & $^2E_2$  & $A$\\
$^2E_1$ & $^2E_1$ & $B$ & $^1E_1$ & $^2E_2$  & $A$  & $^1E_2$\\
$B$ & $B$ & $^1E_1$ & $^2E_2$  & $A$  & $^1E_2$  & $^2E_1$\\
$^1E_1$ & $^1E_1$ & $^2E_2$  & $A$  & $^1E_2$  & $^2E_1$ & $B$\\
$^2E_2$ & $^2E_2$  & $A$  & $^1E_2$  & $^2E_1$ & $B$  & $^1E_1$
\end{tabular}.\label{eq:lambda-6-NSOC-NTRS}
\end{equation}
We consider continuous tunning of $\lambda$ from $1$ to $\lambda_0 = \exp(-i \frac{2\pi}{n} p)$ ($p=1\cdots n-1$).
We emphasize that the path of $\lambda$ does not matter.
\BAB{If the path is complex, then it is not TRS invariant?}
\SZD{Exactly. But this section deals with the case without TRS, thus complex path is allowed. TRS will be discussed in next subsubsection.}
Suppose the irrep $\rho^\pr$ is mapped to $\rho$ by the transformation $\hV(\lambda_0)$, and $\rho$ is mapped to $\rho^\prpr$ ($\neq \rho$) by $\hV(\lambda_0)$, then the change of the multiplicity of $\rho$ during this process is $m(\rho^\pr) - m(\rho)$.
For example, for PG $4$, ${}^2E$ is mapped to $A$ by $\hV(-i)$, and $A$ is mapped to ${}^1E$ by $\hV(-i)$; then if we continuously tune $\lambda$ from $1$ to $-i$, the change of multiplicity of $A$ in the occupied levels will be $\Delta m(A) = m({}^2E) - m(A) = \dt_3$.
(See \cref{tab:RSI-formula} for the definitions of $\delta$'s.)
Following the same calculation, we obtain
\begin{equation} 4:\quad
\begin{tabular}{c|c|c|c}
$\lambda$ & $1\to -i$ & $1\to -1$ & $1\to i$\\\hline\hline
$\Delta m(A)$ & $\dt_3$ & $\dt_2$ & $\dt_1$\\
$\Delta m({}^1E)$ & $- \dt_1$ & $\dt_3-\dt_1$ & $\dt_2-\dt_1$\\
$\Delta m(B)$ & $\dt_1-\dt_2$ & $-\dt_2$ & $\dt_3-\dt_2$\\
$\Delta m({}^2E)$ & $\dt_2-\dt_3$ & $\dt_1-\dt_3$ & $-\dt_3$\\
\end{tabular}, \label{eq:path-4-NSOC-NTRS}
\end{equation}
\begin{equation} 3:\quad
\begin{tabular}{c|c|c}
$\lambda$ & $1\to e^{-i\frac{2\pi}{3}}$ & $1\to e^{i\frac{2\pi}{3}}$ \\\hline\hline
$\Delta m(A_1)$ & $\dt_2$ & $\dt_1$ \\
$\Delta m({}^1E)$ & $-\dt_1$ & $\dt_2-\dt_1$\\
$\Delta m({}^2E)$ & $\dt_1-\dt_2$ & $-\dt_2$ \\
\end{tabular}, \label{eq:path-3-NSOC-NTRS}
\end{equation}
\begin{equation} 6:\quad
\begin{tabular}{c|c|c|c|c|c}
$\lambda$ & $1\to e^{-i\frac{\pi}{3}}$ & $1\to e^{-i\frac{2\pi}{3}}$ & $1\to -1$ & $1\to e^{i\frac{2\pi}{3}}$ & $1\to e^{i\frac{\pi}{3}}$  \\\hline\hline
$\Delta m(A)$ & $\dt_5$ & $\dt_4$ & $\dt_3$ & $\dt_2$ & $\dt_1$ \\
$\Delta m(^1E_2)$ & $-\dt_1$ & $\dt_5-\dt_1$ & $\dt_4-\dt_1$ & $\dt_3-\dt_1$ & $\dt_2-\dt_1$ \\
$\Delta m(^2E_1)$ & $\dt_1-\dt_2$ & $-\dt_2$ & $\dt_5-\dt_2$ & $\dt_4-\dt_2$ & $\dt_3-\dt_2$ \\
$\Delta m(B)$ & $\dt_2-\dt_3$ & $\dt_1-\dt_3$ & $-\dt_3$ & $\dt_5-\dt_3$ & $\dt_4-\dt_3$ \\
$\Delta m(^1E_1)$ & $\dt_3-\dt_4$ & $\dt_2-\dt_4$ & $\dt_1-\dt_4$ & $-\dt_4$ & $\dt_5 - \dt_4$ \\
$\Delta m(^2E_2)$ & $\dt_4-\dt_5$ & $\dt_3-\dt_5$ & $\dt_2-\dt_5$ & $\dt_1-\dt_5$ & $-\dt_5$
\end{tabular}. \label{eq:path-6-NSOC-NTRS}
\end{equation}
\BAB{Can you make this in a table?}
\SZD{Only the results without TRS can be written in such a concise form. The results with TRS are complicated and does not suit a table. If we only make tables for the cases without TRS but do not make tables for cases with TRS, readers may think we don't have complete results.}

\subsubsection{Without SOC, with TRS}\label{sec:detect-2D-NSOC-TR}

As shown in \cref{tab:RSI-formula} and discussed in the following, states having nonzero RSIs of PGs with TRS always evolve to states having nonzero RSIs of PGs without TRS if we break the TRS.
Therefore, in general, all the RSIs of PGs with TRS can be detected by the TBC discussed in the last subsubsection.
To be specific, a TRS-symmetric representation decomposing into $\bigoplus_j m(\rho_G^j) \rho_G^j$, where $\rho_G^j$ are TRS irreps, can be thought as a representation  $\bigoplus_j m^\pr(\rho_{G_U}^j) \rho_{G_U}^j$, where $G_U$ is the unitary subgroup of $G$, and the multiplicity $m^\pr(\rho_{G_U})$ can be calculated as
\begin{equation}
    m^\pr(\rho_{G_U}) = \sum_j f(\rho_{G_U}\ |\ \rho_G^j \down G_U) m(\rho_G^j). \label{eq:mp-m-TRS}
\end{equation}
Then we can apply the TBC in \cref{eq:TBC-2-NSOC-NTRS,eq:TBC-4-NSOC-NTR,eq:TBC-3-NSOC-NTR,eq:TBC-6-NSOC-NTR} and choose the corresponding paths of $\lambda$ to detect the no-TRS RSIs.
In general, these TBC break the TRS.
In this subsection, we will discuss how to choose \uemph{TRS-symmetric} paths of $\lambda$.
We will show that, all the RSIs \uemph{except} those protected by $C_3$ can be detected in TRS-symmetric processes.

\textbf{PGs $2$, $m$.}
The irreps of PGs $2$ and $m$ are real, thus the co-irreps of $2$ and $m$ with TRS are the irreps of $2$ and $m$ without TRS.
As a consequence, the RSIs of PG $2$ and PG $m$ with TRS are also same as the RSIs of PG $2$ and PG $m$ without TRS.
Thus we can also use \cref{eq:TBC-2-NSOC-NTRS} to detect the RSIs.
The only question to check is whether \cref{eq:TBC-2-NSOC-NTRS} fulfills the TRS.
We consider an orbital $a_1$ in I and an orbital $a_2$ in II.
Since
\begin{align}
\bra{a_1} \hH(\lambda) \ket{a_2} = \lambda \bra{a_1} \hH(1) \ket{a_2} 
\end{align}
and
\begin{align}
\bra{a_1} \TRS \hH(\lambda) \TRS^{-1} \ket{a_2} = \bra{\TRS a_1} \hH(\lambda) \ket{ \TRS a_2}^* = \lambda^* \bra{\TRS a_1} \hH(1) \ket{\TRS a_2}^* = \lambda^* \bra{a_1} \TRS  \hH(1)\TRS^{-1} \ket{a_2} = \lambda^* \bra{a_1} \hH(1)\ket{a_2}, \label{eq:TRS-lambda}
\end{align}
$\hH(\lambda)$ respects the TRS iff $\lambda$ is a real number, which is also required by the $C_2$ or mirror symmetry (\cref{eq:C2-real-lambda}.)
Therefore, following the same procedure used without TRS, as we continuously tune $\lambda\in \mbb{R}$ from 1 to -1, the gap of the system closes $|\delta|$ times.

\textbf{PG $4$.}
The irreps $^1E$ and $^2E$ of $4$ are mutually conjugated and hence form the co-irrep $^1E^2E$ due to the TRS, whereas the irreps $A$ and $B$ of $4$ are real and hence themselves form the co-irreps.
Thus $4$ with TRS has three irreps: $A$, $B$, $^1E^2E$.
According to \cref{tab:RSI-formula}, PG $4$ with TRS has two $\mbb{Z}$-type RSIs: $\dt_1 = -m(A) + m(^1E^2E)$ and $\dt_2 = -m(A) + m(B)$, and PG $4$ without TRS has three $\mbb{Z}$-type RSIs, $\dt_1^\pr = m^\pr(^1E) - m^\pr(A)$,  $\dt_2^\pr = m^\pr(B) - m^\pr(A)$,  $\dt_3^\pr = m(^2E) - m(A)$.
Here we use prime in the superscript to denote quantities of PG $4$ without TRS.
Since $^1E^2E$ decomposes into $^1E \oplus {}^2E$, the no-TRS RSIs can be calculated as $\dt_1^\pr = \dt_3^\pr= m(^1E^2E) - m(A) = \dt_1$, $\dt_2^\pr = m(B) - m(A) = \dt_2$.
Thus a state having the RSI $(\dt_1,\dt_2)$ can be thought as having the no-TRS RSIs $(\dt_1^\pr, \dt_2^\pr, \dt_3^\pr) = (\dt_1,\dt_2,\dt_1)$ if we do not consider the TRS.
To detect these RSIs, we consider the TBC in \cref{eq:TBC-4-NSOC-NTR} and the paths of $\lambda$ in \cref{eq:path-4-NSOC-NTRS}.
We only consider the TRS symmetric processes.
To keep the TRS, all the factors on the hoppings in \cref{eq:TBC-2-NSOC-NTRS} must be real numbers.
(To prove this we can repeat the arguments used around \cref{eq:TRS-lambda}.)
Thus, among all the three paths of $\lambda$, only $1\to -1$ can be TRS symmetric.
Due to \cref{eq:path-4-NSOC-NTRS} and $\dt_1^\pr=\dt_3^\pr = \dt_1$, $\dt_2^\pr=\dt_2$, during the process from $\lambda=1$ to $\lambda=-1$, the irrep $A$ cross with the irrep $B$ $|\dt_2|$ times, leading to $|\dt_2|$ gap closings.
However, nonzero $\dt_1$ does not lead to gap closing in this process, since $\dt_1^\pr-\dt_3^\pr=0$ in our case of TRS.

To detect $\dt_1$, we consider another adiabatic tuning process where the $C_4$ symmetry is broken but the $C_2$ symmetry and TRS are kept.
Since the $C_2$ characters of $A$, $B$, and $^1E^2E$ are $1$, $1$, and $-2$, the three irreps reduce to the irreps $A$, $A$ and $2B$ of the PG $2$, respectively; and the $C_2$ RSI $\dt^\pr = m^\pr(B) - m^\pr(A)$ is given by $\dt^\pr = 2m(^1E^2E)-m(A)-m(B) = 2\dt_1 - \dt_2$.
Then we can use the TBC \cref{eq:TBC-2-NSOC-NTRS} to detect $\dt^\pr$.

\textbf{PG $3$.}
The irreps $^1E$ and $^2E$ of $3$ form a pair of conjugated irreps and hence form the co-irrep $^1E^2E$ due to TRS, whereas the irrep $A_1$ of $3$ is real and hence it is itself a co-irrep.
Thus the PG $3$ with TRS has two co-irreps: $A_1$ and $^1E^2E$.
According to \cref{tab:RSI-formula}, PG $3$ with TRS has a single RSI $\dt = m(^1E^2E) - m(A_1)$; PG $3$ without TRs has two RSIs $\dt_1^\pr = m^\pr(^1E) - m^\pr(A)$, $\dt_2^\pr = m^\pr(^2E) - m^\pr(A)$. 
Since $^1E^2E$ decomposes into $^1E \oplus {}^2E$,  a state having the RSI $\dt_1$ can be thought as having no-TRS RSIs $(\dt_1^\pr, \dt_2^\pr) = (\dt_1,\dt_1)$ if we relax the TRS.
We consider the TBC \cref{eq:TBC-3-NSOC-NTR} for PG $3$ without TRS to detect $\dt_1$.
However, all the two paths of $\lambda$ in \cref{eq:path-3-NSOC-NTRS} are not TRS-symmetric.
To preserve the TRS, using the same arguments as in \cref{eq:C2-real-lambda}, we come to the conclusion that the factors on the hoppings must be real numbers.
However, the two paths of $\lambda$ not real.
Therefore, we cannot find a TRS-symmetric TBC to detect RSIs of PG 3 with TRS.


\textbf{PG $6$.}
The irreps $^1E_1$ and $^2E_1$ of PG $6$ are mutually conjugated and hence form the co-irrep $^1E_1{}^2E_1$ due to TRS; similarly,  $^1E_2$ and $^2E_2$ form the co-irrep $^1E_2{}^2E_2$.
$A$ and $B$ of PG $6$ are real and hence themselves form co-irreps.
According to \cref{tab:RSI-formula}, PG $6$ with TRS has three RSIs: $\dt_1 = m(^1E_2{}^2E_2) -m(A) $, $\dt_2 = m(B) -m(A) $, $\dt_3 = m(^1E_1{}^2E_1) -m(A) $; $6$ without TRS has five RSIs: $\dt_1^\pr = m^\pr(^1E_2) - m^\pr(A)$ ,$\dt_2^\pr = m^\pr(^2E_1) - m^\pr(A)$, $\dt_3^\pr = m^\pr(B) - m^\pr(A)$, $\dt_4^\pr = m^\pr(^1E_1) - m^\pr(A)$, $\dt_5^\pr = m^\pr(^2E_2) - m^\pr(A)$.
Since $^1E_i {}^2E_i$ decomposes into $^1E_i \oplus {}^2E_i$ ($i=1,2$),  a state having the RSI $\dt_1$ can be thought as having the no-TRS RSIs $(\dt_1^\pr, \dt_2^\pr, \dt_3^\pr, \dt_4^\pr, \dt_5^\pr) = (\dt_1,\dt_3,\dt_2,\dt_3,\dt_1)$ if we neglect the TRS.
We consider the TBC \cref{eq:TBC-6-NSOC-NTR} to detect these RSIs.
Among the five paths of $\lambda$ in \cref{eq:path-6-NSOC-NTRS}, only $1\to -1$ can respects the TRS.
(See the discussion around \cref{eq:TRS-lambda}.)
After we tune $\lambda$ from $1$ to $-1$, the Hamiltonian is equivalent to the original Hamiltonian up to a transformation $\hV(-1)$.
According to \cref{eq:lambda-6-NSOC-NTRS}, $\hV(-1)$ interchanges $^1E_1{}^2E_1$ with $^1E_2{}^2E_2$ and $A$ with $B$.
Therefore, $m(^1E_2{}^2E_2) - m(^1E_1{}^2E_1) = \dt_1-\dt_3$ gap closings are formed by the level crossings of $^1E_2{}^2E_2$ and $^1E_1{}^2E_1$ irreps, and $m(B)-m(A) = \dt_2$ gap closings are formed by the level crossings of $B$ and $A$ irreps.

However, if $\dt_1=\dt_3\neq 0$ and $\dt_2=0$, no gap closing will appear in the process from $\lambda=1$ to $\lambda=-1$.
This case corresponds to $m(^1E_2{}^2E_2) = m(^1E_1{}^2E_1) \neq m(A)=m(B)$.
We do not find a TRS-symmetric process to detect this case.
TRS and $C_3$ symmetries protect the 2D degeneracy of $^1E_2{}^2E_2$ and $^1E_1{}^2E_1$.
To engineer the level crossings with respect to $C_3$ symmetry and TRS, one needs to interchange the 2D irreps with the 1D irreps, which will change the number of occupied bands.
So far we only discuss the TBC where the final state is a gauge transformation of the initial state, which cannot change band number.
Thus in our framework we cannot design such TBC to interchange the 2D irreps with 1D irreps.
We can remove the two-fold degeneracy of $^1E_2{}^2E_2$ and $^1E_1{}^2E_1$ by breaking $C_3$ and consider the TBC in the subgroup $2$ with TRS.
However, since $A$, $^1E_1{}^ 2E_1$ reduce to $A$ and $A\oplus A$ of $2$ and $B$, $^1E_2{}^ 2E_2$ reduce to $B$ and $B\oplus B$, once $C_3$ is removed, this state will have $m(A)=m(B)$ and hence zero RSI of PG $2$.
Thus we cannot construct TRS-symmetric TBC where gap closing is guaranteed for the case $\dt_1=\dt_3\neq 0$, $\dt_2=0$.

\subsubsection{With SOC, without TRS}

\textbf{PGs $2$, $m$, $4$, $3$, $6$.}
Each of these PGs is generated by a single generator.
We denote the generator as $g_0$ and the order of the PG as $n$.
Since these group are abelian, all the no-SOC and SOC irreps are 1D.
The no-SOC irreps satisfy $[\rho(g)]^n = 1$, thus the representation of the generator  is given by $\rho^p (g) = \exp(-i\frac{2\pi}{n}p)$ ($p=0\cdots n-1$).
The SOC irreps, on the other hand, satisfy $[\rho(g)]^n = -1$, and thus the representation of the generator is given by  $\ovl{\rho}^p(g) = \exp(-i\frac{\pi}{n}(2p-1))$ ($p=0\cdots n-1$).
There is a one-to-one correspondence between the no-SOC irreps and SOC irreps: $\ovl{\rho}^p(g_0) = \exp(i\frac{\pi}n)\rho^p(g_0)$.
From \cref{tab:2D-char}, we obtain the correspondences explicitly:
\begin{enumerate}[label=(\roman*)]
\item PG $2$. $A \leftrightarrow {}^1\ovl{E}$, $B \leftrightarrow {}^2\ovl{E}$.
\item PG $m$. $A^\pr \leftrightarrow {}^1\ovl{E}$, $A^\prpr \leftrightarrow {}^2\ovl{E}$.
\item PG $4$. $A \leftrightarrow {}^1\ovl{E}_1$, $B \leftrightarrow {}^1\ovl{E}_2$, $^1E \leftrightarrow {}^2\ovl{E}_1$, $^2E \leftrightarrow {}^2\ovl{E}_2$.
\item PG $3$. $A_1 \leftrightarrow {}^2\ovl{E}$, $^2E \leftrightarrow \ovl{E}$,  $^1E \leftrightarrow {}^1\ovl{E}$.
\item PG $6$. $A \leftrightarrow {}^2\ovl{E}_3$, $B \leftrightarrow {}^2\ovl{E}_2$, $^1E_2 \leftrightarrow {}^1\ovl{E}_3$, $^2E_2 \leftrightarrow {}^1\ovl{E}_1$, $^1E_1 \leftrightarrow {}^1\ovl{E}_2$, $^2E_1 \leftrightarrow {}^2\ovl{E}_1$.
\end{enumerate}
Correspondingly, there is a one-to-one correspondence between the no-SOC RSIs and the SOC RSIs.
One can verify that the SOC RSIs (the third column of \cref{tab:RSI-formula}) can be obtained from the no-SOC RSIs (the first column of \cref{tab:RSI-formula}) by replacing the no-SOC irreps by corresponding SOC-irreps.
Now we construct the SOC TBC as we did for the no-SOC case, \ie \cref{eq:TBC-2-NSOC-NTRS,eq:TBC-4-NSOC-NTR,eq:TBC-3-NSOC-NTR,eq:TBC-6-NSOC-NTR}.
When $\lambda = \exp(i\frac{2\pi}{n}q)$ ($q=1,2,\cdots n-1$), the Hamiltonian $\hH(\lambda)$ is equivalent to the original Hamiltonian $\hH(1)$ up to a gauge transformation $\hV(\lambda)$ (\cref{eq:V-general}). 
As in the no-SOC case, if $\ket{\psi}$ is an eigenstate of $\hH(1)$ with the symmetry eigenvalue $\xi$, then $\hV(\lambda)\ket{\psi}$ will be an eigenstate of $\hH(\lambda)$ with eigenvalue $\lambda\xi$.
If $\hV(\lambda)$ maps the no-SOC irrep $\rho$ to $\rho^\pr$, then $\hV(\lambda)$ will map the SOC irrep $\exp(i\frac{\pi}{n})\otimes \rho$ to $\exp(i\frac{\pi}{n})\otimes \rho^\pr$.
Therefore, the change of multiplicities of SOC irreps in the process $\lambda=1\to \exp(i\frac{2\pi}{n}q)$ can be obtained from \cref{eq:path-2-NSOC-NTRS,eq:path-m-NSOC-NTRS,eq:path-4-NSOC-NTRS,eq:path-3-NSOC-NTRS,eq:path-6-NSOC-NTRS} by replacing the no-SOC irreps with the corresponding SOC irreps.
\BAB{Are you just rewriting \cref{eq:V-2-NSOC-NTRS,eq:path-m-NSOC-NTRS,eq:TBC-4-NSOC-NTR,eq:RSI-3-NSOC-NTR,eq:TBC-6-NSOC-NTR}? Say this so people know you are just rewriting those tables.}
\SZD{I think I said that. See last sentence: ... can be obtained from \cref{eq:path-2-NSOC-NTRS,eq:path-m-NSOC-NTRS,eq:path-4-NSOC-NTRS,eq:path-3-NSOC-NTRS,eq:path-6-NSOC-NTRS} by replacing the no-SOC irreps with the corresponding SOC irreps.}
We obtain
\begin{equation} 2:\quad
\begin{tabular}{c|c}
$\lambda$ & $1\to -1$\\\hline\hline
$\Delta m(^1\ovl{E})$ & $\dt_1$ \\
$\Delta m(^2\ovl{E})$ & $-\dt_1$\\
\end{tabular}, \label{eq:path-2-SOC-NTRS}
\end{equation}
\begin{equation} m:\quad
\begin{tabular}{c|c}
$\lambda$ & $1\to -1$\\\hline\hline
$\Delta m(^1\ovl{E})$ & $\dt_1$ \\
$\Delta m(^2\ovl{E})$ & $-\dt_1$\\
\end{tabular}, \label{eq:path-m-SOC-NTRS}
\end{equation}
\begin{equation} 4:\quad
\begin{tabular}{c|c|c|c}
$\lambda$ & $1\to -i$ & $1\to -1$ & $1\to i$\\\hline\hline
$\Delta m(^1\ovl{E}_1)$ & $\dt_3$ & $\dt_2$ & $\dt_1$\\
$\Delta m(^2\ovl{E}_1)$ & $- \dt_1$ & $\dt_3-\dt_1$ & $\dt_2-\dt_1$\\
$\Delta m(^1\ovl{E}_2)$ & $\dt_1-\dt_2$ & $-\dt_2$ & $\dt_3-\dt_2$\\
$\Delta m(^2\ovl{E}_2)$ & $\dt_2-\dt_3$ & $\dt_1-\dt_3$ & $-\dt_3$\\
\end{tabular}, \label{eq:path-4-SOC-NTRS}
\end{equation}
\begin{equation} 3:\quad
\begin{tabular}{c|c|c}
$\lambda$ & $1\to e^{-i\frac{2\pi}{3}}$ & $1\to e^{i\frac{2\pi}{3}}$ \\\hline\hline
$\Delta m(^2\ovl{E})$ & $\dt_2$ & $\dt_1$ \\
$\Delta m(^1\ovl{E})$ & $-\dt_1$ & $\dt_2-\dt_1$\\
$\Delta m(\ovl{E})$ & $\dt_1-\dt_2$ & $-\dt_2$ \\
\end{tabular}, \label{eq:path-3-SOC-NTRS}
\end{equation}
\begin{equation} 6:\quad
\begin{tabular}{c|c|c|c|c|c}
$\lambda$ & $1\to e^{-i\frac{\pi}{3}}$ & $1\to e^{-i\frac{2\pi}{3}}$ & $1\to -1$ & $1\to e^{i\frac{2\pi}{3}}$ & $1\to e^{i\frac{\pi}{3}}$  \\\hline\hline
$\Delta m(^2\ovl{E}_3)$ & $\dt_5$ & $\dt_4$ & $\dt_3$ & $\dt_2$ & $\dt_1$ \\
$\Delta m(^1\ovl{E}_3)$ & $-\dt_1$ & $\dt_5-\dt_1$ & $\dt_4-\dt_1$ & $\dt_3-\dt_1$ & $\dt_2-\dt_1$ \\
$\Delta m(^2\ovl{E}_1)$ & $\dt_1-\dt_2$ & $-\dt_2$ & $\dt_5-\dt_2$ & $\dt_4-\dt_2$ & $\dt_3-\dt_2$ \\
$\Delta m(^2\ovl{E}_2)$ & $\dt_2-\dt_3$ & $\dt_1-\dt_3$ & $-\dt_3$ & $\dt_5-\dt_3$ & $\dt_4-\dt_3$ \\
$\Delta m(^1\ovl{E}_2)$ & $\dt_3-\dt_4$ & $\dt_2-\dt_4$ & $\dt_1-\dt_4$ & $-\dt_4$ & $\dt_5 - \dt_4$ \\
$\Delta m(^1\ovl{E}_1)$ & $\dt_4-\dt_5$ & $\dt_3-\dt_5$ & $\dt_2-\dt_5$ & $\dt_1-\dt_5$ & $-\dt_5$
\end{tabular}. \label{eq:path-6-SOC-NTRS}
\end{equation}

\subsubsection{With SOC and TRS}
In \cref{sec:detect-2D-NSOC-TR} we show that, in absence of SOC, nonzero RSIs in the groups with TRS always reduce to nonzero RSIs in the groups without TRS.
A representation of $G= G_U + \TRS\cdot G_U$ described by the multiplicities $m(\rho_G^j)$ can be thought as a representation of $G_U$ described by the multiplicities $m^\pr(\rho_{G_U}^j)$, where $G_U$ is the unitary subgroup of $G$.
$m^\pr(\rho_{G_U}^j)$ can be determined from $m(\rho_G^j)$ through \cref{eq:mp-m-TRS}.
Then we can use the TBC designed for $G_U$ to detect the RSIs of $G_U$ in terms of $m(\rho_G^j)$.
In this subsection, we generalize this method to the case with SOC.
Similar with the no-SOC case, we find that the RSIs protected by $C_3$-rotation \uemph{cannot} be detected through TRS-symmetric TBC.

\textbf{PGs $2$, $m$.}
With SOC and TRS, these two groups do not have $\mbb{Z}$-type RSIs.

\textbf{PG $4$.}
PG $4$ with TRS has only one $\mbb{Z}$-type RSI: $\dt_1 = m({}^1\ovl{E}_2 {}^2\ovl{E}_2) - m({}^1\ovl{E}_1 {}^2\ovl{E}_1)$, and $4$ without TRS has three RSIs $\dt_1^\pr = m^\pr({}^2\ovl{E}_1) - m^\pr({}^1\ovl{E}_1)$, $\dt_2^\pr = m^\pr({}^1\ovl{E}_2) - m^\pr({}^1\ovl{E}_1)$, $\dt_3^\pr = m^\pr({}^2\ovl{E}_2) - m^\pr({}^1\ovl{E}_1)$.
Here we use the prime in the superscript to denote the quantities in the PG $4$ without TRS.
Since ${}^1\ovl{E}_i {}^2\ovl{E}_i$ ($i=1,2$) decomposes into ${}^1\ovl{E}_i \oplus {}^2\ovl{E}_i$, the multiplicities $m^\pr$ are related to the multiplicities $m$ as $m^\pr(^1\ovl{E}_i) = m^\pr(^2\ovl{E}_i) = m({}^1\ovl{E}_i {}^2\ovl{E}_i)$.
Expressing $\dt_{1,2,3}^\pr$ in terms of $m$ and then in terms of $\dt_{1}$, we obtain $(\dt_1^\pr,\dt_2^\pr,\dt_3^\pr) = (0,\dt_1,\dt_1)$.
Thus a state having the RSI $\dt_1$ can be thought as a state having the no-TRS RSI $(0,\dt_1,\dt_1)$.
We consider the TBC \cref{eq:TBC-4-NSOC-NTR} and the paths of evolution in \cref{eq:path-4-SOC-NTRS}.
Among the three paths only the path $\lambda=1\to -1$ keeps the TRS.
Due to \cref{eq:path-4-SOC-NTRS} and the fact $(\dt_1^\pr,\dt_2^\pr,\dt_3^\pr) = (0,\dt_1,\dt_1)$, we have $\Delta m({}^1\ovl{E}_1 {}^2\ovl{E}_1) = \dt_1$ and $\Delta m({}^1\ovl{E}_2 {}^2\ovl{E}_2) =-\dt_1$.
Thus the RSI $\dt_1$ can be detected through the process $\lambda=1\to -1$.
$|\dt_1|$ number of gap closings will be formed by the level crossings between the irreps ${}^1\ovl{E}_1 {}^2\ovl{E}_1$ and ${}^1\ovl{E}_2 {}^2\ovl{E}_2$.

\textbf{PG $3$.}
PG $3$ with TRS has only one RSI $\dt_1 = 2m(\ovl{E}\ovl{E}) - m(^1\ovl{E} {}^2\ovl{E})$; and PG $3$ without TRS has two RSIs $\dt_1^\pr = m^\pr(^1\ovl{E}) - m^\pr({}^2\ovl{E})$, $\dt_2^\pr = m^\pr(\ovl{E}) - m^\pr({}^2\ovl{E})$.
Since $\ovl{E}\ovl{E}$ decomposes into $\ovl{E}\oplus\ovl{E}$ and $^1\ovl{E} {}^2\ovl{E}$ decomposes into $^1\ovl{E} \oplus{}^2\ovl{E}$, we have $m^\pr(\ovl{E}) = 2m(\ovl{E}\ovl{E})$, $ m^\pr(^1\ovl{E}) = m^\pr({}^2\ovl{E}) = m(^1\ovl{E} {}^2\ovl{E})$. 
Expressing $\dt_{1,2}^\pr$ in terms of $m$ and then in terms of $\dt_{1}$, we obtain $(\dt_1^\pr,\dt_2^\pr) =(0,\dt_1)$.
Thus a state having the RSI $\dt_1$ can be thought as a state having the no-TRS RSI $(0,\dt_1)$ if we remove the TRS.
We first consider the TBC designed for $3$ without TRS (\cref{eq:TBC-4-NSOC-NTR}) to detect the RSIs. 
However, the two paths of evolution of $\lambda$ in \cref{eq:path-3-SOC-NTRS}  break the TRS.

We do not find any TRS-symmetric TBC that can detect the RSI of PG 3 with TRS.
Here we present another \uemph{failed} attempt. 
We first consider a spinfull model preserving $C_3$, TRS, and $\hat{s}_z$ symmetry.
We introduce a new TBC where the spin up part of the system evolve in the path $\lambda=1\to \exp(i\frac{2\pi}3)$, whereas the spin down part evolve in another path $\lambda=1\to \exp(-i\frac{2\pi}3)$, such that the TRS is preserved.
At the end of the evolution, the deformed Hamiltonian is equivalent to the original Hamiltonian up to a gauge transformation.
However, as we show below, when $\hat{s}_z$ is not a good quantum number, this deformation must break the $C_3$ symmetry.
We denote the spin up/down local orbitals as $\ket{\RR,\alpha,\up/\down}$, then the gauge transformation can be written as
\begin{equation}
\hV \ket{\RR,\alpha,\up} = (e^{-i\frac{2\pi}3})^{\mrm{idx}(\RR+\tt_\alpha)-1} \ket{\RR,\alpha,\up},\qquad
\hV \ket{\RR,\alpha,\down} = (e^{i\frac{2\pi}3})^{\mrm{idx}(\RR+\tt_\alpha)-1} \ket{\RR,\alpha,\down},
\end{equation}
where $\mrm{idx}(\rr)=$I, II, III represents the region to which $\rr$ belongs (\cref{fig:TBC-2D-open}(d)).
Since the regions I, II, III transform  to each other in turn under the $C_3$ rotation, we obtain
\begin{equation}
\hV^\dagger \hat{C}_3 \hV \ket{\RR,\alpha,\up} = e^{i\frac{2\pi}3} \hat{C}_3\ket{\RR,\alpha,\up},\qquad
\hV^\dagger \hat{C}_3 \hV \ket{\RR,\alpha,\down} = e^{-i\frac{2\pi}3} \hat{C}_3\ket{\RR,\alpha,\down}. \label{eq:C3-V-SOC}
\end{equation}
Now let us check whether the relation $\hat{C}_3 \hV \hH \hV^\dagger \hat{C}_3^\dagger =  \hV \hH \hV^\dagger$ is satisfied.
For convenience, we write $\hat{H}$ as $\begin{pmatrix} \hH_{\up\up} & \hH_{\up\down}\\ \hH_{\down\up} & \hH_{\down\down} \end{pmatrix}$ and check whether $[\hV^\dagger\hat{C}_3\hV,\hH]=0$.
According to \cref{eq:C3-V-SOC}, we can write $\hV^\dagger\hat{C}_3\hV$ as
\begin{equation}
\hV^\dagger\hat{C}_3\hV = \begin{pmatrix}
    e^{i\frac{2\pi}3} & 0 \\ 0 & e^{-i\frac{2\pi}3}
\end{pmatrix}.
\end{equation}
We obtain
\begin{equation}
(\hV^\dagger\hat{C}_3\hV) \hH (\hV^\dagger\hat{C}_3\hV)^\dagger = 
\begin{pmatrix} 
\hH_{\up\up} & e^{-i\frac{2\pi}3}\hH_{\up\down}\\ 
e^{i\frac{2\pi}3}\hH_{\down\up} & \hH_{\down\down} 
\end{pmatrix}.
\end{equation}
Therefore, in the general case where $\hH_{\up\down}\neq0$, the newly designed TBC breaks the $C_3$-symmetry.
\BAB{I need to if we need other complicated twistings.}
\SZD{I have no idea how to systematically find all possible twistings.}

\textbf{PG $6$.}
PG $6$ with TRS has two $\mbb{Z}$-type RSI: $\dt_1 = m({}^1\ovl{E}_1 {}^2\ovl{E}_1) - m({}^1\ovl{E}_3 {}^2\ovl{E}_3)$, $\dt_2 = m({}^1\ovl{E}_2 {}^2\ovl{E}_2) - m({}^1\ovl{E}_3 {}^2\ovl{E}_3)$.
$6$ without TRS has five RSIs $\dt_1^\pr = m^\pr({}^1\ovl{E}_3) - m^\pr({}^2\ovl{E}_3)$, $\dt_2^\pr = m^\pr({}^2\ovl{E}_1) - m^\pr({}^2\ovl{E}_3)$, $\dt_3^\pr = m^\pr({}^2\ovl{E}_2) - m^\pr({}^2\ovl{E}_3)$, $\dt_4^\pr = m^\pr({}^1\ovl{E}_2) - m^\pr({}^2\ovl{E}_3)$, $\dt_5^\pr = m^\pr({}^1\ovl{E}_1) - m^\pr({}^2\ovl{E}_3)$.
Since ${}^1\ovl{E}_i {}^2\ovl{E}_i$ ($i=1,2$) decomposes into ${}^1\ovl{E}_i \oplus {}^2\ovl{E}_i$, the multiplicities $m^\pr$ are related to the multiplicities $m$ as $m^\pr(^1\ovl{E}_i) = m^\pr(^2\ovl{E}_i) = m({}^1\ovl{E}_i {}^2\ovl{E}_i)$.
Expressing $\dt_{1,2,3,4,5}^\pr$ in terms of $m$ and then in terms of $\dt_{1,2}$, we obtain $(\dt_1^\pr,\dt_2^\pr,\dt_3^\pr,\dt_4^\pr,\dt_5^\pr) =(0,\dt_1,\dt_2,\dt_2,\dt_1)$.
Thus a state having the RSI $\dt_1$ can be thought as a state having the no-TRS RSI $(0,\dt_1,\dt_2,\dt_2,\dt_1)$ breaking the TRS.
We consider the TBC \cref{eq:TBC-6-NSOC-NTR} and the paths of evolution in \cref{eq:path-6-SOC-NTRS}.
Among the five paths only $\lambda=1\to -1$ keeps the TRS.
Due to \cref{eq:path-6-SOC-NTRS} and the fact $(\dt_1^\pr,\dt_2^\pr,\dt_3^\pr,\dt_4^\pr,\dt_5^\pr) =(0,\dt_1,\dt_2,\dt_2,\dt_1)$, we have $\Delta m({}^1\ovl{E}_3 {}^2\ovl{E}_3) = \dt_2$ and $\Delta m({}^1\ovl{E}_2 {}^2\ovl{E}_2) =-\dt_2$.
Thus the RSI $\dt_2$ can be detected through the process $\lambda=1\to -1$.
$|\dt_2|$ gap closings will be formed by the level crossings between the irreps ${}^1\ovl{E}_3 {}^2\ovl{E}_3$ and ${}^1\ovl{E}_2 {}^2\ovl{E}_2$.

$\dt_1$ cannot be detected through the TRS symmetric TBC of PG $6$. 
Now we check whether $\dt_1$ can be detected through TBC breaking some crystalline symmetries but keeping TRS.
The PG $6$ has two nontrivial subgroups: PG $2$ and PG $3$.
We will not consider $2$ because it does not have $\mbb{Z}$-type RSI.
Hence we only consider the subgroup $3$.
However, as we discuss above, the RSI of $3$ with TRS cannot be detected through TRS symmetric TBC.
Thus we conclude that $\dt_1$ cannot be detected through any TRS symmetric TBC.
\BAB{What can it be detected through? We are missing.}
\SZD{No idea. Basically the TBC itself form an 1D irrep of the PG. But PG 3 does not have TRS-invariant 1D irrep. I think that's the reason we cannot find TRS-symmetric TBC.}

\subsection{Model}\label{sec:p4-model}

\begin{figure}
\begin{centering}
\includegraphics[width=0.8\linewidth]{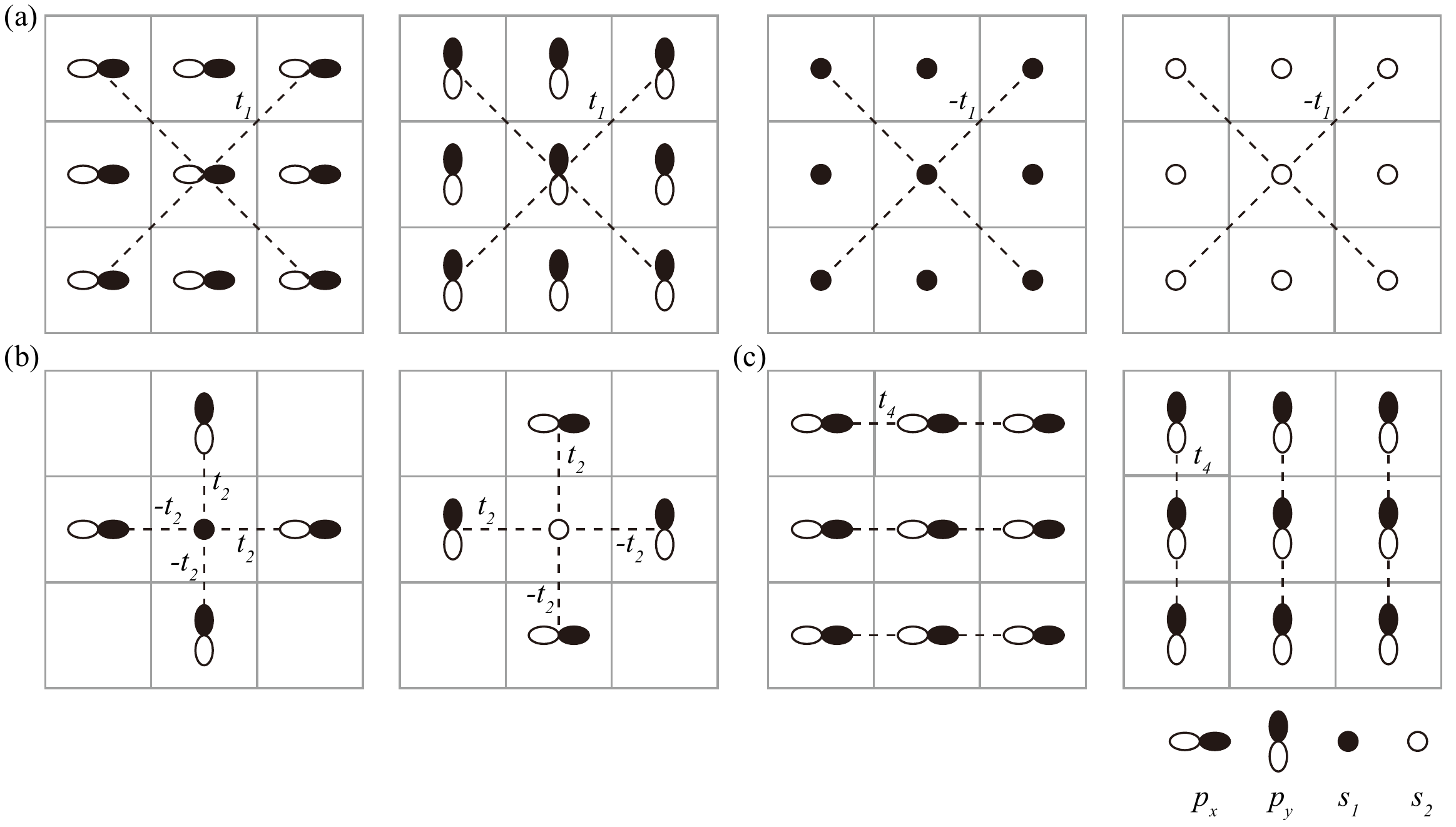}
\par\end{centering}
\protect\caption{The tight-binding model for the spinless EFP in wallpaper group $p4$ with TRS.
\label{fig:TB-p4}}
\end{figure}

\subsubsection{Details of the tight-binding model}
Here we build a spinless model with $C_4$ symmetry and TRS to verify the gapless spectrum under TBC.
The wallpaper group $p4$ with TRS has 12 EFP roots, as shown in \cref{tab:roots}.
We consider the root $2\Gamma_1 + 2M_2 + 2X_2$.
We can write this representation as difference of EBRs as $2(A)_b \up G + ({^1E^2E})_b \up G - ({^1E^2E})_a \up G$, where $b=(1/2,1/2)$, $a=(0,0)$ are the two $C_4$ invariant Wyckoff positions, and
\begin{equation}
    ({^1E^2E})_a \up G = \Gamma_3\Gamma_4 + M_3M_4 + 2 X_1,
\end{equation}
\begin{equation}
    (A)_b \up G = \Gamma_1 + M_2 + X_2,
\end{equation}
\begin{equation}
    (^1E^2E)_b \up G= \Gamma_3\Gamma_4 + M_3M_4 + 2 X_2.
\end{equation}
(One can find the complete EBRs on the \href{http://www.cryst.ehu.es/cgi-bin/cryst/programs/bandrep.pl}{Band Representations of the Double Space Groups} section \cite{Elcoro2017} on the BCS \cite{BCS1}.)
The $^1E^2E$ irreps at $a,b$ are formed by $p_{x,y}$ orbitals at $a,b$, respectively, and the $A$ irrep at $b$ is formed by the $s$ orbital at $b$. (See \cref{fig:TB-p4}.)
We set the onsite energy of the $p$ orbitals as $E$ and the onsite energy of the two $s$ orbitals as $-E$.
To create band inversion at $X$ such that the lower two bands become the EFP,  we introduce the next nearest hopping $t_1$ between $p$ orbitals and the next nearest hopping $-t_1$ between $s$ orbitals, as shown in \cref{fig:TB-p4}(a).
The dispersion of the $p$ orbital bands is $E+2t_1(\cos(k_x+k_y)+\cos(k_x-k_y))$ (two-fold degenerate), and the dispersion of the $s$ orbital bands is $-E-2t_1(\cos(k_x+k_y)+\cos(k_x-k_y))$ (two-fold degenerate).
We choose $E=1$ and $t_1=1/2$ such that the $p$ bands cross with the $s$ bands at two rings surrounding $X$ and $Y$.
We introduce the $s$-$p$ hybridization hopping $t_2$ (\cref{fig:TB-p4}(b)) to open a gap.
The Hamiltonian in momentum space is given by
\begin{equation}
H(\kk) = \tau_z\sigma_0 (E + 2t_1\cos(k_x+k_y) + 2t_1\cos(k_x-k_y)) + 
\tau_y \sigma_z t_2 \sin(k_x) + \tau_y \sigma_x t_2 \sin(k_y). \label{eq:TB-p4}
\end{equation}
Here $\tau_i$ and $\sigma_i$ ($i=1,2,3$) are the three Pauli matrices, $\tau_0$ and $\sigma_0$ are $2\times 2$ identity matrices, $\tau_i\sigma_j$ represents the tensor product of $\tau_i$ and $\sigma_j$, and the order of orbitals are $p_x$, $p_y$, $s_1$, and $s_2$.
The $C_4$ symmetry operator is
\begin{equation}
    \hat{C}_4 = \begin{pmatrix}
    -i\sigma_y & 0\\ 0 & \sigma_0
    \end{pmatrix},
\end{equation}
and the TRS operator is $\TRS = K$, with $K$ the complex conjugate.
The Hamiltonian (\ref{eq:TB-p4}) commutes with the matrix $\tau_z\sigma_y$.
To break this accidental symmetry, we introduce an onsite coupling ($t_3$) of the $s_1$ and $s_2$ orbitals, and a nearest neighbor hopping between $p$ orbitals (\cref{fig:TB-p4})(c).
The momentum space form of this perturbation term is
\begin{equation}
    \Delta H(\kk) = \begin{pmatrix}
        2t_4 \cos(k_x) & 0 & 0 & 0\\
        0 & 2t_4\cos(k_y) & 0 & 0\\
        0 & 0 & 0 & t_3 \\
        0 & 0 & t_3 & 0
    \end{pmatrix}. \label{eq:DHk}
\end{equation}

We choose the parameters as $E=1$, $t_1=t_2=t_3=1/2$, $t_4=1/5$.
The band structure is shown in \cref{fig:TB-TBC}(b).
The two occupied bands form the irreps $2\Gamma_1$, $2M_2$, $2X_2$ at $\Gamma$, $M$, $X$, respectively.
Now we show that the two bands do not form a BR.
All the EBRs are tabulated \cref{tab:EBR-p4}.
There are only three BRs that give the correct irreps at $\Gamma$ ($2\Gamma_1$): twice $(A)_a \up G$ or twice $(A)_b \up G$ or $(A)_a \up G \oplus (A)_b \up G$.  Here $(\rho)_w$ represents the irrep $\rho$ of the site-symmetry group of the Wyckoff position $w$, and $a$, $b$, $c$ represent the positions $(0,0)$, $(\frac12, \frac12)$, and $(0,\frac12)$. respectively. Among these three options, only two times $(A)_a \up G$ has the correct irreps at $X$ ($2X_1$); however at $M$ ($2M_1$) the irreps are different from the irreps of the root. Thus there is no way to decompose the EFP root into sum of EBRs. Instead, we can write the EFP root as a difference of EBRs as $ 2 (A)_b \up G  \oplus ({^1E^2E})_b \up G \ominus ({^1E^2E})_a \up G$.

We put the tight-binding model on a finite (30$\times$30) square lattice.
In \cref{sec:model} we showed that there were 1798 occupied levels forming the representation $450 A \oplus 450 B \oplus 449 ({}^1E{}^2E)$, 4 partially occupation levels forming the representation $A\oplus B\oplus {}^1E{}^2E$, and 1798 empty levels forming the representation $449 A \oplus 449 B \oplus 450 ({}^1E{}^2E)$.
Here we give the energies of the levels around around Fermi level.
They are
\begin{equation}
\underbrace{\cdots B (-0.19646)}_{\text{occupied}},\quad \underbrace{A (0.11035),\quad {}^1E{}^2E (0.11283), \quad B (0.11720)}_{\text{partially occupied}}, \quad \underbrace{B (0.12681), \cdots}_{\text{empty}}
\end{equation} 
where the numbers in the parethenesses represent the energies.
The gap between the occupied levels and the four partially occupied levels is about $0.3$, whereas the gap between the four partially occupied levels and the empty levels is very small ($\sim 0.01$).

\subsubsection{The RSI formulae in the wallpaper group $p4$}\label{app:p4-model-RSI}

One can immediately verify that \cref{eq:da1-main,eq:da2-main,eq:db1-main,eq:db2-main,eq:dc1-main} give correct RSIs for all the EBRs.
For example, from the above equations, the EBR $(A)_a\up G = \Gamma_1 + M_1 + X_1$ has $ \delta_{a1}=-1$, $\dt_{a2}=-1$ and $\dt_{b1}=\dt_{b2}=\dt_{c1}=0$, which are consistent with the real space calculation, where $\dt_{b1,2}$ and $\dt_c$ are zero because there is no state at $b$ or $c$, and the RSIs at $a$ are $\dt_{a1}=m({}^1E{}^2E)-m(A)=-1$ and $\dt_{a1}=m(B)-m(A)=-1$.
For the EFP root $2\Gamma_1 + 2M_2 + 2 X_1$ considered in \cref{sec:model}, the above equations give the RSIs $\dt_{a1}=-1$, $\dt_{a2}=0$, $\dt_{b1}=-1$, $\dt_{b2}=-2$, $\dt_{c1}=0$. We can also understand this result from the EBR decomposition. Since the EFP root can be written as $2 (A)_b \up G  \oplus ({^1E^2E})_b \up G \ominus ({^1E^2E})_a \up G$, the RSIs at $a$ are $\delta_{a1}=m(({}^1E{}^2E)_a)-m((A)_a)=1$, $\delta_{b1}=m((B)_a)-m((A)_a)=1$, the RSIs at $b$ are $\delta_{b1}=m(({}^1E{}^2E)_b)-m((A)_b)= -1$, $\delta_{b1}=m((B)_b)-m((A)_b)= -2$, and the RSIs at $c$ are zero.

\subsubsection{The TBCs of the model}\label{app:p4-model-TBC}
We divide the sites into four parts and implement the TBC \cref{eq:TBC-main2} (\cref{fig:TB-TBC}(c)).
First we explain why the factor between the $\mu$th part and the $(\mu+2)$th part should be real.
Assume $\bra{\mrm{III},\alpha} \hH(\lambda) \ket{\mrm{I},\beta} = c \bra{\mrm{III},\alpha} \hH(1) \ket{\mrm{I},\beta}$. 
Denote $C_2 \ket{\mrm{III},\alpha}$ as $\ket{\mrm{I},\alpha^\pr}$ and $C_2 \ket{\mrm{I},\beta}$ as $\ket{\mrm{III},\beta^\pr}$. 
By the definition of TBC, there is $\bra{\mrm{I},\alpha^\pr} \hH(\lambda) \ket{\mrm{III},\beta^\pr} = c^*\bra{\mrm{I},\alpha^\pr} \hH(1) \ket{\mrm{III},\beta^\pr}$. 
On the other hand, due to the $C_2$ symmetry, we have $\bra{\mrm{I},\alpha^\pr} \hH(\lambda) \ket{\mrm{III},\beta^\pr} =\bra{\mrm{III},\alpha} \hH(\lambda) \ket{\mrm{I},\beta} = c\bra{\mrm{III},\alpha} \hH(1) \ket{\mrm{I},\beta} = c\bra{\mrm{I},\alpha^\pr} \hH(1) \ket{\mrm{III},\beta^\pr} $.Thus there must $c=c^*$.
Then we verify that $\hH(\lambda)$ defined in \cref{eq:TBC-main2} preseves the $C_4$ symmetry. 
For example, since $\hat{C}_4 \ket{\mrm{I},\beta}$ belongs to part II and $\hat{C}_4 \ket{\mrm{II},\beta}$ belongs to part III, we obtain $\bra{\mrm{II},\alpha} \hat{C}^\dagger_4 \hat{H}(\lambda) \hat{C}_4 \ket{\mrm{I},\beta} = \lambda^* \bra{\mrm{II},\alpha}\hat{C}^\dagger_4 \hat{H}(1) \hat{C}_4 \ket{\mrm{I},\beta} = \lambda^* \bra{\mrm{II},\alpha} \hat{H}(1) \ket{\mrm{I},\beta} = \bra{\mrm{II},\alpha} \hat{H}(\lambda) \ket{\mrm{I},\beta}$ according to \cref{eq:TBC-main2}. Similarly one can verify $\hat{C}^\dagger_4 \hH(\lambda) \hat{C}_4 = \hH(\lambda)$ for all matrix elements. $\hH(i)$ breaks TRS unless $\lambda$ is a real number, but this is unimportant as  $2\Gamma_1 + 2M_2 + 2X_1$ is still a fragile phase even after the TRS is broken: it reduces to the 47th EFP root of $p4$ without TRS in \cref{tab:roots}.

We have discussed the evolution path $\lambda=1\to i$ of the TBC in the maintext.
Here we discuss the other two paths: $\lambda= 1\to -i$ and $\lambda=1 \to -1$. 
The path $\lambda=1\to -i$ is similar with the path $\lambda=1\to i$ and a level crossing protected by $C_4$ symmetry appears during the evolution process.
The path $\lambda=1\to -1$ does not give symmetry-protected level crossings. The gauge transformation at $\lambda=-1$ is $\hV \ket{\mu,\alpha} = (-1)^{\mu-1}\ket{\mu,\alpha}$, which anti-commutes with the $C_4$ rotation and hence interchanges $A$ with $B$ and $^1E$ with $^2E$. 
Therefore, the representation of the finial occupied (empty) states is the same as the initial occupied (empty) states, and there is no symmetry-protected level crossings in the process $\lambda=1\to -1$. 
(As discussed in detail in \cref{sec:TBC-2D}, $\lambda=1\to -1$ can lead to gap closing in other EFPs.)
We summarize the results as 
\begin{equation}  \begin{tabular}{c|c|c|c} $\lambda$ & $1\to -i$ & $1\to -1$ & $1\to i$\\\hline\hline $\Delta m(A)$ & $\dt_1$ & $\dt_2$ & $\dt_1$\\ $\Delta m({}^1E)$ & $- \dt_1$ & $0$ & $\dt_2-\dt_1$\\ $\Delta m(B)$ & $\dt_1-\dt_2$ & $-\dt_2$ & $\dt_1-\dt_2$\\ $\Delta m({}^2E)$ & $\dt_2-\dt_1$ & $0$ & $-\dt_1$\\ 
\end{tabular},
\end{equation} 
which is same as \cref{eq:path-4-NSOC-NTRS}.)
For the $C_2$ and TRS-symmetric TBC (\cref{fig:TB-TBC}(d)) we  do the same analysis and obtain  
\begin{equation} \begin{tabular}{c|c} $\lambda \in \mbb{R}$ & $1\to -1$ \\ \hline\hline $\Delta m(A)$ & $2\delta_1-\dt_2$ \\ $\Delta m(B)$ & $-2\delta_1+\dt_2$
\end{tabular},
\end{equation} 
which is same as \cref{eq:path-2-NSOC-NTRS}.
In all cases, the types and numbers of protected level crossings under TBCs are completely determined by the RSIs.


\clearpage
\LTcapwidth=0.97\textwidth
\begin{longtable}{|c|c|L{12cm}|}
\caption{\label{tab:WGRSI} RSIs in wallpaper groups in terms of momentum space irreps. 
            Only the RSIs that can be determined from irreps are tabulated. }\\
\hline
\hline
\hline
WG & WP &  RSI  \\
\hline
\endfirsthead
\multicolumn{3}{r}{\cref{tab:WGRSI} -- continued} \\
\hline
WG & WP &  RSI  \\
\hline
\endhead
\hline
\multicolumn{3}{r}{continued on next page} \\
\endfoot
\hline
\hline
\endlastfoot
\multicolumn{3}{c}{ NSOC, NTRS }\\
\hline
$ p1 $ & & \\
\hline
\multirow{ 4 }{*}{$ p2 $}
& \multirow{ 1 }{*}{$ a $ ($ 2 $)}
& $ \delta_{1} = \frac{1}{2} m(\mathrm{A_{2}}) -\frac{1}{2} m(\mathrm{B_{1}}) -\frac{1}{2} m(\mathrm{\Gamma_{1}}) + \frac{1}{2} m(\mathrm{Y_{2}}) $\\
\cline{2-3}
& \multirow{ 1 }{*}{$ b $ ($ 2 $)}
& $ \delta_{1} = -\frac{1}{2} m(\mathrm{A_{2}}) + \frac{1}{2} m(\mathrm{B_{1}}) -\frac{1}{2} m(\mathrm{\Gamma_{1}}) + \frac{1}{2} m(\mathrm{Y_{2}}) $\\
\cline{2-3}
& \multirow{ 1 }{*}{$ c $ ($ 2 $)}
& $ \delta_{1} = m(\mathrm{A_{1}}) + \frac{1}{2} m(\mathrm{A_{2}}) -\frac{1}{2} m(\mathrm{B_{1}}) -\frac{1}{2} m(\mathrm{\Gamma_{1}}) -\frac{1}{2} m(\mathrm{Y_{2}}) $\\
\cline{2-3}
& \multirow{ 1 }{*}{$ d $ ($ 2 $)}
& $ \delta_{1} = \frac{1}{2} m(\mathrm{A_{2}}) + \frac{1}{2} m(\mathrm{B_{1}}) -\frac{1}{2} m(\mathrm{\Gamma_{1}}) -\frac{1}{2} m(\mathrm{Y_{2}}) $\\
\hline
\multirow{ 2 }{*}{$ p1m1 $}
& \multirow{ 1 }{*}{$ a $ ($ m $)}
& $ \delta_{1} = m(\mathrm{C_{2}}) - m(\mathrm{Y_{1}}) $\\
\cline{2-3}
& \multirow{ 1 }{*}{$ b $ ($ m $)}
& $ \delta_{1} = - m(\mathrm{C_{2}}) + m(\mathrm{\Gamma_{2}}) $\\
\hline
\end{longtable}

\clearpage
\twocolumngrid
\LTcapwidth=0.97\textwidth
\begin{longtable}{|c|l|}
\caption{\label{tab:FC} Fragile criteria in wallpaper groups in terms of RSIs.}\\
\hline
\hline
\hline
 & Criteria \\
\hline
\endfirsthead
\multicolumn{2}{r}{ \cref{tab:FC} -- continued} \\
\hline
 & Criteria \\
\hline
\endhead
\hline
\multicolumn{2}{r}{continued on next page} \\
\endfoot
\hline
\hline
\endlastfoot
\multicolumn{2}{c}{$ p1 $,  NSOC, NTRS }\\
\hline
\multicolumn{2}{c}{$ p2 $,  NSOC, NTRS }\\
\hline
1 & $ \delta_{a1} - \delta_{b1} - \delta_{c1} - \delta_{d1}>N_{\mathrm{occ}} $ \\
\hline
2 & $ - \delta_{a1} + \delta_{b1} - \delta_{c1} - \delta_{d1}>N_{\mathrm{occ}} $ \\
\hline
3 & $ \delta_{a1} + \delta_{b1} - \delta_{c1} + \delta_{d1}>N_{\mathrm{occ}} $ \\
\hline
4 & $ - \delta_{a1} - \delta_{b1} - \delta_{c1} + \delta_{d1}>N_{\mathrm{occ}} $ \\
\hline
5 & $ - \delta_{a1} - \delta_{b1} + \delta_{c1} - \delta_{d1}>N_{\mathrm{occ}} $ \\
\hline
6 & $ - \delta_{a1} + \delta_{b1} + \delta_{c1} + \delta_{d1}>N_{\mathrm{occ}} $ \\
\hline
\end{longtable}

\LTcapwidth=0.97\textwidth
\begin{longtable}{|c|l|}
\caption{\label{tab:roots} Fragile roots in wallpaper groups.}\\
\hline
\hline
\hline
 & Roots \\
\hline
\endfirsthead
\multicolumn{2}{r}{ \cref{tab:roots} -- continued} \\
\hline
 & Roots \\
\hline
\endhead
\hline
\multicolumn{2}{r}{continued on next page} \\
\endfoot
\hline
\hline
\endlastfoot
\multicolumn{2}{c}{$ p1 $,  NSOC, NTRS }\\
\hline
\multicolumn{2}{c}{$ p2 $,  NSOC, NTRS }\\
\hline
1 & $ 2 \mathrm{A_{2}} + 2 \mathrm{B_{1}} + 2 \mathrm{\Gamma_{2}} + 2 \mathrm{Y_{2}} $ \\
\hline
2 & $ 2 \mathrm{A_{2}} + 2 \mathrm{B_{1}} + 2 \mathrm{\Gamma_{1}} + 2 \mathrm{Y_{1}} $ \\
\hline
3 & $ 2 \mathrm{A_{1}} + 2 \mathrm{B_{2}} + 2 \mathrm{\Gamma_{2}} + 2 \mathrm{Y_{2}} $ \\
\hline
4 & $ 2 \mathrm{A_{2}} + 2 \mathrm{B_{2}} + 2 \mathrm{\Gamma_{2}} + 2 \mathrm{Y_{1}} $ \\
\hline
5 & $ 2 \mathrm{A_{2}} + 2 \mathrm{B_{2}} + 2 \mathrm{\Gamma_{1}} + 2 \mathrm{Y_{2}} $ \\
\hline
6 & $ 2 \mathrm{A_{1}} + 2 \mathrm{B_{1}} + 2 \mathrm{\Gamma_{2}} + 2 \mathrm{Y_{1}} $ \\
\hline
\end{longtable}

\end{document}